\documentclass[11pt]{article}
\pdfoutput=1
\usepackage{jcappub,natbib}
\usepackage{framed}
\input{colordvi.tex}

\def\ga{\mathrel{\raise.3ex\hbox{$>$\kern-.75em\lower1ex\hbox{$\sim$}}}}
\def\la{\mathrel{\raise.3ex\hbox{$<$\kern-.75em\lower1ex\hbox{$\sim$}}}}

\title{Nonlinear perturbations from axion-gauge fields dynamics during inflation
}

\author[a]{Alexandros Papageorgiou,}
\author[b,c]{Marco Peloso,}
\author[d]{Caner  \"Unal}

\affiliation[a]{School of Physics and Astronomy, and Minnesota Institute for Astrophysics, University of Minnesota, Minneapolis, 55455 (USA)}
 
\affiliation[b]{Dipartimento di Fisica e Astronomia G. Galilei, Universit`a degli Studi di Padova, via Marzolo 8, I-35131, Padova (Italy)}

\affiliation[c]{INFN, Sezione di Padova, via Marzolo 8, I-35131 Padova, Italy} 

\affiliation[d]{ CEICO, Institute of Physics of the Czech Academy of Sciences, Na Slovance 1999/2, 182 21 Prague, Czechia}

\abstract{
We study a variant of the Chromo-Natural Inflation (CNI) mechanism in which the inflaton interacts only gravitationally with the CNI fields. Integrating out all the non-dynamical scalar fields of the model results in a coupling between the perturbations of the inflaton and of the CNI pseudo-scalar which is significantly greater than the one obtained in the absence of the gauge CNI dynamics. We compute how this greater coupling impacts the power spectrum of the inflaton perturbations that are sourced nonlinearly by the unstable (tensor) gauge CNI modes, and we require that the amplitude of these modes is well below that of the linear perturbations. Combining this result with various constraints, including backreaction effects, the requirement of having observable and dominant sourced gravitational waves (GW), and the current upper bound on the tensor-to-scalar ratio, significantly constrains the range of parameter space where this model can produce an interesting GW signal. 
}

\begin{document}

\maketitle
\flushbottom

\section{Introduction }
 \label{sec:intro} 
In axion, or natural, inflation the flatness of the inflaton potential is maintained by an approximate shift symmetry \cite{Freese:1990rb}(see \cite{Pajer:2013fsa} for a review). In the minimal realizations, compatibility with observations requires a trans-Planckian axion scale which seems to be at odds with quantum gravity and string theory \cite{Banks:2003sx}. Several ways have been proposed in the literature to overcome this problem (see \cite{ArkaniHamed:2003wu,Kim:2004rp,Dimopoulos:2005ac,Silverstein:2008sg,McAllister:2008hb,Kaloper:2008fb,Marchesano:2014mla,Bachlechner:2014gfa,Kappl:2015esy,Choi:2015aem,Parameswaran:2016qqq}). The works \cite{Anber:2009ua,Adshead:2012kp} considered the possibility that a sub-Planckian inflaton range can be due to the interactions of the inflaton with a gauge field.~\footnote{See ref.  \cite{Maleknejad:2012fw} for a review on the role of vector fields during inflation.} 

The case of a pseudo-scalar inflaton coupling to an Abelian U(1) gauge field with a non-vanishing vacuum expectation value (vev) was first proposed in \cite{Anber:2009ua}. The coupling $\chi F {\tilde F}$ (where $\chi$ is the axion inflaton, $F$ the gauge field strength, and ${\tilde F}$ its dual) leads to a very rich phenomenology. As the inflaton evolves, one polarization of the gauge field is amplified while the other one remains small. This amplified polarization in turn, before being diluted away due to the expansion of the universe, sources both scalar and tensor perturbations, through its nonlinear interaction $\delta A + \delta A \rightarrow \delta \chi$ with the inflaton field and $\delta A + \delta A \rightarrow \delta g$ with the metric \cite{Barnaby:2010vf}. Some of the effects that arise from the above interactions include CMB non-Gaussianity  \cite{Barnaby:2010vf,Barnaby:2011vw}, large scalar power spectrum at CMB scales   \cite{Meerburg:2012id}, gravitational waves  at interferometer scales ~\cite{Cook:2011hg,Barnaby:2011qe,Domcke:2016bkh,Garcia-Bellido:2016dkw,Bartolo:2016ami}, parity violation in the CMB~\cite{Sorbo:2011rz} and in interferometers~\cite{Crowder:2012ik}, primordial black holes~\cite{Linde:2012bt,Bugaev:2013fya,Erfani:2015rqv,Cheng:2015oqa,McDonough:2016xvu,Domcke:2017fix,Garcia-Bellido:2017aan}, and large and parity violating tensor bispectra~\cite{odd-TTT}. The limits of  validity of perturbation theory for these models is also well studied (see \cite{Ferreira:2015omg,Peloso:2016gqs}). These studies are performed in the regime of negligible backreaction of the gauge fields on the background evolution. In this regime perturbativity is respected \cite{Peloso:2016gqs}.

A similar model where the Abelian U(1) field is replaced by a non-Abelian SU(2) triplet of vector fields was proposed in \cite{Adshead:2012kp}. The vector fields have non-vanishing spatial vevs arranged in such a way as to lead to isotropic expansion \cite{Maleknejad:2011jw,Maleknejad:2011sq} and they are interacting with the inflaton by an identical term to the U(1) case. In order to respect the cosmological principle the spatial vevs have to be orthogonal to each other and of equal magnitude. This model has been assigned the name "Chromo-Natural Inflation" (CNI) and it shares many similarities to "Gauge-Flation" \cite{Maleknejad:2011jw}. More specifically Gauge-Flation arises as a limit of CNI when the inflaton is close to the bottom of its potential and then integrated out \cite{Adshead:2012qe,SheikhJabbari:2012qf}. The theory of cosmological perturbations for this model was initially studied in \cite{Dimastrogiovanni:2012st} in a low-energy effective description of the model, and then in \cite{Dimastrogiovanni:2012ew,Adshead:2013qp,Adshead:2013nka,Papageorgiou:2018rfx,Maleknejad:2018nxz} in the full model. In (\ref{params}) the quantity $m_Q$ is defined. This parameter can be viewed as a type of "particle production parameter" (analogous to the parameter $\xi$ in the Abelian U(1) case; in fact the two parameters are equal to each other in the large $m_Q$ limit) and it quantifies the strength of the particle production of the gauge field during inflation. The study of the perturbations at the linear level shows that this model is unstable for $m_Q<\sqrt{2}$ and it is outside the allowed $n_s - r$ region in the complementary regime (where $n_s$ is the spectral tilt, and $r$ the tensor-to-scalar ratio).~\footnote{As a consequence, one should expect that also Gauge-Flation is incompatible with data, as confirmed by the analysis of \cite{Namba:2013kia}.}  

Since the original CNI model appears to be incompatible with the data, there has been a number of attempts to build models that share similar favorable features as the original model but are different enough that they are not in tension with experimental observations. Such models include the presence of a second axion inflaton \cite{Obata:2014loa} or a dilaton  \cite{Obata:2016tmo}, a different inflation potential \cite{Obata:2016xcr,Caldwell:2017chz,DallAgata:2018ybl,Fujita:2018ndp}, realizations in which the axion field is not the inflaton \cite{Dimastrogiovanni:2016fuu,McDonough:2018xzh}, and a spontaneous breaking of the SU(2) symmetry \cite{Adshead:2016omu}. 

\begin{figure}[tbp]
\centering 
\includegraphics[width=1\textwidth,angle=0]{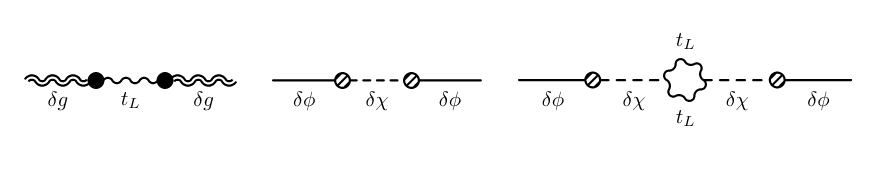}
\caption{Diagrammatic representation of the GW power spectrum sourced by the enhanced mode $t_L$ (left diagram), inflaton power spectra sourced by the linear perturbations of the axion field  (middle diagram) (which, as we show, can be neglected) and inflaton power spectra sourced by the nonlinear perturbations of the axion (produced by the enhanced tensor mode $t_L$) (right diagram). The evaluation of the right diagram is the new result obtained in this work. 
}
\label{fig:diagrams}
\end{figure}

In this work we will focus on the model proposed in \cite{Dimastrogiovanni:2016fuu}. In this model, the axion $\chi$ that couples to the SU(2) gauge field is a different field from the inflaton $\phi$. The axion $\chi$ and the gauge fields $A^a_\mu$ are spectator fields for the purpose of the background inflationary dynamics. However, as we also show in this work, they can impact the primordial tensor and scalar perturbations of this model.~\footnote{As we discuss in Appendix \ref{app:curva}, we also assume that the axion becomes massive before the end of inflation, so that its energy density redshifts away, and it provides a negligible direct contribution to the observed curvature perturbation.}  Adding a separate inflationary sector (coupled to the CNI sector only through gravity) releases the tension with the acceptable $n_s - r$ range that is found for the original CNI model. In Ref. \cite{Dimastrogiovanni:2016fuu} it is shown that for a range of the model parameters one can generate a chiral power spectrum of gravitational waves that is greater than the standard vacuum result (the one obtained in absence of the gauge fields) without disturbing the dynamics of inflation. This production allows to violate the standard relation 
\begin{equation}
\left( \frac{r}{0.01} \right)^{1/4} \Big\vert_{\rm standard} \simeq   \left( \frac{V^{1/4}}{10^{16} \, {\rm GeV}} \right) \;, 
\label{r-V}
\end{equation} 
where $V$ is the potential energy during inflation, that is valid under the assumption of standard vacuum GW \cite{Baumann:2008aq}. The possible production of this additional and chiral GW background is an extremely interesting aspect of this class of models~\cite{Adshead:2013qp,Maleknejad:2016qjz,Agrawal:2017awz,Caldwell:2017chz,Thorne:2017jft,Agrawal:2018mrg}. In these models one of the polarizations of the gauge tensor perturbations of a definite chirality, denoted by $t_L$, is amplified and it in turn sources gravitational waves linearly as is shown in the left diagram of Figure \ref{fig:diagrams}. This mechanisms differs substantially from the one in the U(1) case where the generation of the chiral GW background happens nonlinearly through the channel $\delta A_L + \delta A_L \rightarrow \delta g_L$ \cite{Barnaby:2010vf,Sorbo:2011rz}.~\footnote{See also  \cite{Adshead:2018doq} for the production of gravitational waves at preheating in this class of models.}

To gain a full knowledge of the phenomenology of the model requires the computation also of the scalar perturbations. In particular, we want to understand how the enhanced  $t_L$ mode can impact the inflaton perturbations through nonlinear interactions. A number of steps towards understanding the nonlinear dynamics of the model  \cite{Dimastrogiovanni:2016fuu}  have been taken in the recent literature  \cite{Agrawal:2017awz,Agrawal:2018mrg,Agrawal:2018gzp,Fujita:2018vmv,Dimastrogiovanni:2018xnn}, that computed the tensor and scalar-tensor mixed bispectra. 

The impact of nonlinearities on the scalar spectra has not yet been computed, as it requires the more complicated evaluation of a one-loop diagram. This is the computation performed in the present work. 
The computation is heavily based on our previous work  \cite{Papageorgiou:2018rfx} in which we computed the 
analogous one-loop production of the axion perturbations in the original CNI model. The most substantial addition with respect to that work is that, in the present case, the produced axion modes are not external, but they in turn propagate and source the inflaton perturbations. This is diagrammatically represented by the third diagram in Figure  \ref{fig:diagrams}. The inflaton and axion fields are not directly coupled to each other, and, at the technical level, the coupling arises by integrating out the non-dynamical scalar modes of the model. In inflationary models without gauge fields the coupling arises by integrating out the $\delta g_{00}$ metric perturbation. This results in a $\sqrt{\epsilon_\chi \, \epsilon_\phi} H^2 \delta \chi \delta \phi$ interaction, where $H$ is the Hubble rate, and where $\epsilon_{\chi,\phi}$ are the two standard slow-roll parameters associated to the motion of the fields. This coupling was present, for example, in the analogous U(1) spectator models studied in \cite{Ferreira:2014zia,Namba:2015gja}. This coupling is present also for the model of our interest, and it is the only coupling included in the existing analysis of the model. However, the dynamics of the CNI sector is strongly influenced by the gauge fields. As a consequence, also the scalar part of the nondynamical $\delta A_0^a$ modes should be included in the computation, and integrated out. This introduces additional couplings between the axion and the inflation perturbations, that can be up to order ${\cal O}(10^3)$ times larger than the one considered so far. 

This increased coupling results in a greater production of the inflaton perturbations, that we compute through the third diagram in Figure \ref{fig:diagrams}.~\footnote{In addition, integrating out the non-dynamical scalar perturbations produces also a direct $t_L \, t_L \, \delta \phi$ interaction. In Appendix \ref{app:subdominant} we show that this interaction results gives a subdominant contribution to the inflaton perturbations.} In turn, requiring that these sourced scalar perturbations are significantly smaller than the linear inflaton perturbations reduces the allowed region of parameter space of the model. We discuss this constraint, together with other requirements on the model. 

The plan of this paper is the following. In Section \ref{sec:background} we give a brief overview of the model and its background equations of motion, and we define the relevant parameters. In section \ref{sec:linearpertSCNI} we summarize the  differences in the theory of cosmological perturbations between the model of this paper and the original CNI model, and we derive the complete linear coupling between the axion and the inflaton perturbations. Section \ref{sec:nonlinear} is a detailed application of the in-in formalism in order to compute the rightmost diagram in Figure \ref{fig:diagrams}. Next, in Section \ref{sec:results} we combine the results of this paper with the study of the linear production of chiral gravitational waves carried out in \cite{Dimastrogiovanni:2016fuu}, and we plot the emerging limits on the parameter space of the model. Finally we conclude our work with a summary of the computation as well as a discussion about possible future work on this model and related ones. The paper is supplemented by several Appendices, where we confine the most technical aspects of our computations.

\section{The model, and the background evolution} 
\label{sec:background} 

In this Section we review the background evolution of a model in which an SU(2) gauge field 
carrying a nonvanishing vacuum expectation value (vev) is coupled to a rolling axion field which is not the inflaton, and which gives a negligible contribution to the inflationary expansion \cite{Dimastrogiovanni:2016fuu}. The model is characterized by the action 
\begin{equation}
S= \int d^4x \sqrt{-g} \left[ \frac{M_p^2}{2}R  -\frac{1}{2}\left(\partial \phi \right)^2 -V(\phi)  \, -\frac{1}{4}F^a_{\mu \nu}F^{a , \mu \nu} -\frac{1}{2}\left(\partial \chi \right)^2 -U(\chi)- \frac{\lambda}{8 \sqrt{-g} \, f} \chi \, \epsilon^{\mu \nu \alpha \beta} F^a_{\mu \nu}F^{a}_{\alpha \beta}  \right] \;, 
\label{SCNI}
\end{equation}
where $M_p = \sqrt{\frac{1}{8 \pi G_N}}$ is the reduced Planck mass, and where $\phi$ denotes the inflaton field, with a potential $V$ (which we do not need to specify in this work), while $\chi$ is the pseudo-scalar axion, with potential 
\begin{equation}
U \left( \chi \right) = \mu^4 \left[ 1 + \cos\left( \frac{\chi}{f} \right) \right] \;, 
\end{equation}
which is coupled to a SU(2) gauge field of field strength $F^a_{\mu \nu}=\partial_\mu \, A^a_\nu - \partial_\nu \, A^a_\mu  + g \, \epsilon^{abc} A^b_\mu \, A^c_\nu \,$. In the coupling term, the tensor $\epsilon^{\mu \nu \alpha \beta}$ is totally anti-symmetric, and it is normalized to $\epsilon^{0123}=1$. The vector field has the vev 
\begin{equation}
\left\langle A^a_0 \left( t \right) \right\rangle =0, \quad \left\langle A^a_i \left( t \right) \right\rangle = \delta^a_i \, a \left( t \right) \,  Q \left( t \right)  \;, 
\end{equation}
which is compatible with an isotropic expansion. In this expression, $a = \left\{ 1 ,\, 2 ,\, 3 \right\}$ is the SU(2) index, while the indices $0$ and $i= \left\{ 1 ,\, 2 ,\, 3 \right\}$ refer to the time and space components, respectively. 
We take the line element $ds^2 = - d t^2 + a^2 \left( t \right) d \vec{x}^2 = a^2 \left( \tau \right) \left[ - d \tau^2 + d \vec{x}^2 \right]$. The scale factor $a \left( t \right)$ has been included in the parametrization of  the vector vev, since  $Q \left( t \right)$ is the quantity that is slowly evolving during inflation. 

The 00 component of the Einstein equations for the model reads, 
\begin{eqnarray}
&& 3H^2 M_p^2  = \frac{1}{2}\dot \chi^2 + U(\chi)+ \frac{1}{2}\dot \phi^2 + V(\phi) + \frac{3}{2} \left[  \left( \dot Q + HQ \right)^2+ g^2 Q^4  \right] \;,  
\end{eqnarray} 
with dot denoting derivative with respect to the time $t$. In addition, we have the following evolution equations for the inflaton, the axion, and the gauge field
\begin{eqnarray}
&& \ddot \phi + 3 H \dot \phi + V'(\phi) = 0\;, \nonumber\\
&& \ddot \chi + 3 H \dot \chi + U'(\chi)  +  \frac{3 \,\lambda \, g}{f} \,Q^2  \left( \dot Q  + H Q  \right) =0 \;, \nonumber\\
&& \ddot Q + 3 H \dot Q + (\dot H +2 H^2) Q + g\, Q^2 \left( 2 g Q - \frac{\lambda \dot \chi}{f} \right) = 0 \;, 
\label{eq-bck}
\end{eqnarray}
where a prime on a potential term denotes a derivative with respect to its argument. Combining the 00 and the ii components of the Einstein equations so to eliminate the potential terms, we can also write the exact relation 
\begin{equation}
\epsilon_H =\epsilon_\chi+\epsilon_\phi+\epsilon_B+\epsilon_E \;, 
\label{slow-relation}
\end{equation}
where all the quantities
\begin{equation}
\epsilon_H \equiv -\frac{\dot{H}}{H^2} \;,\;\;
\epsilon_\chi \equiv \frac{\dot{\chi}^2}{2H^2 M_p^2} \;,\;\;
\epsilon_\phi \equiv \frac{\dot{\phi}^2}{2H^2 M_p^2} \;,\;\;
\epsilon_B \equiv \frac{g^2 Q^4}{H^2 M_p^2} \;,\;\; 
\epsilon_E \equiv \frac{\left(H Q+ \dot{Q}\right)^2}{H^2 M_p^2} \;, 
\label{slow}
\end{equation}
are  much smaller than unity during inflation. 

The slow roll parameters (\ref{slow}) modify $\dot{H}$ as shown by eq. (\ref{slow-relation}). This in turn affects the spectral tilt of the linear scalar perturbations \cite{Fujita:2017jwq} 
\begin{equation}
n_s - 1 = 2 \left( \eta_\phi - 3 \epsilon_\phi -  \epsilon_B - \epsilon_E - \epsilon_\chi \right) \simeq 
2 \left( \eta_\phi - 3 \epsilon_\phi -  \epsilon_B \right) \;, 
\label{ns}
\end{equation} 
(with $\eta_ \phi \equiv \frac{V''}{3 H^2}$; an analogous definition applies to $\eta_\chi$) where in the last step we have used the fact that  $\epsilon_B$ is the dominant CNI slow roll parameter. One can impose  \cite{Dimastrogiovanni:2016fuu} that its contribution to $n_s$ is negligible by assuming that $\epsilon_\phi$ is the dominant parameter in (\ref{slow}). The condition $ \epsilon_\phi \gg \epsilon_B$ restricts significantly the allowed range for $\epsilon_B$ (this has an impact on the phenomenological range that we study in Figure \ref{figparameterspace} below). Ref. \cite{Fujita:2017jwq} instead only requested that $\epsilon_B$ is smaller than about $0.02$, not to introduce any tuning in eq. (\ref{ns}) (given that the measured value for $n_s=1$ is about $-0.04$). We follow this second approach, as it is less restrictive on the allowed region for $\epsilon_B$.~\footnote{Analogous considerations were made in  \cite{Peloso:2016gqs} for the abelian case.}

We also impose the  slow-roll requirements 
\begin{equation}
\ddot{\phi}\ll H\dot{\phi} \;,\;\; \ddot{\chi}\ll H\dot{\chi} \;,\;\; \ddot{Q}\ll H\dot{Q} \;.
\end{equation}
Finally, we require that the  inflaton dominates the energy density of the universe. 

The axion-gauge field sector of the model is the one of CNI \cite{Adshead:2012kp}. It is therefore convenient to introduce the usual CNI parameters 
\begin{equation}
\Lambda \equiv \frac{\lambda}{f} Q \; , \;\; m_Q \equiv g \frac{Q}{H} \;, 
\label{params}
\end{equation}
and, as done for the CNI case, to restrict the analysis to the regime $\Lambda\gg \sqrt{2} \;,\; \Lambda \gg \frac{\sqrt{3}}{m_Q}$. While this choice is mandatory in the CNI model, in the present context it has the purpose of simplifying the analysis \cite{Dimastrogiovanni:2016fuu}. As long as these conditions hold, one finds 
\begin{equation}
Q_{min}\simeq\left(\frac{-f U'}{3g\lambda H}\right)^{1/3} \;,\;\; \frac{\lambda}{f H}\dot{\chi}\simeq 2\left(m_Q+\frac{1}{m_Q}\right) \;, 
\label{chiSR}
\end{equation}
as in the CNI case \cite{Adshead:2012kp}.

\section{Linear perturbations}
\label{sec:linearpertSCNI} 

In this section we give a brief summary of the linear perturbations of the model (\ref{SCNI}). As usual, the linear perturbations can be decomposed into the three tensor, vector, and scalar sectors (decoupled from each other at the linear level). The vector and tensor modes behave as in CNI. The vector sector does not play any relevant role in our discussion, and we refer the interested reader to the analysis done in \cite{Dimastrogiovanni:2012ew}. The tensor sector will be briefly described in the following Subsection \ref{sec:TensorSCNI}. One tensor perturbation (that originates from the SU(2) multiplet) is unstable in a given regime of parameters, and it sources at the linear level one metric tensor polarization. The scalar sector differs from that of CNI due to the fact that the axion ($\chi$) is not the inflaton ($\phi$) in this case. We review this sector in  Subsection \ref{sec:ScalarSCNI}. We pay particular attention to the coupling between the inflaton perturbation and the scalar perturbations of the CNI fields (the axion, and the gauge field). While this coupling affects the scalar perturbations of the CNI field in a negligible manner, it is a crucial ingredient to find how the CNI modes (sourced at the nonlinear level by the  gauge field tensor perturbations, see Section \ref{sec:nonlinear}) affect the inflaton perturbation. 

We use the convention 
\begin{equation}
\delta \left( t ,\, \vec{x} \right) = \int \frac{d^3 k}{\left( 2 \pi \right)^{3/2}} \, {\rm e}^{i \vec{k} \cdot \vec{x}} \, \delta \left( t ,\, \vec{k} \right) \;, 
\label{FT} 
\end{equation} 
for the Fourier transform of any perturbation $\delta$.

\subsection{Tensor sector} 
\label{sec:TensorSCNI} 
Tensor perturbations in this model are identical to those in the original CNI model. (The study of tensor perturbations in the CNI model was carried out first in \cite{Dimastrogiovanni:2012ew}, and then in more details in \cite{Adshead:2013qp}.) At the linearized level, modes of different momentum are not coupled to each other, and so we can orient the momentum of a mode  along the $z-$axis, without loss of generality. Doing so, the tensor perturbations of the model can be written as 
\begin{eqnarray} 
&& \delta A_\mu^1 = a \left( \tau \right) \left( 0 ,\, t_+ \left( \tau ,\, z \right) ,\, t_\times \left( \tau ,\, z \right) ,\, 0 \right) \;\;,\;\; 
\delta A_\mu^2 = a \left( \tau \right) \left( 0 ,\, t_\times \left( \tau ,\, z \right) ,\, - t_+ \left( \tau ,\, z \right) ,\, 0 \right) \;\;, \nonumber\\ 
&& \delta g_{11} = - \delta g_{22} = \frac{a^2 \left( \tau \right)}{\sqrt{2}} h_+ \left( \tau ,\, z \right) \;\;,\;\; 
\delta g_{12} = \delta g_{21} = \frac{a^2 \left( \tau \right)}{\sqrt{2}} h_\times \left( \tau ,\, z \right) \;, 
\label{tensor-modes}
\end{eqnarray} 
where $\tau$ is conformal time. Starting from these modes, we define the left handed (+) and right handed (-) helicity variables 
\begin{equation}
{\hat h}^{\pm} \equiv \frac{a M_p}{2} \frac{ h_+  \mp i h_\times }{\sqrt{2}} \equiv \frac{a M_p}{2} h_{L,R}  \;\;\;\;\;\;,\;\;\;\;\;\; 
{\hat t}^{\pm} \equiv \sqrt{2} a \, \frac{ t_+  \mp i t_\times }{\sqrt{2}} \equiv \sqrt{2} a \, t_{L,R}  \;. 
\label{tensor-canonical}
\end{equation} 
The ``hatted'' variables are canonically normalized. Moreover, the two subsets $\left\{ {\hat t}^+ ,\, {\hat h}^+ \right\}$ and $\left\{ {\hat t}^- ,\, {\hat h}^- \right\}$ are decoupled from each other at the linearized level. 

The gauge field modes ${\hat t}^{\pm}$ obey the equation  
\begin{equation} 
\frac{d^2}{d x^2} \, {\hat t}^{\pm} + \left[ 1 + \frac{2 \left( 1 + m_Q^2 \right)}{x^2} \mp \frac{2 \left( 2 m_Q + \frac{1}{m_Q} \right)}{x} \right] t^{\pm} = 0 \;\;,\;\; x \equiv - k \tau \;, 
\label{eq-t}
\end{equation} 
with negligible corrections from their interaction with the metric modes  ${\hat h}^{\pm}$. 

Treating the parameter $m_Q$ as constant, eq. (\ref{eq-t}) admits an analytic solution in terms of Whittaker functions \cite{Adshead:2013qp}. The crucial point is that, as seen from eq. (\ref{eq-t}), the mode ${\hat t}^+$ experiences a tachyonic instability for a range of times. The mode function  ${\hat t}^+$ grows during the unstable regime, and it then oscillates back around zero when this regime is over. Therefore, the time evolution of the mode function shows a bump associated with this unstable growth (see Figure 8 of \cite{Papageorgiou:2018rfx}). The bump is very well fitted by a log-normal shape. In \cite{Papageorgiou:2018rfx}, an accurate fitting function is given in the regime $m_Q<4$ which we are making use of in the present work as further elaborated in the Appendix \ref{app:semianalytic}.
\begin{equation}
 \vert t_L(x) \vert = \vert t_L \vert_{\rm peak} \cdot e^{- m_Q \log^2 \left(\frac{x}{x_{\rm p}}\right)} \; ,  \qquad \vert t_L \vert_{\rm p} \simeq  \frac{8}{3} \,\,  \sqrt{m_Q} \,\,  e^{\frac{\pi}{2} m_Q}  \qquad {\rm and} \qquad x_{\rm p} \equiv \frac{4}{9}m_Q \,. 
\label{tensorIR-main} 
\end{equation}
We found in  \cite{Papageorgiou:2018rfx}  that using the fitting relation for the bump, rather than the full Whittaker solution, speeds up considerably the numerical integration necessary to compute the nonlinear scalar perturbations (see the next section). We verified in \cite{Papageorgiou:2018rfx} that the results obtained with the fitting relation reproduce very well those obtained with the  Whittaker solution. 

The enhanced mode  ${\hat t}^+$ sources the metric perturbations ${\hat h}^+$ at the linear level.~\footnote{It can can also source a significant amount of scalar perturbations, as we discuss in the next section.}  An accurate approximate analytic solution for ${\hat h}^+$ can also be found in Ref. \cite{Adshead:2013qp}. As can be seen from eq. (\ref{eq-t}), an analogous tachyonic growth does not occur for  ${\hat t}^-$. Therefore the mode ${\hat h}^-$ is not (linearly) sourced, and it remains at the standard ``vacuum'' value. The generation of a large chiral GW background is probably the most interesting phenomenological aspect of this class of models. Ref. \cite{Dimastrogiovanni:2016fuu} provided a very accurate fitting relation for the ratio between the power of the sourced  ${\hat h}^+$ mode and the GW ``vacuum'' power spectrum, defined to be  the power in the GW modes in absence of the ${\hat t}^+$ enhancement (which is approximately twice the amount of the power spectrum of  ${\hat h}^-$)~\footnote{With this terminology, the ``vacuum modes'' are the solutions of the homogeneous differential equation for the metric tensor modes, while the ``sourced modes'' are the particular solution due to the enhanced ${\hat t}^+$ mode. We make this clarification, as one could also have denoted as ``vacuum modes'' the full solutions of the linearized theory.
}  
\begin{equation}
{\cal R}_{\rm GW} \equiv \frac{P_h^{(s)}}{P_h^{(v)}}  = \epsilon_B \, \frac{{\cal F}^2}{2} \;\;\;,\;\;\;  
{\cal F}^2 \simeq e^{3.6m_Q}  \;. 
\label{RGW}
\end{equation} 
As long as the sourced scalar perturbations can be neglected, this corresponds to the tensor-to-scalar ratio 
\begin{equation}
r= r_{\rm vac} \left(1+ {\cal R}_{\rm GW} \right) \;\;\;,\;\;\; {\rm if \; negligible \; sourced \; scalar \; perturbations} \;. 
\end{equation}

\subsection{Scalar sector} 
\label{sec:ScalarSCNI} 

Let us now discuss the scalar perturbations of the model. In this sector, the inflaton perturbations $\delta \phi$ 
add up with the scalar perturbations of the CNI fields (where now $\chi$ is an axion different from the inflaton). The number of modes is easily obtained by looking at the possible ``scalar tensorial structures''. Let us provisorily list them as $M_1 ,\, M_2 ,\, \dots$. One has: 
\begin{itemize} 

\item The inflaton perturbation $\delta \phi = M_1$ 

\item The axion perturbation $\delta \chi = M_2$ 

\item $4$ modes from the metric: one from $\delta g_{00} = M_3$, one from $\delta g_{0i}$ with tensorial structure $\partial_i M_4$; two from $\delta g_{ij}$, with tensorial structures $\delta_{ij} M_5$ and $\partial_i \partial_j M_6$. 

\item $4$ modes from the $A_\mu^a$ multiplet: one from $\delta A_0^a$ with tensorial structure $\partial_a M_7$, and three from $\delta A_i^a$ with tensorial structures  $\delta_{ia} M_8$,  $\partial_a \partial_i M_9$,  and  $\epsilon_{aib} \partial_b M_{10}$. 

\end{itemize} 

Some of these modes can be eliminated by gauge fixing. Concerning the CNI sector, different gauge choices were taken in the studies \cite{Dimastrogiovanni:2012ew} and \cite{Adshead:2013nka}. We adopt the convention of  \cite{Adshead:2013nka}, that we also used in \cite{Papageorgiou:2018rfx}. The freedom of general coordinate transformations allows to eliminate the two modes from $\delta g_{ij}$, while the SU(2) fixing eliminates one linear combination of the three modes emerging from $\delta A_i^a$. One is left with three non-dynamical modes (the two modes from the metric, and the mode from $\delta A_0^a$) and $4$ dynamical modes. 

As in the tensor sector, we can orient the momentum of the modes along the third axis without loss of generality. We then write the modes as \cite{Adshead:2013nka}.
\begin{eqnarray}
& & \delta g_{00} =  - a^2 \, 2\Phi \;\;,\;\; 
\delta g_{03} = a^2 \partial_z B  \;, \nonumber\\ 
& & \delta \phi = \frac{{\hat \Phi} }{a} \;\;,\;\; 
\delta \chi = \frac{{\hat X} }{a} \nonumber\\ 
& & \delta A_\mu^1 = \left( 0 ,\; \delta \varphi  - Z  ,\; \chi_3  ,\; 0 \right) \;, \nonumber\\ 
& & \delta A_\mu^2 = \left( 0 ,\; - \chi_3  ,\; \delta \varphi  - Z   ,\; 0 \right) \;, \nonumber\\ 
& & \delta A_\mu^3 = \left( \delta A_0^3  ,\; 0 ,\; 0 ,\; \delta \varphi + 2 Z  \right) \;, \;\; {\rm with  } \;\;\;\; \chi_3 = - \partial_z \frac{2 Z+ \delta \varphi}{2 g a Q} \;, 
\end{eqnarray} 
where the final constraint arises from the SU(2) gauge fixing (one should not confuse the non-dynamical metric perturbation $\Phi$ with the rescaled inflaton perturbation ${\hat \Phi}$). The CNI scalar sector is stable only for $m_Q > \sqrt{2}$  \cite{Dimastrogiovanni:2012ew}. This continues to be true also in the present model. We define the combinations 
\begin{equation}
{\hat X} \equiv a \, \delta \chi \;\;,\;\; {\hat Z} \equiv \sqrt{2} \left( Z - \delta \varphi \right) \;\;,\;\; 
{\hat \varphi} \equiv \sqrt{2 + \frac{x^2}{m_Q^2}} \left( \frac{\delta \varphi}{\sqrt{2}} + \sqrt{2} \, Z \right) \;\;, 
\label{scalar-canonical}
\end{equation} 
which are the canonical modes of the CNI sector in the absence of metric perturbations. 

The nondynamical modes $\Phi ,\, B ,\, \delta A_0^3$ enter in the quadratic action of the scalar perturbations without time derivatives. We integrate these modes out, following the formal procedure outlined in Section III of \cite{Himmetoglu:2009qi}.  We end up with a rather lengthy action for the canonical modes, which, in momentum space, is of the type 
\begin{equation}
S = \frac{1}{2} \int d \tau \, d^3 k \sum_{i,j=1}^4 \left[ Y_i^{'*} C_{ij}  Y_j' + \left( Y_i^{'*} K_{ij} Y_j + {\rm h.c.} \right) - 
Y_i^* \Omega_{ij}^2 Y_j \right] \,,  
\label{S2-dynamical} 
\end{equation}
where $Y = \left\{ {\hat Z} ,\, {\hat \varphi} ,\, {\hat X} ,\, {\hat \Phi} \right\} $ is the array made of the four dynamical modes, while $C ,\, K ,\, \Omega^2$ are $4 \times 4 $ matrices, which depend on background quantities as well as on the momentum $k$ of the mode. The matrices $C$ and $\Omega^2$ are hermitian, which ensures that the action is real. The extremization of this action provides the linearized equations of motion for the dynamical scalar perturbations of the model. 

The first three modes in $Y$ are the dynamical modes of the CNI model.  In the present case, the inflaton perturbation adds a fourth component to the multiplet. Let us first discuss the $3 \times 3$ restricted system setting the inflaton perturbation to zero (which coincides with the CNI system). In this case, if we ignore the contributions to the elements of the matrices in (\ref{S2-dynamical}) that arise from integrating out the metric perturbations (namely, if we set to zero the metric perturbations by hand), one obtains the set of equations written for instance in eqs. (3.17) of  \cite{Papageorgiou:2018rfx}.  As explicitly proven in \cite{Dimastrogiovanni:2012ew}, adding and then integrating out the metric perturbations provides additional contributions to the matrices $C ,\, K ,\, \Omega^2$, and then to the equation of motions, which are suppressed by higher order of the slow roll parameters with respect to the leading terms present in eqs. (3.17) of  \cite{Papageorgiou:2018rfx}. Therefore, we can ignore these contributions. 

Let us now discuss the effect of adding the inflaton perturbations to the system. The inflaton couples to the other fields in the model only gravitationally, and so the couplings between ${\hat \Phi}$ (which is also a canonical variable of the system) and the CNI dynamical scalar perturbations arise only when we integrate out the metric perturbations.  We verified that the inflaton perturbation modifies the equation of motion of the other scalar modes with terms that are slow roll suppressed. Therefore, the inflaton perturbation modifies the evolution of the CNI fields only in a negligible manner, in agreement with what already concluded in \cite{Dimastrogiovanni:2016fuu}. 

\begin{figure}[tbp]
\centering 
\includegraphics[width=0.6\textwidth,angle=0]{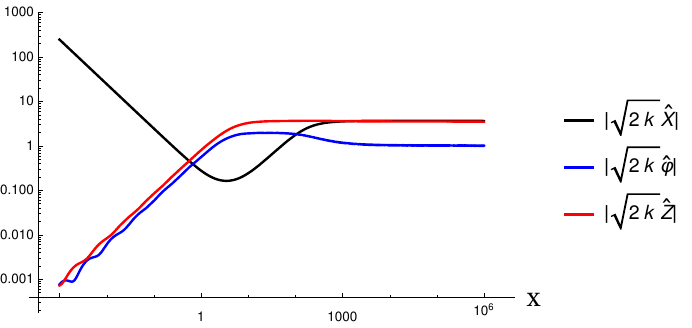}
\caption{
Evolution of the scalar perturbations (\ref{scalar-canonical}) obtained from the linearized equations (3.17) of \cite{Papageorgiou:2018rfx}, and with initial conditions (3.19) of the same reference. The parameters in the evolutions are ${\tilde f} = 0.0164 ,\, \lambda = 500 ,\, \frac{g}{{\tilde \mu}^2} = 17850$. This gives ${\bf \Lambda } \simeq 50$ and $m_Q \simeq 3.46$ at $N=50$ e-folds of inflation.} 
\label{fig:scalar-lin}
\end{figure}

In Figure \ref{fig:scalar-lin} we show the evolution of the CNI modes from the linear theory. The liner evolution is computed using all the CNI fields, as described above. We see that $\vert {\hat X} \vert$ is much greater than the other two modes outside the horizon~\footnote{The super-horizon evolution of ${\hat \varphi}$ and ${\hat Z}$
is strongly dependent on the axion background value. In the evolution shown in Figure \ref{fig:scalar-lin}, $\chi \simeq \frac{f \, \pi}{2}$ has been chosen. A different value for $\chi$ results in a different super-horizon behaviour \cite{Adshead:2013nka}; in any case, however, the amplitudes of these modes is significantly smaller than that of ${\hat X}$ in the super-horizon regime \cite{Adshead:2013nka}.} For this reason, we only studied the coupling between this mode and the inflaton perturbation ${\hat \Phi}$. Concretely, we set the perturbations in the system above to zero,~\footnote{This is done only for the purpose of computing the interaction between the axion and the inflaton perturbations. For the unperturbed mode functions used in our computation (specifically, in eq. (\ref{inin})), and entering in eqs. (\ref{definitions1}) and (\ref{definitions2}),  we employ the solution of the CNI  ${\hat X}$ scalar from the linearized theory, in which all the CNI modes are retained.}  except for the canonical perturbation of the inflaton $\hat{\Phi}$, the canonical perturbation of the axion $\hat{X}$ and the non dynamical perturbations of the metric and gauge fields $\Phi ,\, B ,\, \delta A_0^3$. The resulting action is given in Appendix \ref{app:dsdc}. We integrate out the non-dynamical variables and end up with an action that is contained in (\ref{S2-dynamical}). Specifically, it contains the terms that are present in  (\ref{S2-dynamical}), with the indices $i$ and $j$ restricted to the third and fourth component: 
\begin{equation}
S_{{\hat X}{\hat\Phi}} = \frac{1}{2} \int d \tau \, d^3 k \sum_{i,j=3}^4 \left[ Y_i^{'*} C_{ij}  Y_j' + \left( Y_i^{'*} K_{ij} Y_j + {\rm h.c.} \right) - 
Y_i^* \Omega_{ij}^2 Y_j \right] \,.   
\label{SY} 
\end{equation}

We stress that the matrices $C,K,\Omega^2$ are functions of background quantities, and of the momentum of the mode. In these expressions we assume that the background gauge field is always at the bottom of its potential by setting $\dot{Q}=0$, in agreement with  \cite{Adshead:2012kp,Dimastrogiovanni:2016fuu}. In addition to that, we are using the slow-roll approximated background equations of motion for the axion and inflaton to eliminate the derivatives of their respective potentials, and we are eliminating the second derivatives of the potentials by expressing them in terms of $\eta_\phi$ and $\eta_{\chi} $. The resulting expression is only a function of the parameters 
\begin{equation}
\frac{\lambda}{f}\;\;,\;\;g\;\;,\;\;\dot{\chi}\;\;,\;\;\dot{\phi}\;\;,\;\;Q\;\;,\;\;\eta_\phi\;\;,\;\;\eta_{\chi} \;. 
\end{equation}

This set of parameters can be replaced by the set 
\begin{equation}
\Lambda\;\;,\;\;\epsilon_{B}\;\;,\;\;\epsilon_{E}\;\;,\;\;\epsilon_{\chi}\;\;,\;\;\epsilon_{\phi}\;\;,\;\;\eta_\phi\;\;,\;\;\eta_{\chi} \;, 
\end{equation}
defined in the previous section.  The use of these parameters allows us to expand the mixed action in a well organized way, and to extract the terms that are of lowest order in powers of the slow roll parameters.

The matrix elements entering in (\ref{SY}) can be written as ratios of two polynomials of the physical momentum $p$ of the modes (in each term, the denominator arises from integrating out the nondynamical variables \cite{Himmetoglu:2009qi}). We expand the coefficients of these polynomials in slow roll. Namely, all these entries have exact expressions that are formally of the type 
\begin{equation}
\sum_{k=0}^M c_k \, p^{2 k} \Bigg/ \sum_{k'=0}^N c_{k'} \, p^{2 k'} \;, 
\label{poly-slow}
\end{equation}
and we expand each coefficient $c_k$ and $c_{k'}$ in slow roll, keeping only the leading order term for each coefficient. 

From the action (\ref{SY}), we are interested in the off-diagonal terms that couple $\hat{\Phi}$ with $\hat{X}$. 
We note that the Lagrangian in (\ref{SY}) can be written in many equivalent ways by adding a total derivative to it. We remove this arbitrariness by adding a total derivative that removes the terms that couple the time derivative of the inflaton $\hat{\Phi}'$ to the axion perturbation $\hat{X}$. Following the above procedure we obtain 
\begin{eqnarray}
{\cal L}_{\text{int}}&=& \left[3 \sqrt{\epsilon_{\phi}\epsilon_{}\chi}+ \frac{\Lambda \sqrt{\epsilon_{\phi}\epsilon_{B}}\left(10 H^4 \epsilon_B^2+9 H^2 p^2 \epsilon_B\epsilon_E+p^4\epsilon_E^2\right)}{\sqrt{2}\left(2 H^2 \epsilon_B+p^2 \epsilon_E\right)^2}\right]a^2H^2\; \hat{X}\left(\tau\right)\hat{\Phi}^*\left(\tau\right)\nonumber\\
&&+ \frac{\Lambda H^3 \epsilon^{3/2}_B\sqrt{\epsilon_\phi} a }{\sqrt{2}(2H^2 \epsilon_B+p^2\epsilon_E)} \hat{X}'\left(\tau\right)\hat{\Phi}^*\left(\tau\right)+\text{cc} \;, 
\label{lin-inter}
\end{eqnarray}
where we have disregarded the $\hat{X}' \hat{\Phi}'$ coupling, which is of higher order in slow roll with respect to the terms just written. In terms of the noncanonical variables, this Lagrangian gives the interaction Hamiltonian
\begin{eqnarray}
H_{\text{int},\chi \phi}&=& - \int d^3 x \Bigg\{ \left[3 \sqrt{\epsilon_{\phi}\epsilon_{}\chi}+ \frac{\Lambda \sqrt{\epsilon_{\phi}\epsilon_{B}}\left(12 H^4 \epsilon_B^2+10 H^2 p^2 \epsilon_B\epsilon_E+p^4\epsilon_E^2\right)}{\sqrt{2}\left(2 H^2 \epsilon_B+p^2 \epsilon_E\right)^2}\right]a^4 H^2\; \delta\chi\left(\tau\right)\delta\phi^*\left(\tau\right)\nonumber\\
&&\quad\quad \quad\quad \quad 
+ \frac{\Lambda H^3 \epsilon^{3/2}_B\sqrt{\epsilon_\phi} a }{\sqrt{2}(2H^2 \epsilon_B+p^2\epsilon_E)} \delta\chi'\left(\tau\right)\delta\phi^*\left(\tau\right)+\text{cc} \Bigg\} \;. 
\label{Lint}
\end{eqnarray}

It is interesting to observe that the first term of the first line corresponds to the standard coupling between two scalar fields that interact only gravitationally, arising after integrating out the nondynamical modes of the metric. 
This is the standard term that one would have expected, and it is the only term included in previous analyses and / or discussions of  the nonlinearly produced scalar perturbations in this model.  Our computation shows that, in reality, more terms are present. They arise because, besides the metric perturbations, we have also included and integrated out the nondynamical gauge perturbation $\delta A_0^3$. These additional terms 
are typically  ${\cal O}\left(10^{2}\right)-{\cal O}\left(10^3\right)$ times bigger than the first term. 

One can observe this hierarchy as follows: considering the fact that the inflaton is sourced by the axion mostly in the superhorizon regime, we can set momenta terms to zero. Then the ratio between the dominant coupling that we have found and the one that has been considered so far amounts in 
\begin{equation}
\frac{{\cal O} ( \Lambda \sqrt{\epsilon_B \, \epsilon_\phi})} { {\cal O} ( \sqrt{\epsilon_\chi \, \epsilon_\phi} ) } \sim \Lambda \cdot \sqrt{ \frac{m_Q^2 \, Q^2 /M_p^2}{{\dot \chi}^2 / \left( 2 H^2 \, M_p^2 \right)}} \sim \Lambda^2 \gg 1 \;, 
\end{equation} 
where eqs. (\ref{slow}), (\ref{params}), and (\ref{chiSR}) have been employed. The strong hierarchy between these two slow-roll parameters can be also observed in Figure 2 of  \cite{Dimastrogiovanni:2016fuu}.

\section{A specific nonlinear interaction} 
\label{sec:nonlinear} 

Eq. (\ref{Lint}) encodes the interactions between the inflaton perturbations and the perturbations of the axion. 
Through these interactions, the axion modes source the inflaton perturbations, as diagrammatically shown in the second and third diagram of Figure \ref{fig:diagrams} for the two-point function. 

We can actually disregard the contribution corresponding to the second diagram of Figure \ref{fig:diagrams}. Ref.   \cite{Dimastrogiovanni:2016fuu} studied that contribution, and obtained a ratio of about $10^{-5}$ between the amplitude of the $\delta \phi$ modes obtained from that diagram and the vacuum modes. We recall that this result was obtained  using only the first term in eq. (\ref{Lint}), and that the remaining terms are about  ${\cal O}\left(10^{2}\right)-{\cal O}\left(10^3\right)$ times bigger. The amplitude $\delta \phi$ scales linearly with this coupling, so, when we account for the increase due to the full set of terms in (\ref{Lint}) we still find a highly subdominant contribution from this diagram. For this reason, we disregard it from now on. 

In this section we instead compute the contribution of the third diagram of Figure \ref{fig:diagrams}, in which the inflaton perturbation is sourced by the axion perturbation (through eq. (\ref{lin-inter})), which is enhanced by its nonlinear interaction with the unstable  $t_{L}$ tensor mode. We stress that, as mentioned in the previous section, by looking only at this coupling we are disregarding two of the dynamical scalar perturbations of the CNI sector. We believe that the coupling considered here is the dominant one (for the reasons mentioned in the previous section). We, at the very least, consider the result obtained in this way as a  lower bound on the amplitude of the sourced inflaton perturbation. For an exact computation, one should include all the cubic interactions between the scalar fields and the $t_L$ modes as well as all the linear couplings between all the scalar fields in $Y_i$. It would be hard to imagine that these additional contributions would precisely cancel the contribution that we compute here.

For the computation, we also need the  $\delta \chi\,t_L\,t_L$ interaction. This is given by  \cite{Papageorgiou:2018rfx} 
\begin{equation}
H_{int, \chi\,tt} = - \frac{\lambda}{f} \int d^3 x \left\{ \delta \chi \left[ \frac{g}{2} \left( a Q t_{ab} t_{ab} \right)' 
- \epsilon^{ijk} t_{ai}' \partial_j t_{ak} \right] + \left[  \frac{g^2 a^2 Q^2}{- \partial^2+2 g^2 a^2 Q^2} \delta \chi \right] \, \partial_j \left( \epsilon^{ijk} t_{ia}' t_{ak} \right)  \right\} \,. 
\label{intctt}
\end{equation}

In Appendix \ref{app:B} we re-write the two interaction terms (\ref{Lint}) and (\ref{intctt}) in momentum space. 
We use these expressions to evaluate the third diagram of Figure \ref{fig:diagrams} via the in-in formalism
\begin{eqnarray}
\!\!\!\!\!\!\!\!\!\! \delta\left\langle\delta\phi\left(\tau,\vec{k}_1\right)\delta\phi\left(\tau,\vec{k}_2\right)\right\rangle &=& \int^\tau_{-\infty}d\tau_1 \int^{\tau_1}_{-\infty}d\tau_2 \int^{\tau_2}_{-\infty}d\tau_3 \int^{\tau_3}_{-\infty}d\tau_4\nonumber\\
&&\!\!\!\!\!\!\! \!\!\!\!\!\!\! \!\!\!\!\!\!\! \!\!\!\!\!\!\!  \!\!\!\! \!\!\!\!\!\!\!  \!\!\!\! 
\times\bigg\{\Big\langle\big[\big[\big[\big[\delta\phi^{(0)}\left(\tau,\vec{k}_1\right)\delta\phi^{(0)}\left(\tau,\vec{k}_2\right),H_{int, \chi\phi}\left(\tau_1\right)\big],H_{int, \chi\phi}\left(\tau_2\right)\big],H_{int, \chi\,tt}\left(\tau_3\right)\big],H_{int, \chi\,tt}\left(\tau_4\right)\big]\Big\rangle\nonumber\\
&&\!\!\!\!\!\!\! \!\!\!\!\!\!\! \!\!\!\!\!\!\! \!\!\!\!\!\!\! \!\!\!\!\!\!\!  \!\!\!\! 
+\Big\langle\big[\big[\big[\big[\delta\phi^{(0)}\left(\tau,\vec{k}_1\right)\delta\phi^{(0)}\left(\tau,\vec{k}_2\right),H_{int, \chi\phi}\left(\tau_1\right)\big],H_{int, \chi\,tt}\left(\tau_2\right)\big],H_{int, \chi\phi}\left(\tau_3\right)\big],H_{int, \chi\,tt}\left(\tau_4\right)\big]\Big\rangle\nonumber\\
&&\!\!\!\!\!\!\! \!\!\!\!\!\!\! \!\!\!\!\!\!\! \!\!\!\!\!\!\! \!\!\!\!\!\!\!  \!\!\!\! 
+\Big\langle\big[\big[\big[\big[\delta\phi^{(0)}\left(\tau,\vec{k}_1\right)\delta\phi^{(0)}\left(\tau,\vec{k}_2\right),H_{int, \chi\phi}\left(\tau_1\right)\big],H_{int, \chi\,tt}\left(\tau_2\right)\big],H_{int, \chi\,tt}\left(\tau_3\right)\big],H_{int, \chi\phi}\left(\tau_4\right)\big]\Big\rangle\bigg\} \;, \nonumber\\
\label{inin}
\end{eqnarray}
where the suffix "(0)" remarks that the mode functions entering at the right hand side are the "unperturbed" ones, namely those obtained in the linear theory presented in the previous section. For brevity we omit this suffix from now on.

We note that the the three terms in eq. (\ref{inin}) correspond to all possible permutations of the two interactions
(subject to the fact that the innermost term must be $H_{int, \chi\phi}$). We expect that the final result is dominated by the first term, for the reason that we now explain. The origin of the large correction that we obtain from (\ref{inin}) is the tachyonic growth of the tensor modes $t_L$. These are the modes that source the axion, and, eventually, the inflaton perturbations. We note that the integration extrema enforce $\tau_4 \leq \tau_3  \leq \tau_2  \leq \tau_1  \leq \tau$. In the first term, the tensor modes are evaluated at the two earliest times $\tau_3$ and $\tau_4$. This corresponds to a greater overall integration region for which the tensor modes are first enhanced, and then source the scalar perturbations. A numerical evaluation of the three terms indeed confirmed that the contribution from the second and third term is negligible with respect to the one from the first term. Therefore,  only the first term is kept in the results presented below.  

In the numerical evaluations, it is convenient to express the unperturbed modes in terms of the dimensionless mode functions ${\tilde X}_c$ and ${\tilde t}_c$, defined through
\begin{eqnarray} 
\delta \phi \left(\tau,\;k \right) \equiv\frac{1}{\sqrt{2k}} \frac{\tilde{\Phi}\left(x\right)}{a(\tau)} \;,\;\; 
\delta \chi \left( \tau ,\, k \right) \equiv 
\frac{\sqrt{1+m_Q^2}}{\sqrt{2 k}} \, \frac{{\tilde X}_c \left( x \right)}{a \left( \tau \right)} \;,\;\; 
t_{k,L} \left( \tau  \right) \equiv \frac{\tilde{t}_L \left( x \right)}{\sqrt{2 k}} \,. 
\label{code-var} 
\end{eqnarray} 
They correspond to the canonically normalized variables, times $\sqrt{2 k}$. Therefore, their initial amplitude is $1$ (see \cite{Papageorgiou:2018rfx} for a discussion of the proper normalization of the $\delta \chi$ mode), and they are function of the dimensionless quantity $x \equiv - k \tau$. 

We are interested in the ratio between this nonlinear contribution to the power spectrum and the linear term
\begin{equation}
{\cal R}_{\delta \phi} \equiv \frac{\delta P_\phi \left( \tau ,\, k \right)}{P_\phi \left( \tau ,\, k \right)} = \frac{ \delta \left\langle \delta \phi \left( \tau ,\, \vec{k}_1 \right) \,  \delta \phi \left( \tau ,\, \vec{k}_2 \right) \right\rangle' }{\left\langle \delta \phi \left( \tau ,\, \vec{k}_1 \right) \,  \delta \phi \left( \tau ,\, \vec{k}_2 \right) \right\rangle' }  \;, 
\label{Rdsigma-def}
\end{equation} 
where the prime on the left hand side denotes the correlator without the corresponding $\delta \left( \vec{k_1} + \vec{k}_2 \right)$ function. The explicit expression for the ratio is evaluated in Appendix \ref{app:B}, where we find

\begin{eqnarray}
{\cal R}_{\delta\phi}&=&\int^x_{\infty}dx_1 \int^{x_1}_{\infty}dx_2 \int^{x_2}_{\infty}dx_3 \int^{x_3}_{\infty}dx_4 \int \frac{d^3 q_1 d^3 q_2}{\left( 2 \pi \right)^{3}} \, 
\frac{\left(\hat{q_1}\cdot\hat{q_2}-1\right)^4}{16 q_1 q_2}\delta^{(3)}\left(\hat{k}_1+\vec{q_1}+\vec{q_2}\right)\nonumber\\
&& \quad\quad\quad\quad  \quad\quad\quad\quad  \quad\quad\quad\quad  \quad\quad\quad\quad 
\times F\left(x,x_1,x_2,x_3,x_4,q_1,q_2\right) \;, 
\label{Rdsigmaresult}
\end{eqnarray}
where we have introduced  $x_i\equiv-k_1 \tau_i ,\; q_i\equiv\frac{p_i}{k_1}$, as well as the real function 
\begin{eqnarray}
&& \!\!\!\!\! \!\!\!\!\! \!\!\!\!\!F\left(x,x_1,x_2,x_3,x_4,q_1,q_2\right)\equiv\frac{H^2\epsilon_\phi\left(1+m_Q^2\right)^2}{16 M_p^2 \vert\tilde{\Phi}\left(x\right)\vert^2}\;\frac{1}{x_1^3 x_2^3}\nonumber\\
&& \!\!\!\!\!\!\times \Big[\frac{6 \sqrt{2}}{m_Q}\left(1+m_Q^2\right)- \frac{\sqrt{2} m_Q^3\Lambda^2\,x_1 }{\left[x_1^2+2 m_Q^2\right]}\frac{d}{dx_{1\chi}}+\frac{\sqrt{2} m_Q \Lambda^2\left[x_1^4+10 m_Q^2 x_1^2 +12 m_Q^4\right]}{\left[x_1^2+2m_Q^2\right]^2}\Big]\nonumber\\
&& \!\!\!\!\!\!\times \Big[\frac{6 \sqrt{2}}{m_Q}\left(1+m_Q^2\right)- \frac{\sqrt{2} m_Q^3\Lambda^2\,x_2 }{\left[x_2^2+2 m_Q^2\right]}\frac{d}{dx_{2\chi}}+\frac{\sqrt{2} m_Q \Lambda^2\left[x_2^4+10 m_Q^2 x_2^2+ 12 m_Q^4\right]}{\left[x_2^2+2m_Q^2\right]^2}\Big]\nonumber\\
&& \!\!\!\!\!\! \times\Bigg\{ i\;x_{1\chi}x_{2\chi}\left[\tilde{\Phi}\left(x\right)\tilde{\Phi}^*\left(x_{1}\right)-\tilde{\Phi}^*\left(x\right)\tilde{\Phi}\left(x_{1}\right)\right]\left[\tilde{\Phi}\left(x\right)\tilde{\Phi}^*\left(x_{2}\right)-\tilde{\Phi}^*\left(x\right)\tilde{\Phi}\left(x_{2}\right)\right]\nonumber\\
&&\!\!\!\!\!\!   \times \left[\tilde{X}_c\left(x_{2\chi}\right)\tilde{X}_c^*\left(x_3\right)-\tilde{X}_c^*\left(x_{2\chi}\right)\tilde{X}_c\left(x_3\right)\right] {\rm Im}\Bigg[\tilde{X}_c\left(x_{1\chi}\right)\tilde{X}_c^*\left(x_4\right)\frac{{\cal W}(x_3,x_4,q_2,q_3)+{\cal W}(x_3,x_4,q_3,q_2)}{2}\Bigg]\nonumber\\ 
&&\!\!\!\!\!\! +\left(x_1\leftrightarrow x_2\right)\Bigg\}\Big|_{x_{1\chi}=x_1,\,x_{2\chi}=x_2} \;. 
\end{eqnarray}

In this expression we have symmetrized over the  two internal momenta $q_1$ and $q_2$, and we have introduced the quantity 
\begin{eqnarray}
&&{\cal W}(x_3,x_4,q_1,q_2)\equiv\Bigg\{\frac{m_Q^2}{x_3 x_4}\tilde{t}_L\left(q_1 x_3\right)\tilde{t}^*_L\left(q_1 x_4\right)\tilde{t}_L\left(q_2 x_3\right)\tilde{t}^*_L\left(q_2 x_4\right)\nonumber\\
&&\quad\quad\quad + 2q_2\left[m_Q -q_1 x_3 + \frac{x_3 m_Q^2}{x_3^2 + 2 m_Q^2} \left( q_1 - q_2 \right) \right] \tilde{t}_L\left(q_1 x_3\right)\tilde{t}^*_L\left(q_1 x_4\right)
\nonumber\\
&&\quad\quad\quad\quad\quad\!\!\times \Big\{ q_2\left[ m_Q- q_1x_4 + \frac{x_4 m_Q^2}{x_4^2 + 2m_Q^2 } \left( q_1 - q_2 \right) \right]\,\tilde{t}'_L\left(q_2 x_3\right)\tilde{t}'^*_L\left(q_2 x_4\right) - \frac{m_Q}{x_4}\tilde{t}'_L\left(q_2 x_3\right)\tilde{t}^*_L\left(q_2 x_4\right)\Big\}\nonumber\\ 
&&\quad\quad\quad+ 2q_2\left[m_Q - q_1x_4 + \frac{x_4m_Q^2}{x_4^2 + 2 m_Q^2} \left( q_1 - q_2 \right) \right]\tilde{t}_L\left(q_2 x_3\right)\tilde{t}'^{*}_L\left(q_2 x_4\right)
\nonumber\\
&&\quad\quad\quad\quad\quad\!\!\times \Big\{ q_1\left[ m_Q - q_2x_3 + \frac{x_3m_Q^2}{x_3^2 + 2m_Q^2 } \left( q_2 - q_1 \right) \right]\tilde{t}'_L\left(q_1 x_3\right)\tilde{t}^*_L\left(q_1 x_4\right) -\frac{m_Q}{x_3}  \tilde{t}_L\left(q_1 x_3\right)\tilde{t}^*_L\left(q_1 x_4\right)\Big\}\Bigg\} \;. \nonumber\\
\end{eqnarray}

We exploit the Dirac $\delta-$function present in (\ref{Rdsigmaresult}) to perform the integration over $d^3 q_2$. 
We use polar coordinates for the remaining  $d^3 q_1$ integration, using a coordinate system for which the external vector $\vec{k}_1$ is oriented along the third axis. In this way, the $\int d\phi$ integration is trivial,
while the $\int d\theta$ integration can be traded back for an integration over $q_2$ (using the identity $q_2^2 = 1 + q_1^2+2 q_1 \, \cos \theta$, which is enforced by the Dirac $\delta-$function). In this way, we are left with 
a $\int d q_1 d q_2$ integration. We further change variables 
\begin{equation}
{\cal X} \equiv \frac{q_1+q_2}{\sqrt{2}} \;\;,\;\;   {\cal Y}   \equiv \frac{q_1-q_2}{\sqrt{2}} \;,  
\label{changeofvrbles}
\end{equation}
In terms of which the integral  becomes 
\begin{eqnarray}
&& {\cal R}_{\delta\phi}= \frac{1}{8\left( 2 \pi \right)^{2}}\int^x_{\infty}dx_1 \int^{x_1}_{\infty}dx_2 \int^{x_2}_{\infty}dx_3 \int^{x_3}_{\infty}dx_4 \int^\infty_{\frac{1}{\sqrt{2}}} d X \int^{\frac{1}{\sqrt{2}}}_0 dY \nonumber\\
&&
\quad\quad\quad\quad \quad\quad\quad\quad \quad\quad \times\left(\frac{1-2X^2}{X^2-Y^2}\right)^4 F\left(x,x_1,x_2,x_3,x_4,\frac{X+Y}{\sqrt{2}},\frac{X-Y}{\sqrt{2}}\right) \,. 
\label{finalresult}
\end{eqnarray}

This expression is ready to be integrated. We want to evaluate the ratio on super-horizon scales, at the moment in which the axion stops sourcing the inflaton perturbations. As we discuss in the next section, we consider the two separate cases in which the production lasts for, respectively, $50$ and $10$ e-folds in the super-horizon regime. This corresponds, respectively, to  $\ln \, x \sim -50$ and $-10$, or, respectively, $x\simeq 2 \cdot 10^{-22}$ and $5 \cdot 10^{-5}$. 

The result of the integral  grows logarithmically with respect to the external time, namely $ {\cal R}_{\delta\phi} \propto \ln^2 x$. This emerges clearly from the analytical study that we present in Appendix \ref{app:semianalytic}. This happens because the inflaton perturbation keeps being sourced, while in the super-horizon regime, by the  mode $\delta \chi$. We note that this also happens in the U(1) version of this model \cite{Namba:2015gja}, in which the inflaton is coupled gravitationally to an axion, that is sourced by an unstable U(1) vector mode. Our numerical evaluations of (\ref{finalresult}), performed in the range  $10^{-5} \la x \la 10^{-1}$ are in excellent agreement with this scaling (see Figure \ref{fig:loggrowth} below).

\section{Constraints, Results and Phenomenology} 
\label{sec:results} 

In this section, we will combine the results obtained in Section \ref{sec:nonlinear} and in Appendix \ref{app:semianalytic} with different constraints to probe the available parameter space of the model (\ref{SCNI}). 

\subsection{Constraints} 
\label{subsec:constraints}

We can impose several constraints on the model (\ref{SCNI}) to ensure that it is consistent with the current data and that it can reveal its characteristic phenomenology in (near) future experiments.
\begin{itemize}
\item This model aims at challenging the robustness of the one-to-one relation (\ref{r-V}) between the energy scale of inflation and  the GWs spectrum (that assumed vacuum GWs) by producing a large amount of additional sourced GWs. (See the footnote before eq. (\ref{RGW}) for the precise distinction between the vacuum and the sourced term.) To violate this relation in a sizeable manner, we must request  ${\cal R}_{\rm GW} \equiv P_{\rm gw}^{s} / P_{\rm gw}^{v} \ga 1$. We reflect this condition in our figures by excluding the regions which have ${\cal R}_{\rm GW}<1$, such that non-vacuum GWs always dominate over the vacuum ones. 

The ratio of sourced GWs to vacuum GWs is estimated as \cite{Dimastrogiovanni:2016fuu,Fujita:2017jwq}

\begin{equation}
{\cal R}_{\rm GW} \simeq \frac{\epsilon_B {\cal F}^2}{2} \,, \quad  \quad  {\rm with} \qquad {\cal F}^2 \simeq {\rm e}^{3.6 m_Q} \quad ,  \quad \epsilon_B=\frac{m_Q^4}{g^2}\frac{H^2}{M_p^2}= \frac{m_Q^4}{g^2} \frac{\pi^2}{2} r_{\rm vac} P_\zeta \,. 
\label{ratiosourcedvacgw}
\end{equation}

\item The copious production of tensor degrees of freedom results in back-reaction in equations of motion of the axion, gauge field and the expansion rate of the inflationary universe \cite{Dimastrogiovanni:2016fuu,Fujita:2017jwq}. The most important backreaction effect is on the last of eqs. (\ref{eq-bck}), where the produced tensor modes add the additional term  \cite{Dimastrogiovanni:2016fuu,Fujita:2017jwq} 
\begin{equation}
{\cal T}_{BR}^Q = \frac{g H^3 \, \xi}{12 \pi^2} \left[ {\cal B} \left( m_Q \right) - {\tilde {\cal B}} \left( m_Q \right) / \xi \right] \;, 
\end{equation} 
(with the correction of a typo appearing in \cite{Dimastrogiovanni:2016fuu}), where $\xi \equiv m_Q + m_Q^{-1}$, and where the functions ${\cal B}$ and ${\tilde {\cal B}}$ are given in eq. (\ref{calB-tildecalB}) below.

With this contribution, the last of eqs. (\ref{eq-bck}) can be written as 
\begin{equation}
\ddot Q + 3 H \dot Q + \dot H \, Q + V_{\rm eff} \left( Q \right) + {\cal T}_{BR}^Q = 0 \;, 
\end{equation} 
where we introduced the ``effective potential'' 
\begin{equation}
V_{\rm eff} \left( Q \right) \equiv  2 \, H^2 \, Q \left( 1 + m_Q^2 \right) - g \, Q^2 \, \lambda \frac{\dot{\chi}}{f} \;. 
\end{equation} 

Ref. \cite{Dimastrogiovanni:2016fuu} imposed that backreaction is negligible by requiring that $ {\cal T}_{BR}^Q $ is smaller  (in absolute value) than the last term in the effective potential. Doing so, one obtains the condition 
$ \frac{g H^3 \, \xi}{12 \pi^2} \left[ {\cal B} \left( m_Q \right) - {\tilde {\cal B}} \left( m_Q \right) / \xi \right] \ll 
 g Q^2 \, \frac{\lambda \dot{\chi}}{f}$. Using the second of (\ref{chiSR}) one can then immediately write  \cite{Dimastrogiovanni:2016fuu} $g \ll \left(  \frac{24\pi^2 m_Q^2}{{\cal B} - { \tilde {\cal B}}/\xi }  \right)^{1/2}$. 
 However, the value for $\dot{\chi}$ given in  (\ref{chiSR}) generates a cancellation among the terms in the effective potential, and it is therefore safer to demand \cite{Maleknejad:2018nxz} that  $ {\cal T}_{BR}^Q $ should be smaller than the smallest term in it.~\footnote{We are indebted to Azadeh Maleknejad for private communications on this issue.} This results in the stronger constraint 
\begin{equation}
g  \ll \left( \frac{24 \, \pi^2}{  {\cal B} \left( m_Q \right) - {\tilde {\cal B}} \left( m_Q \right) / \xi  } \, 
\frac{1}{1+\frac{1}{m_Q^2}}   \right)^{1/2} \,, 
\label{backreaction}
\end{equation}
which is essentially stronger by a factor $\frac{1}{m_Q}$ than the one imposed in  \cite{Dimastrogiovanni:2016fuu}. 

To evaluate the relation (\ref{backreaction}), we recall the definition of ${\cal B}$ and of ${\tilde  {\cal B}}$ from \cite{Dimastrogiovanni:2016fuu} 
\begin{equation}
{\cal B} \equiv \int^{x_{\rm max}}_{x_{\rm min}} dx \, x \,  \left \vert i^{-\alpha} W_{\alpha,\beta} (2ix) \right \vert^2  \quad, \quad {\tilde  {\cal B}} \equiv \int^{x_{\rm max}}_{x_{\rm min}} dx \, x^2 \,  \left \vert i^{-\alpha} W_{\alpha,\beta} (2ix) \right \vert^2 \; . 
\label{calB-tildecalB}
\end{equation}
The extrema of integration $x_{{\rm max}/{\rm min}}= m_Q + \xi  \pm \sqrt{m_Q^2 + \xi^2}$ are the values between which the tensor mode of the gauge field is tachyonic. We numerically fit  the first denominator of the expression (\ref{backreaction}) and found $ \left( {\cal B} - { \tilde {\cal B}}/\xi \right) \simeq 2.3 \cdot {\rm e}^{3.9 \, m_Q}$. We then used it to evaluate the constraint (\ref{backreaction}).

\item One also requires that loop corrections to the adiabatic curvature perturbations that we computed in this work are significantly smaller than the vacuum ones. For definiteness, we require that  
\begin{equation}
{\cal R}_{\delta \phi} \equiv \frac{ \langle  \delta \phi^{(s)} \,  \delta \phi^{(s)}  \rangle}{ \langle  \delta \phi^{(v)} \, \delta \phi^{(v)} \rangle} < 0.1 \;, 
\label{perturbativitycond}
\end{equation}
where $\delta \phi^{(s)}$ indicating the sourced fluctuations that results from quantum loop corrections, while $\delta \phi^{(v)}$ indicating the linear result. There are two reasons for imposing the bound (\ref{perturbativitycond}). Firstly, we impose it so that the total tensor-to-scalar ratio is not decreased by the additional inflaton perturbations. Secondly, even if we did not perform this computation here, we expect the sourced modes to be highly non-Gaussian, so that, if they dominate, they would violate the strong limits on non-Gaussianity of the observed primordial perturbations \cite{Ade:2015ava}. Based on the results from the abelian case \cite{Barnaby:2010vf}, we expect that the sourced perturbations are highly non-Gaussian, with a shape peaked in the equilateral configuration. For the following consideration, let us assume that the amplitude of non-Gaussianity obtained in the present context is comparable to that obtained in the abelian case, for which the value of the nonlinear parameter of the sourced modes alone was found to be $f_{\rm NL,{\rm sourced}} = {\cal O } \left( 10^4 \right)$  \cite{Barnaby:2010vf}. This amplitude is ``diluted'' by the vacuum modes (which have a negligible deviation from Gaussianity), if we assume them to dominate the inflaton two-point function. In this case, the observed $f_{\rm NL}$ scales as ${\cal R}_{\delta \phi}^2 \times f_{\rm NL,{\rm sourced}}$. Therefore, the current ($2 \sigma$) bound on equilateral non-Gaussianity, $f_{\rm NL,equil} \la {\cal O } \left( 100 \right)$ \cite{Ade:2015ava}, leads to the condition (\ref{perturbativitycond}) written above. While a more precise bound awaits a detailed computation of $\left\langle \delta  \phi^{(s) 3} \right\rangle$ in this model, we believe that the condition (\ref{perturbativitycond}) is a reasonable one as an order of magnitude bound.

\item The recent Planck/BICEP 2 / KECK Array  results constrain the tensor-to-scalar-ratio as $r <0.06$ \cite{Ade:2018gkx}. This limit assumes the standard $n_t = - r/8$ relation between the tensor tilt and the tensor-to-scalar ratio, that might be violated in the present context. Ref. \cite{Akrami:2018odb} studied how the limit on $r$ relaxes when $n_t$ is allowed to vary. We do not expect a large variation of $n_t$ in the present model, since $m_Q$ (which controls the sourced tensor modes) is nearly constant at CMB scales \cite{Dimastrogiovanni:2016fuu}. For this reason we continue to apply the limit of \cite{Ade:2018gkx}, which assumes a very small $n_t$.

\item Next generation CMB experiments are expected to measure tensor-to-scalar ratio with $s(r)\ga 10^{-3} $
\cite{Abazajian:2016yjj}, where $s^2$ is the variance of the measurement.  Therefore, we impose that the total tensor-to-scalar ratio satisfies 
\begin{equation}
r= r_{\rm vac} \left(1+ {\cal R}_{\rm GW} \right) > 10^{-3} \,, 
\label{eq-r-rvac}
\end{equation}
where we disregarded the contributions of the nonlinear scalar perturbations, under the assumption of ${\cal R}_{\delta \phi} < 0.1$.

\end{itemize}

\subsection{Results and Phenomenology} 
\label{subsec:resultsandpheno}

In this subsection, we combine the constraints discussed in the previous subsection with our result for the quantum loop corrections to the inflaton fluctuations  
\begin{equation}
{\cal R}_{\delta \phi} \simeq   \, \frac{5 \cdot 10^{-12}}{\left(1+\frac{\epsilon_B}{\epsilon_\phi}\right)^2} \; {\rm e}^{7 m_Q} \; m_Q^{11} \; N_k^2 \;  r_{\rm vac}^2 \;. 
\label{eqfittedRinf}
\end{equation}
This result, derived in Appendix \ref{app:semianalytic}, is based on a semi-analytic fit (valid in the $2.5 \leq m_Q \leq  3.5$ range) to the numerical results.  Figures  \ref{fig:loggrowth}  and \ref{fig:Rinflvsmq} confirm that this result accurately reproduces the values obtained from the numerical evaluation of eq. (\ref{finalresult}). The factor $N_k$ is the number of e-folds during inflation during which the background axion rolls. As discussed at the end of the previous section, during this period the  models $\delta \chi$ acts a source for the inflaton perturbation $\delta \phi$. We assume that the axion reaches the minimum of its potential before the end of inflation. This ensures that the direct contribution of $\delta \chi$ to the curvature perturbation $\zeta$ is negligible with respect to that of $\delta \phi$. We discuss this in Appendix \ref{app:curva}, where we follow a similar study performed in \cite{Fujita:2018vmv}. We present our results under the two different assumptions: (i) the axion rolls for $50$ e-folds after the CMB modes are produced, and (ii) it runs for only  $10$ e-folds (after the CMB modes are produced). The first case results in a greater $\delta \phi$ production, since the field $\delta \chi$ acts as a source for the inflaton modes while $\chi$ is light.

To study the impact of our result (\ref{eqfittedRinf}), we first understand the relevant region of parameter space 
where we need to focus our attention. In  Appendix \ref{app:Pzandrvac} we obtained the slow roll expression 
\begin{equation}
P_\zeta \simeq \frac{g^2}{8 \pi^2 \, m_Q^4} \, \frac{\epsilon_\phi \epsilon_B}{\left( \epsilon_\phi + \epsilon_B \right)^2} \;, 
\end{equation} 
for the scalar power spectrum. We use this expression, and the measured value $P_{\zeta,{\rm measured}} = 2.1 \cdot 10^{-9}$ \cite{Akrami:2018odb}, to relate the ratio $\frac{\epsilon_B}{\epsilon_\phi}$ to the other parameters of the model 
\begin{equation}
\frac{\epsilon_B}{\epsilon_\phi} = \left\{ \begin{array}{l} 
r_- \equiv A - 1 - \sqrt{A^2 - 2 A} \\ \\ 
r_+ \equiv A - 1 + \sqrt{A^2 - 2 A}   
\end{array} \right. \;\;\;,\;\;\; A \equiv \frac{g^2}{16 \pi^2 P_{\zeta,{\rm measured}} m_Q^4} \;. 
\label{branches}
\end{equation}  
These two solutions exist for $A \geq 2$, and they satisfy $0 < r_- \leq 1$ and $r_+ \geq 1$   (where $r_- = r_+ = 1$ corresponds to $A=2$; in the limit of very large $A$, one finds $r_- \simeq \frac{1}{2 A}$ and $r_+ \simeq 2 \, A$). For any fixed $g$, the condition $A \geq 2$ results in an upper bound on $m_Q$. Together with the $m_Q > \sqrt{2}$ bound \cite{Dimastrogiovanni:2012ew}, the allowed interval for $m_Q$ is 
\begin{equation}
\sqrt{2} < m_Q \leq \left( \frac{g^2}{32 \pi^2 \,  P_{\zeta,{\rm measured}} }  \right)^{1/4} \simeq 35 \, \sqrt{g} \,,   
\label{mQ-limits}
\end{equation}
where the last approximate equality has been obtained using the measured value  $P_{\zeta,{\rm measured}} = 2.1 \cdot 10^{-9}$ \cite{Akrami:2018odb}. We see that the allowed interval for $m_Q$ is not empty only for $g \gtrsim 1.6 \cdot 10^{-3}$. 

\begin{figure}[tbp]
\begin{center}
\includegraphics[width=0.48\textwidth,angle=0]{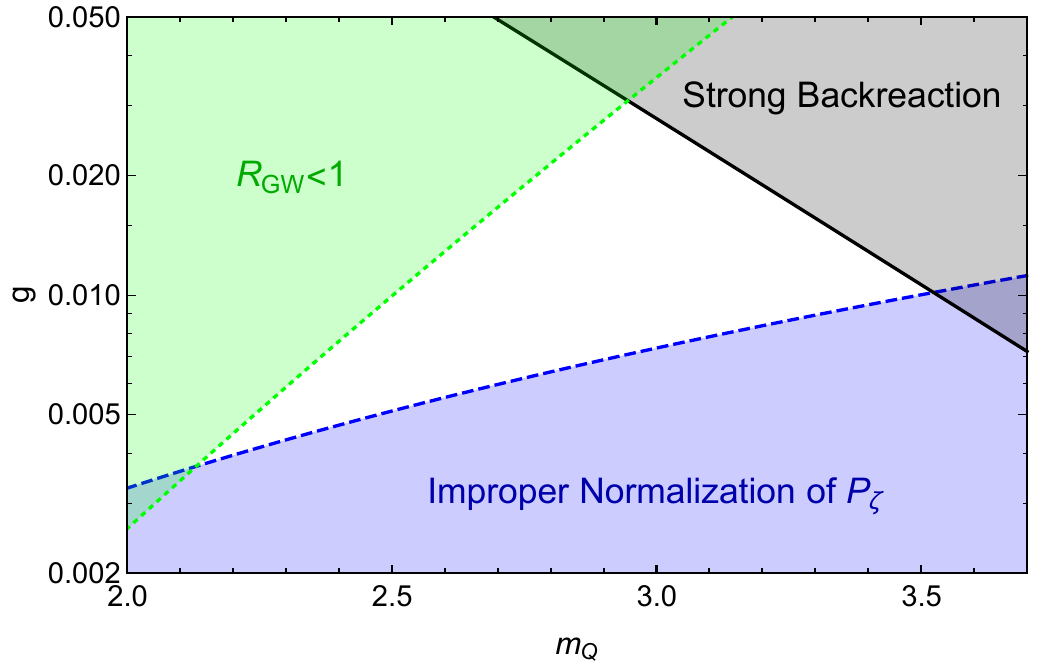}
\end{center}
\vspace{-7mm}
\caption{Condition on the parameters of the model, before studying the impact of our result (\ref{eqfittedRinf}). 
The blue dashed line is the second condition in (\ref{mQ-limits}), which, ultimately, selects the region where the model parameters allow to set the scalar power spectrum to its measured value. The black solid line is the backreaction limit  (\ref{backreaction}). Finally, the Green dotted line is a necessary condition to have a greater amount of sourced vs. vacuum GW. The region in white satisfies these three requirements. 
} 
\label{fig:region}
\end{figure}

As we now show, we can actually obtain a more stringent interval on $g$, and on $m_Q$, by combining the condition $g \ga \left( \frac{m_Q}{35} \right)^2$ that we have just obtained, with the lower limit (\ref{backreaction}) on $g$ imposed by back reaction, and with the relation 
\begin{equation}
g \la 1.8 \cdot 10^{-5} \, m_Q^2 \, {\rm e}^{1.8 m_Q} \;\;,\;\; {\rm necessary \; for } \;\; {\cal R}_{\rm GW} \ga 1 \,.  
\end{equation}
This relation is obtained by maximizing the ratio (\ref{ratiosourcedvacgw}), by setting $r_{\rm vac}$ to its largest possible value ($0.06$). The combination of these conditions is shown in Figure \ref{fig:region}. We see that $0.0037 \la g \la 0.031$ and $2.1 \la m_Q \la 3.5$ in this region.

\begin{figure}[tbp]
\begin{center}
\includegraphics[width=0.48\textwidth,angle=0]{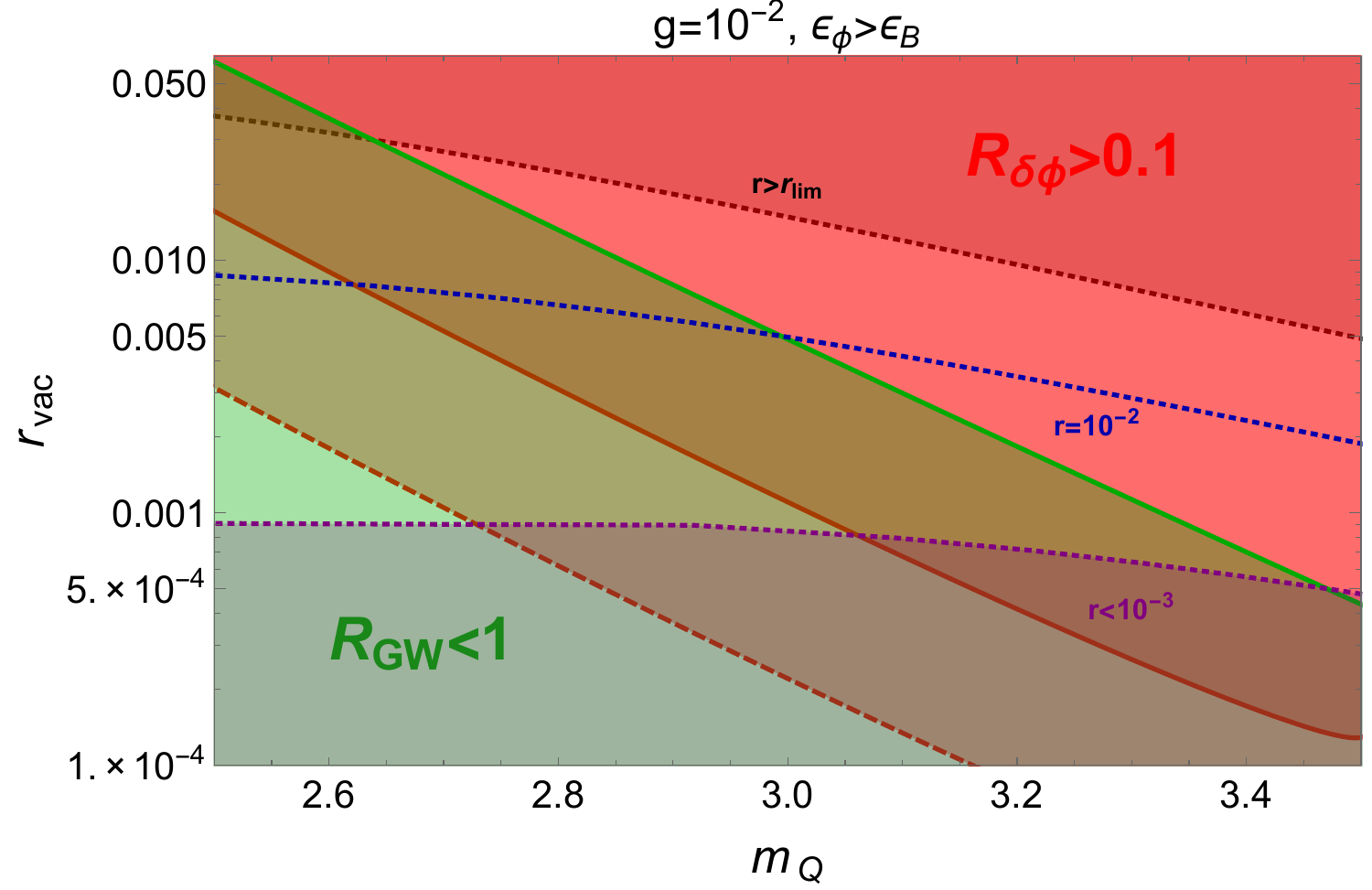}
\includegraphics[width=0.48\textwidth,angle=0]{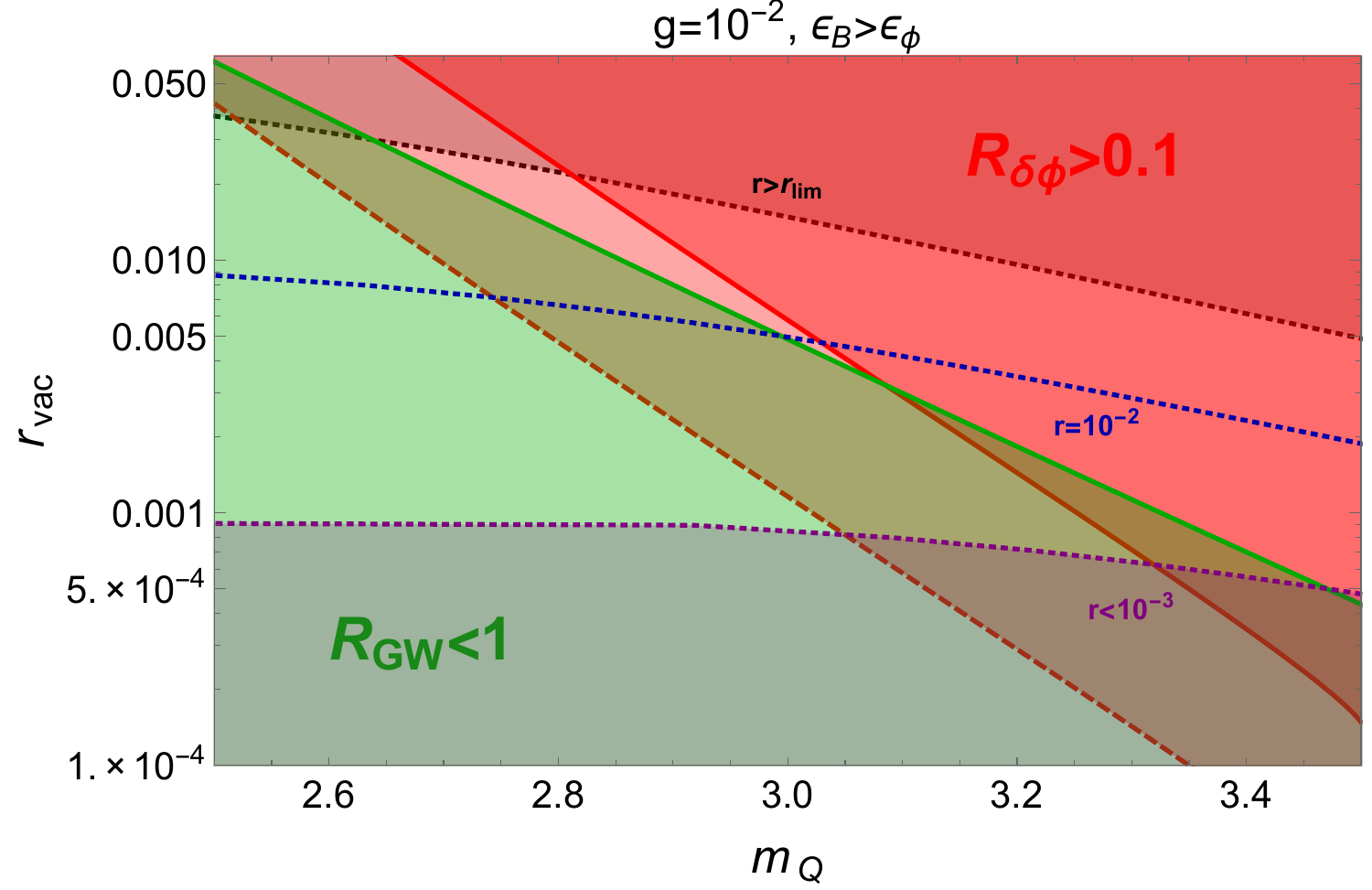}
\end{center}
\vspace{-7mm}
\caption{Parameter space of the spectator CNI model with various constraints. Various regions of the parameter space are ruled out or are phenomenologically disfavoured due to different constraints such as ${\cal R}_{\delta \phi}<0.1$, ${\cal R}_{\rm GW} >1$, $10^{-3}<r<0.06$.  The origin and the corresponding expressions for these constraints are given in the Subsection \ref{subsec:constraints}.} 
\label{figparameterspace}
\end{figure}

With this in mind, we can present our results  in the  $m_Q-r_{\rm vac}$ plane, for the fixed value $g = 10^{-2}$ that lies towards the middle of the allowed (white) region in Figure \ref{fig:region}. Consistently with the results shown in Figure \ref{fig:region}, we focus on the interval $2.5 \leq m_Q \leq 3.5$, and we consider the two branches of eq. (\ref{branches}), namely $\frac{\epsilon_B}{\epsilon_\phi} = r_-$ (for which $\epsilon_\phi > \epsilon_B$) and $\frac{\epsilon_B}{\epsilon_\phi} = r_+$ (for which $\epsilon_B > \epsilon_\phi$). The two branches are shown, respectively, in the left and right panels of  Figure \ref{figparameterspace}. 

Let us now discuss the various lines present in these two panels. The red solid and dashed lines correspond to ${\cal R}_{\delta \phi} = 0.1$ in the case of, respectively, $N_k = 10$ and  $N_k = 50$. The regions above these lines are excluded due to the overproduction of sourced scalar perturbations. The green solid line corresponds to ${\cal R}_{\rm GW}=1$. The region below this line is disregarded since the phenomenological interest in this model is the significant production of sourced chiral GWs. The dotted black line indicated with $r_{\rm lim}$, indicates the most recent experimental bound on the tensor-to-scalar-ratio, $r_{\rm lim}=0.06$ \cite{Ade:2018gkx}. This excludes the region above this line. The dotted blue line indicates $r=0.01$; this is not a current constraint, and we only show it as a reference line to guide the eye. The dotted purple line indicates $r=10^{-3}$, which is the best sensitivity that the next generation CMB experiments are expected to reach \cite{Abazajian:2016yjj}.

From the two panels presented we can observe that the simultaneous requirement ${\cal R}_{\rm GW}>1$ and ${\cal R}_{\delta\phi}<0.1$ rules out all of the parameter space for the red dashed line which corresponds to $N_k=50$. For the case $N_k=10$ there is a small region of parameter space in the right panel that is compatible with the constraints imposed. We recall that in this panel the branch  $\frac{\epsilon_B}{\epsilon_\phi}>1$ is considered. We see from eq.  (\ref{eqfittedRinf}) that, at equal value of $r_{\rm vac}$, having $\epsilon_B$ greater than $\epsilon_\phi$ corresponds to a smaller value of ${\cal R}_{\delta\phi}$, which ultimately results in the presence of an allowed region for  $N_k=10$ in this branch. 

We also studied other values of $g$ in the allowed region $0.0037 \la g \la 0.031$ shown in Figure \ref{fig:region}. These leads to very similar results to those presented in Figure  \ref{figparameterspace} (for this reason, we do not present them here). For all these cases, a narrow region of parameters is allowed only for  $N_k=10$ and for the   $\frac{\epsilon_B}{\epsilon_\phi}>1$ branch.

\section{Conclusions}
\label{sec:conclusions} 

This work is a step towards the full understanding of the phenomenology of an interesting class of inflationary models based on the mechanism of  Chromo-Natural Inflation (CNI)  \cite{Adshead:2012kp}. In this mechanism, an SU(2) triplet with a nonvanishing vacuum expectation value backreacts on the evolution for the inflaton, directly coupled to the triplet via the axial interaction $\chi F {\tilde F}$. This allows for a slow roll evolution of the inflaton even if the potential is too steep to allow for inflation in absence of the vector field. 
The linearized study of the perturbations of this model showed that one SU(2) mode becomes unstable  in an intermediate range of momenta (shortly after horizon crossing). This mode transforms as a rank 2 tensor under a combination of spacial rotations and SU(2) transformations that leaves the background and the gauge fields vev unaffected, and it is therefore coupled to tensor metric perturbations at the linearized level. This results in a strong enhancement of one polarization of the GW signal of one definite polarization (depending on whether $\dot{\chi}$ is positive or negative during inflation; this motion of the axion is the - spontaneous - origin of CP breaking in these models). While the original CNI model is now ruled out by the phenomenology of the (linearized) perturbations, several variants have been proposed in the literature that use the same mechanism and that appear to be compatible with data. In this work we studied the phenomenology of the spectator CNI model of \cite{Dimastrogiovanni:2016fuu}, in which the inflaton is a different scalar field that is coupled only gravitationally to the CNI sector. 

The fact that the unstable mode $t_L$ is linearly coupled to the tensor modes but not to the scalar perturbations suggests the possibility that these models might have enhanced tensor perturbations, without a corresponding increase of the scalar modes. In this sense the situation appears to improve over the one encountered in the case of an axion coupled to a U(1) field, in which the tensor mode production (sourced at the nonlinear level by the vector fields generated by the rolling axion) typically reaches an interesting level only for couplings that are anyhow ruled out by the overproduction of non-Gaussian scalar perturbations \cite{Barnaby:2010vf}, with the possible exception of very special constructions \cite{Namba:2015gja} in which the axion is different from the inflaton, and rolls only for a limited number of e-folds during inflation. 

This conclusion, however, does not account for the inflaton perturbations that can be produced at the nonlinear level by the $t_L$ mode. The power spectrum of these nonlinear scalar perturbations was computed only recently in the CNI model \cite{Papageorgiou:2018rfx}, where it was shown that these perturbations can be greater than the ones obtained by the linearized theory for a wide range of parameters. In this work we have repeated the same computation for the spectator model of  \cite{Dimastrogiovanni:2016fuu}. Let us for a moment consider a model of only two scalar fields $\chi$ and $\phi$ that are coupled to each other only gravitationally, and that are slowly rolling. The perturbations of these two fields are coupled to each other via an $\sqrt{\epsilon_\chi \, \epsilon_\phi} H^2 \delta \chi \delta \phi$ interaction, where $\epsilon_{\chi,\phi}$ are the two standard slow-roll parameters associated to the motion of the fields. If the coupling between the of the perturbations of the CNI axion and of the inflaton was also of this type, one could conclude that the inflaton perturbations are sufficiently screened by the instability of the CNI sector. However, the perturbations of the CNI sector behave very differently from those of a single scalar field. This results in a greater coupling between the inflaton perturbations and the CNI sector, see eq. (\ref{Lint}) and the following discussion. This realization motivated the present work. 

Evaluating the nonlinear scalar perturbations requires extensive numerical computations. However, our earlier study of the CNI axion perturbations  \cite{Papageorgiou:2018rfx}, and the fact that the inflaton modes are mostly sourced in the super-horizon regime, makes it possible to obtain a very simple semi-analytical formula for the produced modes. This formula, eq. (\ref{eqfittedRinf}), is based on the analytic knowledge of the mode function solutions for the CNI linearized perturbations, and on fits of the integrals involved in the loop computation. Remarkably, the scalar perturbations in the CNI sector are mostly controlled by a single parameter $m_Q$, that is parametrically related to the ratio between the mass of the fluctuations of the vector field and the Hubble rate $H$, and that, in the large $m_Q$ regime, coincides with the parameter $\xi$ responsible for the gauge field production in the U(1) case. The effect of these modes on the inflaton perturbations are then controlled by four additional parameters: the gauge coupling $g$ in the CNI sector, the Hubble rate during inflation, the ratio of the two slow roll parameters $\frac{\epsilon_B}{\epsilon_\phi}$ and the number of e-folds during which the CNI axion rolled during inflation. 

To understand the acceptable range of values for the gauge coupling we imposed a number of constraints which appear in Figure \ref{fig:region}, namely i) that the ratio between the sourced GW mode and the one that is not sourced is significantly greater than one (so that the mechanism characteristic of CNI is at work), ii) that the backreaction of the unstable modes are negligible in the background equations and iii) that the power spectrum of scalar perturbations is normalized to the measured value. 

For definiteness, we fixed the number of e-folds for which the CNI axion rolls to either $N_k = 10$ or $N_k=50$, and we showed results for one value of the gauge coupling $g$, namely $10^{-2}$. By an educated guess of the amount of scalar non-Gaussianity produced in this model we imposed that the ratio ${\cal R}_{\delta \phi}$ between the nonlinear and the linear scalar perturbations should be smaller than about 0.1. We also imposed that the total tensor-to-scalar ratio is smaller than the current bound \cite{Ade:2018gkx} and greater than $10^{-3}$ (so that the GWs can be measured in the reasonably near future \cite{Abazajian:2016yjj}). The set of constraints define the phenomenologically favoured region of parameter space for the model, that is illustrated in Figure \ref{figparameterspace}. We see that only a very small portion of the parameter space in the model respects all these requirements and only a small duration of axion rolling is allowed as in the U(1) case \cite{Namba:2015gja}. An identical result is obtained for other values of $g$ in the allowed $0.0037 \la g \la 0.031$ range. 

An immediate impact of our results on the recent literature is in the computation of mixed scalar-tensor correlators performed for the model of  \cite{Dimastrogiovanni:2016fuu}. The additional, and dominant, couplings present in (\ref{Lint}) result in an increased of the amplitude of the scalar modes, and so of the correlators. 

We conclude by mentioning two less immediate additional computations that we believe should be performed to complete the results presented here. Firstly, the CNI model has three dynamical scalar perturbations. We only considered the coupling between the one originating from the axion field (in the specific gauge considered here) and the inflaton. This is motivated by the fact that the mode we have included is much greater than the other two in the super-horizon regime, which is where the inflaton mode is mostly sourced. We do not expect that the inclusion of other perturbations will change the order of magnitude of the results presented here. In particular, it is hard to imagine that the inclusion of additional couplings would lead to a decreased production.~\footnote{ This appears to be confirmed by the findings of  \cite{Fujita:2018vmv}, that included all the scalar perturbations in the computation of the tree level $\left\langle \delta \phi \, t \, t \right\rangle$ correlator, without finding a strong change with respect to a diagram in which only $\delta \chi$ is included (this argument is independent of the different $\delta \phi - \delta \chi$ coupling considered in   \cite{Fujita:2018vmv} and in the present work).} However, this should be checked.  Secondly, we expect the sourced inflaton perturbations to be highly non-Gaussian, as in the corresponding abelian case \cite{Barnaby:2010vf}. The amount of non-Gaussianity obtained in that case led us to impose the bound  ${\cal R}_{\delta \phi} \la 0.1$ as an order of magnitude limit on the sourced scalar perturbations. However, a full computation is needed in order to obtain a more precise bound.  We hope to return to these issues in future work.

\vskip.25cm
\section*{Acknowledgements} 

We would like to thank Tomohiro Fujita, Eiichiro Komatsu, Ryo Namba, and Angelo Ricciardone, for fruitful discussions and comments. We particularly thank Azadeh Maleknejad for very useful discussions on backreaction and on the generalized slow-roll relations for the scalar power spectrum. 

C\"U is supported by European Structural and Investment Funds and the Czech Ministry of Education, Youth and Sports (Project CoGraDS - CZ.$02.1.01/0.0/0.0/15\_003/0000437$).

\vskip.25cm

\appendix

\section{$\delta \phi - \delta \chi $ interaction lagrangian}
\label{app:dsdc}

We follow the formalism of \cite{Himmetoglu:2009qi}. The quadratic action for the scalar perturbations is of the form 
\begin{eqnarray} 
S &=& \int d^3 k\, d\tau \Bigg[ a_{ij}D^{*'}_i {D^{'}}_j +\left(b_{ij}N^*_i D^{'}_j+{\text h.c.}\right)+c_{ij}N^*_i N_j+\left(d_{ij}D^{*'}_i D_j+{\text h.c.}\right) \nonumber\\ 
& & \quad\quad\quad\quad +e_{ij}D^*_i D_j + \left(f_{ij}N^*_i D_j+{\text h.c.}\right) \Bigg] \;. 
\label{S2-sig-chi-nondyn}
\end{eqnarray} 

The vector $D$ contains the dynamical perturbations while the vector $N$ contains the non-dynamical perturbations, 
\begin{equation}
D\equiv\begin{pmatrix}
    \hat{X}        \\
    \hat{\Phi}     
\end{pmatrix} \;\;\;,\;\;\; 
N\equiv\begin{pmatrix}
    \delta A^3_0        \\
    \Phi                \\
    B
\end{pmatrix} \;\;. 
\end{equation} 

Under the above conventions the matrices $a_{ij},\dots,f_{ij}$ read

\begin{eqnarray} 
& & 
a=\begin{pmatrix}
    \frac{1}{2}  & 0     \\
    0   &  \frac{1}{2} 
\end{pmatrix} \;\;\;,\;\;\; 
b=\begin{pmatrix}
    0 & 0     \\
    -\frac{1}{2} a \chi' &  -\frac{1}{2} a \phi' \\
    0 & 0
\end{pmatrix} \;\;\;, \nonumber\\ 
& & 
c=\begin{pmatrix}
    \frac{k^2}{2}+g^2 \left(a Q\right)^2 & \frac{1}{2} i k \frac{d}{d\tau}\left(a Q\right) & i g^2 k \left(a Q\right)^3    \\
    -\frac{1}{2} i k \frac{d}{d\tau}\left(a Q\right) & \;\;\;\;\frac{1}{2}\left[(-6 M_p^2+3 Q^2){a'}^2+6a Q a' Q'+ a^2(3{Q'}^2+{\phi'}^2+{\chi'}^2)\right] & \;\;\;\;k^2 M_p^2  a\, a' \\
    -i g^2 k \left(a Q\right)^3  &  k^2 M_p^2  a\, a' & g^2 k^2 \left( a Q\right)^4
\end{pmatrix} \;\;\;, \nonumber\\ 
& & 
d=\begin{pmatrix}
    -\frac{a'}{2 a} & 0    \\
    0  &   -\frac{a'}{2 a}
\end{pmatrix} \;\;\;,\;\;\; 
e=\begin{pmatrix}
    -\frac{k^2}{2} +\frac{{a'}^2}{2a^2}-\frac{1}{2}a^2 \frac{d^2 U}{d\chi^2} & 0    \\
    0  &  -\frac{k^2}{2} +\frac{{a'}^2}{2a^2}-\frac{1}{2}a^2 \frac{d^2 V}{d\phi^2}
\end{pmatrix} \;\;\;, \nonumber\\ 
& & 
f=\begin{pmatrix}
    \frac{i g k \lambda a Q^2}{2f} & 0   \\
    \frac{1}{2}\left(a'\chi'-a^3 \frac{d U}{d\chi}\right) &\;\;\;\; \frac{1}{2}\left(a'\phi'-a^3 \frac{d V}{d\phi}\right)\\
    -\frac{1}{2}k^2 a \chi'  &  -\frac{1}{2}k^2 a \phi'
\end{pmatrix} \;\;\;.
\end{eqnarray} 

Integrating out the non dynamical variables results in an action that is formally of the type (\ref{SY}), where the matrices $C,K,\Omega^2$ are related to the matrices $a_{ij},\dots,f_{ij}$ by 
\begin{eqnarray}
C_{2+i\;\;2+j}&\equiv& a_{ij}-\left(b^\dagger\right)_{ik}\left(c^{-1}\right)_{km}b_{mj} \;, \nonumber\\
K_{2+i\;\;2+j}&\equiv& d_{ij}-\left(b^\dagger\right)_{ik}\left(c^{-1}\right)_{km}f_{mj} \;, \nonumber\\
\Omega^2_{2+i\;\;2+j}&\equiv& -e_{ij}+\left(f^\dagger\right)_{ik}\left(c^{-1}\right)_{km}b_{mj} \;, 
\end{eqnarray}
and also $Y_{2+i\;\;2+j}=D_{ij}$.

\section{Evaluation of $\delta \left\langle \delta \phi^2 \right\rangle$ through the in-in formalism}
\label{app:B}

In this Appendix we present the derivation of eq. (\ref{Rdsigmaresult}), starting from eqs.  (\ref{intctt}), (\ref{Hcoupling}) and (\ref{inin}). 
We decompose the inflaton, axion and the tensor perturbations as 
\begin{eqnarray}
&& \delta\phi =\int\frac{d^3k}{(2\pi)^{3/2}}e^{i\vec{k}\cdot\vec{x}}\,\delta\phi^{(0)}\left(\tau,\vec{k}\right) \;\;,\;\; 
\delta\phi^{(0)}(\tau,\vec{k})=\delta\phi_k\left(\tau\right)\, a\left(\vec{k}\right)+\delta\phi^*_k\left(\tau\right)\, a^\dagger\left(-\vec{k}\right) \,, \nonumber\\ 
&& \delta\chi =\int\frac{d^3k}{(2\pi)^{3/2}}e^{i\vec{k}\cdot\vec{x}}\,\delta\chi^{(0)}\left(\tau,\vec{k}\right) \;\;,\;\; 
\delta\chi^{(0)}(\tau,\vec{k})=\delta\chi_k\left(\tau\right)\, a\left(\vec{k}\right)+\delta\chi^*_k\left(\tau\right)\, a^\dagger\left(-\vec{k}\right) \,, \nonumber\\ 
&& t_{ab} = \int \frac{d^3 k}{\left( 2 \pi \right)^{3/2}} \, {\rm e}^{i \vec{k} \cdot \vec{x}} \, \sum_{\lambda=\pm 1} \Pi^*_{ab,\lambda} \left( {\hat k} \right) \,  t_\lambda \left( \tau ,\, \vec{k} \right) \;\;,\;\; 
t_\lambda \left( \tau ,\, \vec{k} \right) = t_{\lambda,k} \left( \tau  \right) a_\lambda \left( \vec{k} \right) +  t_{\lambda,k}^* \left( \tau  \right) a_\lambda^\dagger \left( - \vec{k} \right) \,. \nonumber\\ 
\label{tlr}
\end{eqnarray} 
The annihilation / creation operators satisfy the nonvanishing relations $\left[ a_\chi \left( \vec{k} \right) ,\,  a_\chi^\dagger \left( \vec{k}'  \right) \right] =\delta^{(3)} \left( \vec{k} - \vec{k}' \right)$, and analogously for $\phi$, and $\left[ a_\lambda \left( \vec{k} \right) ,\,  a_{\lambda'}^\dagger \left( \vec{k}'  \right) \right] =\delta^{(3)} \left( \vec{k} - \vec{k}' \right) \delta_{\lambda \lambda'}$. The sum in the last line is over the left L ($\lambda = +1$) and right R ($\lambda = -1$) handed helicities.  The transverse, traceless, and symmetric tensor operators can be written as 
\begin{equation}
{\Pi_{ab,\lambda}}^*  \left( {\hat k} \right) \equiv \epsilon_{a,\lambda} \left( {\hat k} \right)  \epsilon_{b,\lambda} \left( {\hat k} \right) \;, 
\end{equation} 
where the  vector circular polarization operators satisfy $\vec{k} \cdot \vec{\epsilon}_\lambda \left( {\hat k} \right)= 0 ,\, \vec{k} \times \vec{\epsilon}_\lambda \left( {\hat k} \right) = - i \lambda k \vec{\epsilon}_\lambda \left( {\hat k } \right) $, $ \vec{\epsilon}_\lambda \left( - {\hat k } \right) =  \vec{\epsilon}_\lambda^{\;*} \left(  {\hat k } \right) ,\,  \vec{\epsilon}_\lambda^{\;*} \left( {\hat k } \right) \cdot  \vec{\epsilon}_{\lambda'} \left( {\hat k } \right) = \delta_{\lambda \lambda'} $, in addition to $\vec{\epsilon}_+ \left( {\hat k} \right) \cdot \vec{\epsilon}_+ \left( {\hat k} \right) = \vec{\epsilon}_- \left( {\hat k} \right) \cdot \vec{\epsilon}_- \left( {\hat k} \right) = 0$. Thanks to these properties, the mode functions $t_{\pm}$ are canonically normalized, and coincide with those introduced in eq. (\ref{tensor-canonical}), see \cite{Papageorgiou:2018rfx} for the explicit check. In the following, we only consider the enhanced $t_L$ mode, and therefore keep only the term $\lambda = +1$ in eq. (\ref{tlr}). 

Keeping all this into account, the two interaction Hamiltonian (\ref{Lint}) and (\ref{intctt}) can be written as 
\begin{eqnarray}
H_{int,\chi\phi}(\tau)&=&-\int d^3 k \sqrt{\epsilon_\phi} \;a^4\Big[6 H^2 \sqrt{\epsilon_\chi}+ \frac{\sqrt{2} g^3\lambda a\, Q^5\, }{f M_p \left(k^2+2g^2 a^2 Q^2\right)}\frac{d}{d\tau_\chi}\nonumber\\
&&+ \frac{\sqrt{2} g H\lambda Q^3 \left(k^4+10 g^2 k^2 a^2 Q^2+12 g^4 Q^4 a^4\right)}{f M_p\left(k^2+2 g^2 a^2 Q^2\right)^2}\Big]\; \delta\chi^{(0)}\left(\tau_\chi,\vec{k}\right)\,\delta\phi^{(0)}\left(\tau,-\vec{k}\right)\Big\vert_{\tau\chi=\tau} \;, \nonumber\\
\label{Hcoupling}
\end{eqnarray}
and
\begin{eqnarray} 
H_{int,\chi\,tt} \left( \tau \right) &=& - \frac{\lambda}{2 f} \int \frac{d^3 p_1 d^3 p_2}{\left( 2 \pi \right)^{3/2}} \, 
\left[ \vec{\epsilon}_+ \left( {\hat p}_1 \right) \cdot  \vec{\epsilon}_+ \left( {\hat p}_2 \right) \right]^2 
\Bigg\{ \left[ g a Q - p_2 + \frac{g^2 a^2 Q^2}{\vert \vec{p}_1 + \vec{p}_2 \vert^2 + 2 g^2 a^2 Q^2 } \left( p_2 - p_1 \right) \right] \partial_{\tau'} \nonumber\\ 
& & \quad\quad  \quad\quad  \quad\quad  \quad\quad  \quad\quad 
+ \left[ g a Q - p_1 + \frac{g^2 a^2 Q^2}{\vert \vec{p}_1 + \vec{p}_2 \vert^2 + 2 g^2 a^2 Q^2 } \left( p_1 - p_2 \right) \right] \partial_{\tau''} + g \left( a Q \right)' \Bigg\} \nonumber\\ 
& & \quad\quad  \quad\quad  \quad\quad  \quad\quad  \quad\quad  \quad\quad 
 \times \delta  \chi^{(0)} \left( \tau ,\, - \vec{p}_1 - \vec{p}_2 \right) \, {\hat t}_+ \left( \tau' ,\, \vec{p}_1 \right) 
{\hat  t}_+ \left( \tau'' ,\, \vec{p}_2 \right) \Big\vert_{\tau' = \tau'' = \tau} \;. 
\end{eqnarray} 

After neglecting the last two permutations of the Hamiltonian terms (see the discussion after eq. (\ref{inin}) in the main text), the correction (\ref{inin}) to the inflaton $2-$point function acquires the form 
\begin{eqnarray}
&&  \!\!\!\!\!\!\!\!   \!\!\!\!\!\!\!\!   \!\!\!\!\!\!\!\! 
\int^\tau_{-\infty}d\tau_1 \int^{\tau_1}_{-\infty}d\tau_2 \int^{\tau_2}_{-\infty}d\tau_3 \int^{\tau_3}_{-\infty}d\tau_4\nonumber\\
&&\!\!\!\!\!\!  \!\!\!\!\!\!\!\!   \!\!\!\!\!\!\!\!   \!\!\!\!\!\!\!\! 
 \times \int\,d^3p_1 \sqrt{\epsilon_\phi} \;a(\tau_1)^4\Big[6 H^2 \sqrt{\epsilon_\chi}+ \frac{\sqrt{2} g^3\lambda a(\tau_1)\, Q^5\, }{f M_p \left[p_1^2+2g^2 a(\tau_1)^2 Q^2\right]}\frac{d}{d\tau_{1\chi}}
+\frac{\sqrt{2} g H \lambda Q^3 \left[p_1^4+10 g^2 p_1^2 a(\tau_1)^2 Q^2+12 g^4 a(\tau_1)^4Q^4\right]}{f M_p\left[p_1^2+2 g^2 a(\tau_1)^2 Q^2\right]^2}\Big]\nonumber\\
&&\!\!\!\!\!\!  \!\!\!\!\!\!\!\!   \!\!\!\!\!\!\!\!   \!\!\!\!\!\!\!\! 
\times \int\,d^3p_6\sqrt{\epsilon_\phi} \;a(\tau_2)^4\Big[6 H^2 \sqrt{\epsilon_\chi}+ \frac{\sqrt{2} g^3\lambda a(\tau_2)\, Q^5\, }{f M_p \left[p_6^2+2g^2 a(\tau_2)^2 Q^2\right]}\frac{d}{d\tau_{2\chi}}
+\frac{\sqrt{2} g H\lambda Q^3 \left[p_6^4+10 g^2 p_6^2 a(\tau_2)^2 Q^2+12 g^4 a(\tau_2)^4 Q^4\right]}{f M_p\left[p_6^2+2 g^2 a(\tau_2)^2 Q^2\right]^2}\Big] \nonumber\\
&&\!\!\!\!\!\!  \!\!\!\!\!\!\!\!   \!\!\!\!\!\!\!\!   \!\!\!\!\!\!\!\! 
\times \frac{\lambda}{2 f} \int \frac{d^3 p_2 d^3 p_3}{\left( 2 \pi \right)^{3/2}} \, 
\left[ \vec{\epsilon}_+ \left( {\hat p}_2 \right) \cdot  \vec{\epsilon}_+ \left( {\hat p}_3 \right) \right]^2 
\Bigg\{ \left[ g a(\tau_3) Q - p_3 + \frac{g^2 a(\tau_3)^2 Q^2}{\vert \vec{p}_2 + \vec{p}_3 \vert^2 + 2 g^2 a(\tau_3)^2 Q^2 } \left( p_3 - p_2 \right) \right] \partial_{\tau_3'} \nonumber\\ 
& & \quad\quad  \quad\quad  \quad\quad  \quad\quad  \quad\quad 
+ \left[ g a(\tau_3) Q - p_2 + \frac{g^2 a(\tau_3)^2 Q^2}{\vert \vec{p}_2 + \vec{p}_3 \vert^2 + 2 g^2 a(\tau_3)^2 Q^2 } \left( p_2 - p_3 \right) \right] \partial_{\tau_3''} + g \left[ a(\tau_3) Q \right]' \Bigg\} \nonumber\\ 
&&\!\!\!\!\!\!  \!\!\!\!\!\!\!\!   \!\!\!\!\!\!\!\!   \!\!\!\!\!\!\!\!  
\times \frac{\lambda}{2 f} \int \frac{d^3 p_4 d^3 p_5}{\left( 2 \pi \right)^{3/2}} \, 
\left[ \vec{\epsilon}_+ \left( {\hat p}_4 \right) \cdot  \vec{\epsilon}_+ \left( {\hat p}_5 \right) \right]^2 
\Bigg\{ \left[ g a(\tau_4) Q - p_5 + \frac{g^2 a(\tau_4)^2 Q^2}{\vert \vec{p}_4 + \vec{p}_5 \vert^2 + 2 g^2 a(\tau_4)^2 Q^2 } \left( p_5 - p_4 \right) \right] \partial_{\tau_4'} \nonumber\\ 
& & \quad\quad  \quad\quad  \quad\quad  \quad\quad  \quad\quad 
+ \left[ g a(\tau_4) Q - p_4 + \frac{g^2 a(\tau_4)^2 Q^2}{\vert \vec{p}_4 + \vec{p}_5 \vert^2 + 2 g^2 a(\tau_4)^2 Q^2 } \left( p_4 - p_5 \right) \right] \partial_{\tau_4''} + g \left[ a(\tau_4) Q \right]' \Bigg\} \nonumber\\ 
&&\!\!\!\!\!\!\times 2\Bigg\{ G_\phi\left(\tau,\tau_{1};k_1\right)\,G_{\phi}\left(\tau,\tau_{2};k_2\right)\delta^{(3)}\left(\vec{k}_1-\vec{p}_1\right)\delta^{(3)}\left(\vec{p}_6-\vec{p}_2-\vec{p}_3\right)\delta^{(3)}\left(\vec{k}_2-\vec{p}_6\right)\delta^{(3)}\left(-\vec{p}_4-\vec{p}_5+\vec{p}_1\right)\nonumber\\
&&\quad G_\chi\left(\tau_{2\chi},\tau_3;p_6\right)\times{\rm Im}\Big\{ C_\chi\left(\tau_{1\chi},\tau_4;p_1\right)\Big[ C_L\left(\tau_3',\tau_4'';p_2\right)C_L\left(\tau_3'',\tau_4';p_3\right)\delta^{(3)}\left(\vec{p}_3+\vec{p}_4\right)\delta^{(3)}\left(\vec{p}_2+\vec{p}_5\right)\nonumber\\
&&\quad\quad\quad\quad\quad \quad\quad\quad\quad\quad \quad\quad\quad\quad +C_L\left(\tau_3'',\tau_4'';p_3\right)C_L\left(\tau_3',\tau_4';p_2\right)\delta^{(3)}\left(\vec{p}_2+\vec{p}_4\right)\delta^{(3)}\left(\vec{p}_3+\vec{p}_5\right)\Big]\Big\}\nonumber\\
&&+ G_\phi\left(\tau,\tau_{1};k_1\right)\,G_{\phi}\left(\tau,\tau_{2};k_2\right)\delta^{(3)}\left(\vec{k}_1-\vec{p}_1\right)\delta^{(3)}\left(\vec{p}_1-\vec{p}_2-\vec{p}_3\right)\delta^{(3)}\left(\vec{k}_2-\vec{p}_6\right)\delta^{(3)}\left(-\vec{p}_4-\vec{p}_5+\vec{p}_6\right)\nonumber\\
&& \quad\;\; G_\chi\left(\tau_{1\chi},\tau_3;p_1\right)\times{\rm Im}\Big\{C_\chi\left(\tau_{2\chi},\tau_4;p_6\right) \Big[C_L\left(\tau_3',\tau_4'';p_2\right)C_L\left(\tau_3'',\tau_4';p_3\right)\delta^{(3)}\left(\vec{p}_3+\vec{p}_4\right)\delta^{(3)}\left(\vec{p}_2+\vec{p}_5\right)\nonumber\\
&&\quad\quad\quad\quad\quad \quad\quad\quad\quad\quad \quad\quad\quad\quad +C_L\left(\tau_3'',\tau_4'';p_3\right)C_L\left(\tau_3',\tau_4';p_2\right)\delta^{(3)}\left(\vec{p}_2+\vec{p}_4\right)\delta^{(3)}\left(\vec{p}_3+\vec{p}_5\right)\Big]\Big\}\nonumber\\
&&+\left(\vec{k}_1 \leftrightarrow \vec{k}_2\right)\Bigg\}\Bigg\vert_{\tau_3'=\tau_3''=\tau_3 \, ,\, \tau_4'=\tau_4''=\tau_4} \;, 
\end{eqnarray}
where we defined the commutators 
\begin{eqnarray}
\left[\delta\chi^{(0)}\left(\tau,\vec{k}\right),\delta\chi^{(0)}\left(\tau',\vec{k}'\right)\right]&=&\left[\delta\chi_k\left(\tau\right)\delta\chi_{k}^*\left(\tau'\right)-\delta\chi_k^*\left(\tau\right)\delta\chi_{k}\left(\tau'\right)\right]\delta^{(3)}\left(\vec{k}+\vec{k}'\right)
\equiv  i\,G_\chi(\tau,\tau';k)\delta^{(3)}\left(\vec{k}+\vec{k}'\right)\;,\nonumber\\
\left[\delta\phi^{(0)}\left(\tau,\vec{k}\right),\delta\phi^{(0)}\left(\tau',\vec{k}'\right)\right]&=&\left[\delta\phi_k\left(\tau\right)\delta\phi_{k}^*\left(\tau'\right)-\delta\phi_k^*\left(\tau\right)\delta\phi_{k}\left(\tau'\right)\right]\delta^{(3)}\left(\vec{k}+\vec{k}'\right) 
\equiv  i\,G_\phi(\tau,\tau';k)\delta^{(3)}\left(\vec{k}+\vec{k}'\right) \;, \nonumber\\ 
\left[t_L\left(\tau,\vec{k}\right),t_L\left(\tau',\vec{k}'\right)\right]&=&\left[t_{L,k}\left(\tau\right)t_{L,k}^*\left(\tau'\right)-t_{L,k}^*\left(\tau\right)t_{L,k}\left(\tau'\right)\right]\delta^{(3)}\left(\vec{k}+\vec{k}'\right)
\equiv  i\,G_L(\tau,\tau';k)\delta^{(3)}\left(\vec{k}+\vec{k}'\right) \;,\nonumber\\
\label{definitions1}
\end{eqnarray}
and the expectation values 
\begin{eqnarray}
\Big\langle\delta\chi^{(0)}\left(\tau,\vec{k}\right)\delta\chi^{(0)}\left(\tau',\vec{k}'\right)\Big\rangle&=&\delta\chi_k\left(\tau\right)\delta\chi_{k}^*\left(\tau'\right)\delta^{(3)}\left(\vec{k}+\vec{k}'\right)\equiv C_\chi(\tau,\tau';k)\delta^{(3)}\left(\vec{k}+\vec{k}'\right) \;,\nonumber\\
\Big\langle\delta\phi^{(0)}\left(\tau,\vec{k}\right)\delta\phi^{(0)}\left(\tau',\vec{k}'\right)\Big\rangle&=&\delta\phi_k\left(\tau\right)\delta\phi_{k}^*\left(\tau'\right)\delta^{(3)}\left(\vec{k}+\vec{k}'\right)\equiv C_\phi(\tau,\tau';k)\delta^{(3)}\left(\vec{k}+\vec{k}'\right) \;,\nonumber\\
\Big\langle t_L\left(\tau,\vec{k}\right) t_L\left(\tau',\vec{k}'\right)\Big\rangle&=&t_{L,k}\left(\tau\right)t_{L,k}^*\left(\tau'\right)\delta^{(3)}\left(\vec{k}+\vec{k}'\right) \equiv C_L(\tau,\tau';k)\delta^{(3)}\left(\vec{k}+\vec{k}'\right)\;. \nonumber\\
\label{definitions2}
\end{eqnarray}
 
After carrying out the $\vec{p}_1$, $\vec{p}_6$, $\vec{p}_4$, $\vec{p}_5$ integrations using the delta functions and observing that some of the lines obey a symmetry under the simultaneous exchange $\tau_3' \leftrightarrow \tau_3''$ and $\vec{p}_2 \leftrightarrow \vec{p}_3$ we can ignore some of the terms while multiplying the entire expression by a factor of two. The result becomes

\begin{eqnarray}
&& \delta\left(\vec{k}_1+\vec{k}_2\right) \int^\tau_{-\infty}d\tau_1 \int^{\tau_1}_{-\infty}d\tau_2 \int^{\tau_2}_{-\infty}d\tau_3 \int^{\tau_3}_{-\infty}d\tau_4\;\frac{\epsilon_\phi\;\lambda^2\; \;a(\tau_1)^4\;a(\tau_2)^4}{f^2}\nonumber\\
&& \!\!\!\!\!\!\times \Big[6 H^2 \sqrt{\epsilon_\chi}+ \frac{\sqrt{2} g^3\lambda a(\tau_1)\, Q^5\, }{f M_p \left[k_1^2+2g^2 a(\tau_1)^2 Q^2\right]}\frac{d}{d\tau_{1\chi}}+\frac{\sqrt{2} g H \lambda Q^3 \left[k_1^4+10 g^2 k_1^2 a(\tau_1)^2 Q^2+12 g^4 a(\tau_1)^4 Q^4\right]}{f M_p\left[k_1^2+2 g^2 a(\tau_1)^2 Q^2\right]^2}\Big]\nonumber\\
&& \!\!\!\!\!\!\times \Big[6 H^2 \sqrt{\epsilon_\chi}+ \frac{\sqrt{2} g^3\lambda a(\tau_2)\, Q^5\, }{f M_p \left[k_1^2+2g^2 a(\tau_2)^2 Q^2\right]}\frac{d}{d\tau_{2\chi}}+\frac{\sqrt{2} g H \lambda Q^3 \left[k_1^4+10 g^2 k_1^2  a(\tau_2)^2 Q^2+12 g^4 a(\tau_2)^4 Q^4\right]}{f M_p\left[k_1^2+2 g^2 a(\tau_2)^2 Q^2\right]^2}\Big] \nonumber\\
&&\!\!\!\!\!\! \times \int \frac{d^3 p_2 d^3 p_3}{\left( 2 \pi \right)^{3}} \, 
\left[ \vec{\epsilon}_+ \left( {\hat p}_2 \right) \cdot  \vec{\epsilon}_+ \left( {\hat p}_3 \right) \right]^2 
\Bigg\{ \left[ g a(\tau_3) Q - p_3 + \frac{g^2 a(\tau_3)^2 Q^2}{\vert \vec{p}_2 + \vec{p}_3 \vert^2 + 2 g^2 a(\tau_3)^2 Q^2 } \left( p_3 - p_2 \right) \right] \partial_{\tau_3'} \nonumber\\ 
&& \quad\quad  \quad\quad  \quad\quad  \quad\quad  \quad\quad 
+ \left[ g a(\tau_3) Q - p_2 + \frac{g^2 a(\tau_3)^2 Q^2}{\vert \vec{p}_2 + \vec{p}_3 \vert^2 + 2 g^2 a(\tau_3)^2 Q^2 } \left( p_2 - p_3 \right) \right] \partial_{\tau_3''} + g \left[ a(\tau_3) Q \right]' \Bigg\} \nonumber\\ 
&&\quad\quad \quad\quad \times \left[ \vec{\epsilon}_+ \left( -{\hat p}_3 \right) \cdot  \vec{\epsilon}_+ \left( -{\hat p}_2 \right) \right]^2 
\Bigg\{ \left[ g a(\tau_4) Q - p_2 + \frac{g^2 a(\tau_4)^2 Q^2}{\vert \vec{p}_3 + \vec{p}_2 \vert^2 + 2 g^2 a(\tau_4)^2 Q^2 } \left( p_2 - p_3 \right) \right] \partial_{\tau_4'} \nonumber\\ 
&& \quad\quad  \quad\quad  \quad\quad  \quad\quad  \quad\quad 
+ \left[ g a(\tau_4) Q - p_3 + \frac{g^2 a(\tau_4)^2 Q^2}{\vert \vec{p}_3 + \vec{p}_2 \vert^2 + 2 g^2 a(\tau_4)^2 Q^2 } \left( p_3 - p_2 \right) \right] \partial_{\tau_4''} + g \left[ a(\tau_4) Q \right]' \Bigg\} \nonumber\\ 
&&\!\!\!\!\!\! \!\!\!\!\!\! \!\!\!\!\!\!\times \Bigg\{ G_\phi\left(\tau,\tau_{1};k_1\right)\,G_{\phi}\left(\tau,\tau_{2};k_1\right)\delta^{(3)}\left(\vec{k}_1+\vec{p}_2+\vec{p}_3\right)G_\chi\left(\tau_{2\chi},\tau_3;k_1\right){\rm Im}\Big[C_\chi\left(\tau_{1\chi},\tau_4;k_1\right) C_L\left(\tau_3',\tau_4'';p_2\right)C_L\left(\tau_3'',\tau_4';p_3\right)\Big]\nonumber\\
&&\!\!\!\!\!\! \!\!\!\!\!\! \!\!\!\!\!\! \quad + G_\phi\left(\tau,\tau_{1};k_1\right)\,G_{\phi}\left(\tau,\tau_{2};k_1\right)\delta^{(3)}\left(\vec{k}_1-\vec{p}_2-\vec{p}_3\right)G_\chi\left(\tau_{1\chi},\tau_3;k_1\right){\rm Im}\Big[C_\chi\left(\tau_{2\chi},\tau_4;k_1\right) C_L\left(\tau_3',\tau_4'';p_2\right)C_L\left(\tau_3'',\tau_4';p_3\right)\Big]\nonumber\\
&&+\left(\vec{k}_1 \leftrightarrow -\vec{k}_1\right)\Bigg\}\Bigg\vert_{\tau_3'=\tau_3''=\tau_3 \, ,\, \tau_4'=\tau_4''=\tau_4} \;. 
\end{eqnarray}

The expression before the last line is symmetric under changing the sign of $\vec{k}_1$ (if, at the same time, one also reverses the sign of the dummy variables $\vec{p}_2$ and $\vec{p}_3$). Therefore the operation in the last line simply amounts in multiplying the previous lines by two. We then substitute $g a Q \left( \tau \right) = - \frac{m_Q}{\tau}$ and $\Lambda=\frac{\lambda}{f}Q$, and we neglect the evolution of $m_Q$ during inflation as it is slow-roll suppressed. We then use the identity $\vert \vec{\epsilon}_L \left( {\hat p}_1 \right) \cdot  \vec{\epsilon}_L \left( {\hat p}_2 \right) \vert^2 = \frac{\left( {\hat p}_1 \cdot {\hat p}_2 - 1 \right)^2}{4} $.
This gives 
\begin{eqnarray}
&& \delta\left(\vec{k}_1+\vec{k}_2\right) \int^\tau_{-\infty}d\tau_1 \int^{\tau_1}_{-\infty}d\tau_2 \int^{\tau_2}_{-\infty}d\tau_3 \int^{\tau_3}_{-\infty}d\tau_4\;\frac{2\epsilon_\phi}{H^4 \tau_1^4\tau_2^4}\nonumber\\
&& \!\!\!\!\!\!\times \Big[\frac{6\lambda \sqrt{\epsilon_\chi}}{f}- \frac{\sqrt{2} m_Q^3\Lambda^2\,\tau_1 }{M_p\left[k_1^2\tau_1^2+2 m_Q^2\right]}\frac{d}{d\tau_{1\chi}}+\frac{\sqrt{2} m_Q \Lambda^2 \left[k_1^4\tau_1^4+10 m_Q^2 k_1^2 \tau_1^2 +12 m_Q^4 \right]}{M_p\left[k_1^2\tau_1^2+2m_Q^2\right]^2}\Big]\nonumber\\
&& \!\!\!\!\!\!\times \Big[\frac{6\lambda \sqrt{\epsilon_\chi}}{f}- \frac{\sqrt{2} m_Q^3\Lambda^2\,\tau_2 }{M_p\left[k_1^2\tau_2^2+2 m_Q^2\right]}\frac{d}{d\tau_{2\chi}}+\frac{\sqrt{2} m_Q \Lambda^2 \left[k_1^4\tau_2^4+10 m_Q^2 k_1^2 \tau_2^2 +12 m_Q^4\right]}{M_p\left[k_1^2\tau_2^2+2m_Q^2\right]^2}\Big]\nonumber\\
&& \int \frac{d^3 p_2 d^3 p_3}{\left( 2 \pi \right)^{3}} \, 
\frac{\left(\hat{p}_2\cdot\hat{p}_3-1\right)^4}{16} \times \Bigg\{ \left[G_\phi\left(\tau,\tau_{1};k_1\right)\,G_{\phi}\left(\tau,\tau_{2};k_1\right)\delta^{(3)}\left(\vec{k}_1+\vec{p}_2+\vec{p}_3\right)G_\chi\left(\tau_{2\chi},\tau_3;k_1\right)\right]\nonumber\\
&& \!\!\!\!\!\!\times {\rm Im}\Bigg[C_\chi\left(\tau_{1\chi},\tau_4;k_1\right) \Bigg\{\left[\frac{m_Q}{\tau_3} \right]' \left[\frac{m_Q}{\tau_4} \right]' C_L\left(\tau_3,\tau_4;p_2\right)C_L\left(\tau_3,\tau_4;p_3\right)\nonumber\\
&&\quad\quad\quad + 2\left[ -\frac{m_Q}{\tau_3} - p_2 + \frac{m_Q^2}{k_1^2\tau_3^2 + 2 m_Q^2} \left( p_2 - p_3 \right) \right] C_L\left(\tau_3,\tau_4;p_2\right)
\nonumber\\
&&\quad\quad\quad\quad\quad\times \Big\{ \left[ -\frac{m_Q}{\tau_4} - p_2 + \frac{m_Q^2}{k_1^2\tau_4^2 + 2m_Q^2 } \left( p_2 - p_3 \right) \right]C_L^{(1,1)}\left(\tau_3,\tau_4;p_3\right) - \left[\frac{m_Q}{\tau_4} \right]' C_L^{(1,0)}\left(\tau_3,\tau_4;p_3\right)\Big\}\nonumber\\ 
&&\quad\quad\quad+ 2\left[ -\frac{m_Q}{\tau_4} - p_2 + \frac{m_Q^2}{k_1^2\tau_4^2 + 2 m_Q^2} \left( p_2 - p_3 \right) \right] C_L^{(0,1)}\left(\tau_3,\tau_4;p_3\right)
\nonumber\\
&&\quad\quad\quad\quad\quad\times \Big\{ \left[ -\frac{m_Q}{\tau_3} - p_3 + \frac{m_Q^2}{k_1^2\tau_3^2 + 2m_Q^2 } \left( p_3 - p_2 \right) \right]C_L^{(1,0)}\left(\tau_3,\tau_4;p_2\right) - \left[\frac{m_Q}{\tau_3} \right]' C_L\left(\tau_3,\tau_4;p_2\right)\Big\}\Bigg\}\Bigg]\nonumber\\
&& \!\!\!\!\!\! +\left(\tau_1\leftrightarrow\tau_2\right)\Bigg\} \;. 
\end{eqnarray}

We manipulate this expression by inserting the  dimensionless time and momentum  $x_i\equiv-k_1 \tau_i$ and $q_i\equiv\frac{p_i}{k_1}$, by using the rescaled variables (\ref{code-var}), and by using the expressions  (\ref{definitions1}) and  (\ref{definitions2}). We also use the background equations to simplify the terms that involve $\epsilon_{\chi}$
\begin{equation}
\frac{6\lambda \sqrt{\epsilon_\chi}}{f}=\frac{6 \lambda \dot{\chi}}{\sqrt{2} f H M_p}=\frac{6 \sqrt{2}}{M_p}\left(m_Q+\frac{1}{m_Q}\right)=\frac{6 \sqrt{2}}{m_Q\,M_p}\left(1+m_Q^2\right) \;. 
\end{equation}

We recall that this result is the expression for the correction (\ref{inin}) to the inflaton two-point function. We now divide it by the vacuum correlator as defined in eq. (\ref{Rdsigma-def}), and obtain 
\begin{eqnarray}
&& \!\!\!\!\! \!\!\!\!\! \!\!\!\!\!{\cal R}_{\delta\phi}= \frac{H^2\epsilon_\phi\left(1+m_Q^2\right)^2}{16 M_p^2 \vert\tilde{\Phi}\left(x\right)\vert^2}\int^x_{\infty}dx_1 \int^{x_1}_{\infty}dx_2 \int^{x_2}_{\infty}dx_3 \int^{x_3}_{\infty}dx_4\;\frac{1}{x_1^3 x_2^3}\nonumber\\
&& \!\!\!\!\!\!\times \Big[\frac{6 \sqrt{2}}{m_Q}\left(1+m_Q^2\right)- \frac{\sqrt{2} m_Q^3\Lambda^2\,x_1 }{\left[x_1^2+2 m_Q^2\right]}\frac{d}{dx_{1\chi}}+\frac{\sqrt{2} m_Q \Lambda^2\left[x_1^4+10 m_Q^2 x_1^2 +12 m_Q^4\right]}{\left[x_1^2+2m_Q^2\right]^2}\Big]\nonumber\\
&& \!\!\!\!\!\!\times \Big[\frac{6 \sqrt{2}}{m_Q}\left(1+m_Q^2\right)- \frac{\sqrt{2} m_Q^3\Lambda^2\,x_2 }{\left[x_2^2+2 m_Q^2\right]}\frac{d}{dx_{2\chi}}+\frac{\sqrt{2} m_Q \Lambda^2\left[x_2^4+10 m_Q^2 x_2^2+ 12 m_Q^4\right]}{\left[x_2^2+2m_Q^2\right]^2}\Big]\nonumber\\
&& \int \frac{d^3 q_2 d^3 q_3}{\left( 2 \pi \right)^{3}} \, 
\frac{\left(\hat{q}_2\cdot\hat{q}_3-1\right)^4}{16}\delta^{(3)}\left(\hat{k}_1+\vec{q}_2+\vec{q}_3\right)\nonumber\\
&& \!\!\!\!\!\! \!\!\!\!\!\!  \!\!\!\!\!\!  \!\!\!\!\!\! \times\Bigg\{ \frac{i\;x_{1\chi}x_{2\chi} }{q_2 q_3}\left[\tilde{\Phi}\left(x\right)\tilde{\Phi}^*\left(x_{1}\right)-\tilde{\Phi}^*\left(x\right)\tilde{\Phi}\left(x_{1}\right)\right]\left[\tilde{\Phi}\left(x\right)\tilde{\Phi}^*\left(x_{2}\right)-\tilde{\Phi}^*\left(x\right)\tilde{\Phi}\left(x_{2}\right)\right]\nonumber\\
&&\!\!\!\!\!\! \!\!\!\!\!\!  \!\!\!\!\!\! \!\!\!\!\!\!\times \left[\tilde{X}_c\left(x_{2\chi}\right)\tilde{X}_c^*\left(x_3\right)-\tilde{X}_c^*\left(x_{2\chi}\right)\tilde{X}_c\left(x_3\right)\right] {\rm Im}\Bigg[\tilde{X}_c\left(x_{1\chi}\right)\tilde{X}_c^*\left(x_4\right)\frac{{\cal W}(x_3,x_4,q_2,q_3)+{\cal W}(x_3,x_4,q_3,q_2)}{2}\Bigg]\nonumber\\ 
&&\!\!\!\!\!\! \!\!\!\!\!\! \!\!\!\!\!\! \!\!\!\!\!\! +\left(x_1\leftrightarrow x_2\right)\Bigg\} \;. 
\end{eqnarray}

This is the result that is presented in (\ref{Rdsigmaresult}) in the main text, after relabeling the dummy integrations variables as   $q_2 \rightarrow q_1$ and $q_3 \rightarrow q_2$.

\section{Semi-Analytic Approximation to the Numerical Result}
\label{app:semianalytic}

We devote this appendix to the understanding of the numerical results shown in Section \ref{sec:results} via semi-analytical methods.  We start our discussion with the equation of motion for the inflaton field, $\phi$, and its sourcing via the axion field, $\chi$, which is enhanced by the tensor mode of the gauge field, $t_L$: 
\begin{equation}
\left( \frac{\partial^2}{\partial \tau^2} + k^2 -\frac{2 + {\cal O} ( \epsilon, \eta) }{\tau^2} \right) {\hat \Phi} 
+ \frac{ {\cal O} ( \epsilon, \eta) }{\tau} \, {\hat \Phi}' 
\simeq C_X \,   \frac{{\hat X} }{\tau^2} \;  - \;  C_{X'} \,  \frac{{\hat X}'}{\tau} ,
\label{eqS}
\end{equation}
where  prime indicates derivative with respect to conformal time, and
\begin{eqnarray}
C_X \equiv \frac{\sqrt{2 \, \epsilon_\phi \, \epsilon_B} \,  \Lambda \, \left( k^4 \epsilon_E \, + 9 k^2 \epsilon_E \, \epsilon_B a^2 H^2 + 10 \epsilon_B^2 a^4 H^4  \right) }{\left(  k^2 \epsilon_E + 2 \epsilon_B \, a^2 \, H^2 \right)^2} \;\;,\;\; 
C_{X'} \equiv \frac{\sqrt{2} \sqrt{\epsilon_\phi} \, \epsilon_B^{3/2} \, \Lambda \,  a^2 \, H^2}{k^2 \epsilon_E + 2 \epsilon_B \, a^2 \, H^2} \,. \nonumber\\ 
\end{eqnarray} 

We obtained this expression starting from the quadratic action for the perturbations written in Appendix \ref{app:dsdc}, by integrating out the nondynamical modes, by extremizing the resulting action with respect to the inflaton perturbation, and by performing a slow roll expansion as explained around eq. (\ref{poly-slow}). We disregarded the subdominant term $6\sqrt{\epsilon_\phi \, \epsilon_\chi} $ inside $C_X$. We also disregard the slow-roll suppressed terms at the left hand side of (\ref{eqS}). 

We now consider the rescaled time variable $x=-k\tau$, so that $x > 1$ and $x<1$ correspond, respectively, to the sub-horizon and to the super-horizon regime. We want to solve eq. (\ref{eqS}) in the super-horizon regime. 
We recall that ${\hat X}$ is the canonical variable associated to the axion perturbations. In the de-Sitter limit, $x {\hat X} = \frac{k}{H} \, \delta \chi$. Since the axion perturbations are frozen outside the horizon, they satisfy  ${\hat X} = \frac{{\cal D}^{(s)}_\chi}{x}$, with constant ${\cal D}^{(s)}_\chi$, at $x \la 1$. With this in mind, eq. (\ref{eqS}) in the super horizon regime can be approximated as 
\begin{equation}
\left( \frac{\partial^2}{\partial x^2}  + 1 - \frac{2}{x^2}  \right) {\hat \Phi} \simeq  \frac{ C_X \, {\cal D}^{(s)}_\chi }{x^3} - \frac{ C_{X'} \, {\cal D}^{(s)}_\chi }{x}\partial_x \left(\frac{1}{x} \right) \simeq \frac{ \left(C_X + C_{X'}  \right) \, {\cal D}^{(s)}_\chi }{x^3} \;. 
\label{eqS2} 
\end{equation}

We define $  {\cal C} \equiv C_X + C_{X'} \to  3 \sqrt{ 2 \, \epsilon_\phi \, \epsilon_B} \, \Lambda$ in the super-horizon regime $x \to 0$. This equation is solved by 
\begin{equation}
{\hat \Phi} = \left( 1 + \frac{i}{x} \right) \, \frac{{\rm e}^{ix}}{\sqrt{2 k}} + {\hat \Phi}^{(s)} \;, 
\end{equation}
where the first term is the Bunch–Davies vacuum solution, which gives the vacuum inflaton mode $ {\hat \Phi}^{(s)} \to \frac{i}{\sqrt{2 k} x}$ in the super-horizon regime, while the second term is the one sourced by the right hand side of (\ref{eqS2}). In the super-horizon regime this second term gives 
\begin{equation}
{\hat \Phi}^{(s)} \to  \frac{- {\cal C} \,  {\cal D}^{(s)}_\chi \, \left(  -1 +\gamma + \ln{x}  \right)}{3 x} \simeq \frac{ - {\cal C}  \,  {\cal D}^{(s)}_\chi \, \ln{x} }{3 x} =  \frac{ - {\cal C}  \,  {\cal D}^{(s)}_\chi \, N_k }{3 x} \; ,
\label{eqnanasourcedinfla}
\end{equation}
where in the last step $N_k$ denotes the number of e-folds between the moment in which the mode leaves the horizon during inflation and the moment in which inflation ends. For the modes of our interest, $N_k \gg 1$, which justifies the approximation in the second step. 

Following \cite{Papageorgiou:2018rfx}, we denote by ${\cal R}_{\delta \chi} = \left\vert \frac{{\cal D}^{(s)}_\chi}{{\cal D}^{(v)}_\chi}  \right\vert^2$ the ratio between the power of the sourced vs. the vacuum modes of the axion field.  In eq. (\ref{Rdsigma-def}) we have defined the analogous ratio ${\cal R}_{\delta \phi}$ for the inflaton modes. This gives 
\begin{equation}
{\cal R}_{\delta \phi} =  \frac{{\cal C}^2 \,  N_k^2 \, {\cal R}_{\delta \chi} }{9} \, \left\vert \frac{{\cal D}^{(v)}_\chi}{x \, {\hat \Phi}^{(v)}} \right\vert^2 = 2 \epsilon_\phi \epsilon_B \, \Lambda^2 \, N_k^2 \, {\cal R}_{\delta \chi}  \, \left\vert \sqrt{2 k} \, {\cal D}^{(v)}_\chi \right\vert^2 \;, 
\label{Rds}
\end{equation} 
where we recall that we are working in the super-horizon regime. 

In order to proceed further, we need the expressions for the ratio of the sourced axion fluctuations to the vacuum axion fluctuations, ${\cal R}_{\delta \chi}$, and for the amplitude of vacuum axion fluctuations in the super-horizon regime resulting from nonlinear dynamics with the scalar degrees of freedom of the gauge field, $\sqrt{2k}{\cal D}_\chi^{(v)}$. In ref. \cite{Papageorgiou:2018rfx}, we provided a fit for the dependence of these quantities on $m_Q$, namely
\begin{equation}
 \vert t_L(x) \vert = \vert t_L \vert_{\rm peak} \cdot e^{- m_Q \log^2 \left(\frac{x}{x_{\rm p}}\right)} \; ,  \qquad \vert t_L \vert_{\rm p} \simeq  \frac{8}{3} \,\,  \sqrt{m_Q} \,\,  e^{\frac{\pi}{2} m_Q}  \qquad {\rm and} \qquad x_{\rm p} \equiv \frac{4}{9}m_Q \,, 
\label{tensorIR} 
\end{equation}
(that we also write as eq. (\ref{tensorIR-main}) in the main text) for the behaviour of the unstable SU(2) tensor mode around its maximum value, 
\begin{equation}
\frac{{\rm Im} \left( {\tilde X}_c(x) {\tilde X}_c(x_1)^* \right)}{|{\tilde X}_c(x)|}\bigg|_{x\to0} \simeq \frac{1}{\Lambda}\left(0.22-\frac{0.34}{m_Q}\right)x_1^2  \;, 
\label{greenIR}
\end{equation}
(where ${\tilde X}_c$ is rescaled according to eq. (\ref{code-var})) for the super-horizon behaviour of the axion perturbations sourced by $t_L$, and 
\begin{equation}
\left\vert \sqrt{2 k} \, {\cal D}^{(v)}_\chi \right\vert \simeq  \frac{\sqrt{1+m_Q^2}}{\Lambda}\left(\frac{-12.5}{m_Q}+\frac{255}{m_Q^2}-\frac{897}{m_Q^3}+\frac{1050}{m_Q^4} \right)  \equiv  {\widetilde {\cal D}}_\chi^{(v)} (m_Q) \frac{1}{\Lambda} \;, 
\label{axionIR}
\end{equation}
for the axion perturbations obtained from the linear theory. 

By employing the expressions given in \eqref{tensorIR}, \eqref{greenIR}  and \eqref{axionIR}, one can first evaluate the time integral in (\ref{finalresult}) analytically, and obtain

\begin{eqnarray}
{\cal R}_{\delta \chi} (g,m_Q) \simeq  \frac{1}{256 \pi^2} \frac{g^2(1+m_Q^2)}{m_Q^2}\,   \, \vert t_L(m_Q) \vert^4 \int dp \, \int_{\vert1-p\vert}^{1+p} dq 
\;  \left( 2 \; \frac{1- \left(p+q \right)^2}{p^2-q^2} \right)^4
 \; {\cal T}^2(p,q) \;, \nonumber\\ 
\end{eqnarray}
where 
\begin{eqnarray}
{\cal T}(p ,q) &\equiv & \frac{\sqrt{\pi}\,\, x_p^3\,\, e^{-m_Q \log^2\left(\frac{p}{q}\right)}}{2 \sqrt{2}\,\sqrt{m_Q}\,\,p^3}\Bigg\{\frac{6 \,m_Q\,\, p \,\,e^{\frac{\left[1+m_Q\log\left(\frac{p}{q}\right)\right]^2}{2 m_Q}}}{x_p} \nonumber\\
&& +e^{\frac{\left[3+2 m_Q\, \log\left(\frac{p}{q}\right)\right]^2}{8 \,m_Q}}\left[-3\left(p+q\right)+2\, m_Q \left(p-q\right)\, \log\left(\frac{p}{q}\right)\right]\Bigg\}\;, 
\end{eqnarray}
 is the analytical result of the time integration for each $t_L$ propagator in the one loop diagram that accounts for the production of $\delta \chi$. 

We integrate this expression numerically, and we fit the result as 
\begin{equation}
\!\!\!\!{\cal R}_{\delta \chi} (g,m_Q) \simeq  \frac{1}{256 \pi^2} \frac{g^2(1+m_Q^2)}{m_Q^2} \, \left(0.22-\frac{0.34}{m_Q}\right)^2\left[\frac{8}{3} \,\,  \sqrt{m_Q} \,\,  e^{\frac{\pi}{2} m_Q}  \right]^4 \,  \left(11.3+1.48 m_Q^5\right)  \equiv g^2 \, {\widetilde {\cal R}}_{\delta \chi} \left( m_Q \right) \,, 
\label{eqnRchi}
\end{equation}
where the third factor arises from taking the square of the combination in the second line of the \eqref{greenIR}, while the last factor is the numerical fitting to the numerically evaluated momentum integral. 

Combining the expression above with $ g^2 \, \epsilon_{B} = \frac{g^4 \, Q^4}{H^4} \, \frac{H^2}{M_p^2} = m_Q^4 \, \frac{H^2}{M_p^2}$, we have
\begin{equation}
{\cal R}_{\delta \phi} =   2 \, \epsilon_\phi   \, N_k^2  \, m_Q^4 \, \frac{H^2}{M_p^2}  \,   {\widetilde {\cal D}}_\chi^{(v)2} \left( m_Q \right) \,   {\widetilde {\cal R}}_{\delta \chi} \left( m_Q \right) \bigg\vert_{x \to0} \;, 
\end{equation} 
where we note that the explicit dependence on $\Lambda$ and $g^2$ has dropped. 

To proceed, we express the ratio  $H/M_p$ in terms of the amplitude of the vacuum tensor modes \cite{Baumann:2008aq} 
\begin{equation}
P_h^{(v)} =  \frac{2}{\pi^2} \frac{H^2}{M_p^2} \,, 
\end{equation}
and we define the parameter 
\begin{equation}
r_{\rm vac} \equiv \frac{P_h^{(v)}}{P_{\zeta,{\rm measured}}} \;, 
\end{equation} 
so that 
\begin{equation}
{\cal R}_{\delta \phi} =   2 \, \epsilon_\phi   \, N_k^2  \, m_Q^4 \, 
\frac{\pi^2}{2} \, r_{\rm vac} \, P_{\zeta,{\rm measured}}  \,   {\widetilde {\cal D}}_\chi^{(v)2} \left( m_Q \right) \,   {\widetilde {\cal R}}_{\delta \chi} \left( m_Q \right) \bigg\vert_{x \to0} \;. 
\end{equation} 
Moreover we can use $r_{\rm vac}  =  16 \, \epsilon_\phi \left( 1 + \frac{\epsilon_B}{\epsilon_\phi} \right)^2$ (which we derived in Appendix \ref{app:Pzandrvac}), giving 
\begin{equation}
{\cal R}_{\delta \phi} =  N_k^2  \, m_Q^4 \, 
\pi^2 \, \frac{r_{\rm vac}^2}{16\left(1+\frac{\epsilon_B}{\epsilon\phi}\right)^2} \, P_{\zeta,{\rm measured}}  \,   {\widetilde {\cal D}}_\chi^{(v)2} \left( m_Q \right) \,   {\widetilde {\cal R}}_{\delta \chi} \left( m_Q \right) \bigg\vert_{x \to0} \;, 
\end{equation} 
where we recall that  ${\widetilde {\cal D}}_\chi^{(v)2} \left( m_Q \right)$ and ${\widetilde {\cal R}}_{\delta \chi} \left( m_Q \right)$ are given, respectively, by eqs. (\ref{axionIR}) and (\ref{eqnRchi}).  By plotting the resulting expression, one can see that, in the range $2.5 \la m_Q \la 3.5$,  this product is well fitted by $0.007 \, m_Q^7 \times {\rm e}^{7 \, m_Q}$ (where the final exponent come from  (\ref{eqnRchi}), while the monomial $0.007 \, m_Q^7 $ is a fit). 

Using this, and $P_{\zeta,{\rm measured}} = 2.1 \cdot 10^{-9}$ \cite{Akrami:2018odb}, we arrive to 
\begin{equation}
{\cal R}_{\delta \phi} \simeq \frac{10^{-11}}{\left(1+\frac{\epsilon_B}{\epsilon\phi}\right)^2} \,  m_Q^{11} \, {\rm e}^{7 \, m_Q}   \,  N_k^2  \, r_{\rm vac}^2 \,. 
\label{R-ds-analytical}
\end{equation} 

\begin{figure}[tbp]
\begin{center}
\includegraphics[width=0.76\textwidth,angle=0]{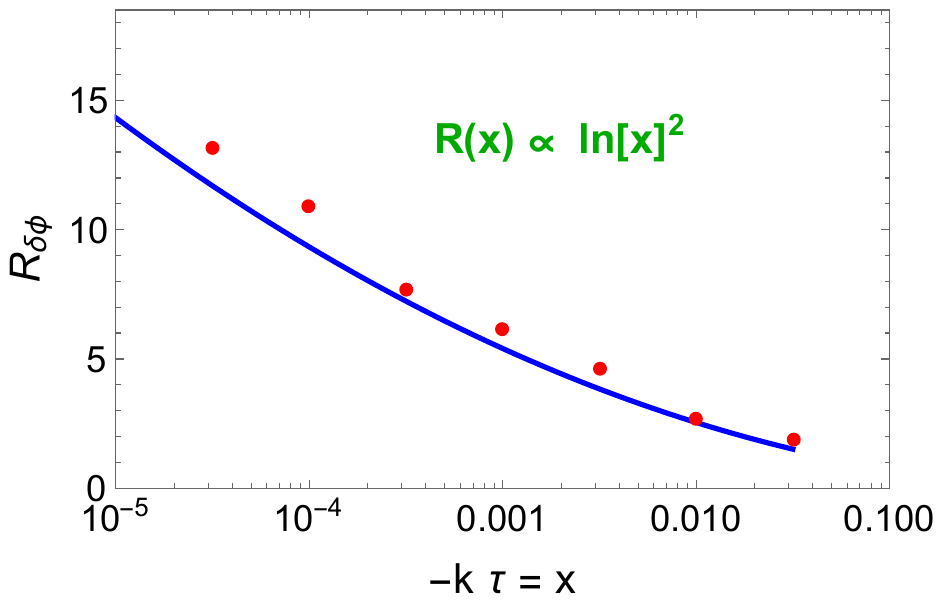}
\end{center}
\vspace{-0.75cm}
\caption{Scaling with time of the ratio ${\cal R}_{\delta \phi}$ in the super-horizon $x \ll 1$ regime. 
The solid line is the analytic result (\ref{R-ds-analytical-resc}), for the choice of $r_{\rm vac} = 10^{-2}$ and $m_Q = 2.986$ (we recall that $N_k = \ln x$). The dots correspond to the numerical evaluation of eq. (\ref{finalresult}). For simplicity in the comparison between the analytic and numerical results we have assumed that $\epsilon_B\ll\epsilon_\phi$.}
\label{fig:loggrowth}
\end{figure}

\begin{figure}[tbp]
\begin{center}
\includegraphics[width=0.76\textwidth,angle=0]{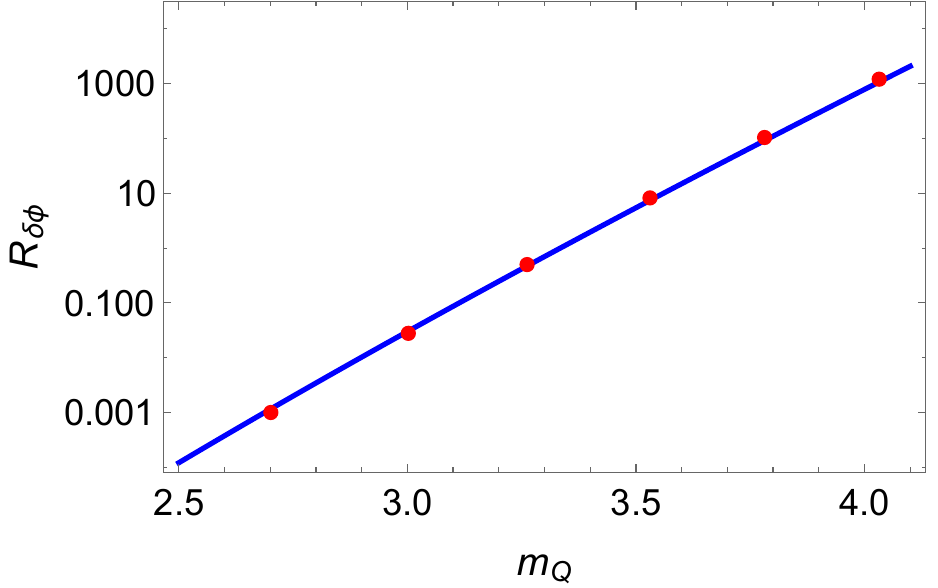}
\end{center}
\vspace{-0.75cm}
\caption{
Scaling with $m_Q$ of the ratio ${\cal R}_{\delta \phi}$. The solid line is the analytic result  (\ref{R-ds-analytical-resc}), for the choice of $r_{\rm vac} = 10^{-3}$ and for $x = 0.01$.  The dots correspond to the numerical evaluation of eq. (\ref{finalresult}). The value $x = 0.01$, corresponding to $N_k \simeq 4.6$, is assumed in this figure. Similarly to the previous figure $\epsilon_B\ll\epsilon_\phi$ has been assumed for the purposes of comparing the numerical and semi-analytic result.
}
\label{fig:Rinflvsmq}
\end{figure}

Remarkably, this expression only depends on $m_Q$, on $r_{\rm vac}$ (or equivalently $H$), on $N_k$ and on the ratio $\frac{\epsilon_B}{\epsilon_\phi}$. As we discussed, the result (\ref{R-ds-analytical}) grows as $\log^2 \left( x \right) = N_k^2$ in the super-horizon regime. We verified this expression against a fully numerical evaluation of eq. (\ref{finalresult}). We found that dividing eq. (\ref{R-ds-analytical}) by two reproduces the numerical results with sufficient accuracy, up to  $\sim 20\% $ corrections. Therefore we modify  (\ref{R-ds-analytical}) with 
\begin{equation}
{\cal R}_{\delta \phi} \simeq \frac{5 \cdot 10^{-12}}{\left(1+\frac{\epsilon_B}{\epsilon\phi}\right)^2} \,  m_Q^{11} \, {\rm e}^{7 \, m_Q}   \,  N_k^2  \, r_{\rm vac}^2 \,. 
\label{R-ds-analytical-resc}
\end{equation} 
which we also report as eq. (\ref{eqfittedRinf}) of the main text. Figures \ref{fig:loggrowth} and  \ref{fig:Rinflvsmq} confirm the accuracy of this expression.

\newpage 
\section{Direct subdominant $\delta \phi t_L t_L$ interaction}
\label{app:subdominant}

Integrating out the non-dynamical perturbations from the action (\ref{S2-sig-chi-nondyn}) we find that the $0-$th component of the gauge field can be related to the inflaton perturbations by 
\begin{eqnarray} 
\delta A_0^a &=& - \frac{\sqrt{2 \epsilon_E \epsilon_\phi} a^2 \, H}{- \Delta \left( - \epsilon_E \, \Delta +2 \epsilon_B \, a^2 \, H^2 \right)} \left[  \epsilon_B H \,  \partial_\tau + \left( - \epsilon_E \, \Delta-2 \epsilon_B \, a^2 \, H^2 \right) \frac{1 }{2 a} \right] \partial_a \Phi + \dots \nonumber\\
&=&  - \frac{\sqrt{2 \epsilon_E \epsilon_\phi} a^2 \, H}{- \Delta \left( - \epsilon_E \, \Delta + 2 \epsilon_B \, a^2 \, H^2 \right)} 
\left( \epsilon_B \, a \, H  \, \partial_\tau  - \frac{\epsilon_E}{2} \, \Delta \,  \right) \partial_a \delta \phi + \dots \;, 
\label{dA-sigma}
\end{eqnarray} 
where $\Delta$ is the spatial Laplacian in comoving coordinates, and where the dots denote the term proportional to the axion perturbations that contribute to the $t_L t_L \delta \chi$ interaction (\ref{intctt}) \cite{Papageorgiou:2018rfx}. Proceeding in the same way, we find that the nondynamical metric perturbations are related to the inflaton perturbations by 
\begin{eqnarray} 
\Phi &=& \frac{\sqrt{\epsilon_\phi}}{\sqrt{2} M_p} \left( 1 - \frac{a^2 \, H \epsilon_E \epsilon_B}{-\epsilon_E \, \Delta + 2 \epsilon_B \, a^2 \, H^2 } \, \frac{\partial_\tau}{a} \right) \delta \phi + \dots \;\;, 
 \nonumber\\ 
B &=& \frac{\sqrt{\epsilon_\phi} \, \partial_\tau}{-\sqrt{2} M_p \, \Delta} \, \delta \phi + \dots \;, 
\label{Phi-B-sigma}
\end{eqnarray} 
(where again the dots denote terms proportional to the CNI axion, which play no role in this discussion). 

The terms (\ref{dA-sigma}) and (\ref{Phi-B-sigma}) can induce interactions between the inflaton mode and $t_L$  (without involving the CNI scalar dynamical modes). We note that these interactions cannot arise from the  pseudo-scalar term in the action of the model. Firstly, this term does not contain the metric tensor. Secondly,  interactions between  $\delta A_0^3$ and the $t_L$ mode cannot be induced by 
\begin{equation}
 -  \frac{\lambda}{8 f} \chi \epsilon^{\mu \nu \alpha \beta} F_{\mu \nu}^a F_{\alpha \beta}^a =   \frac{\lambda}{f} \chi \partial_\phi J^\phi \;\;,\;\; 
J^\phi = \epsilon^{\mu \nu \rho \phi} \left( \frac{1}{2} A_\mu^a \partial_\nu A_\rho^a  + \frac{g}{6} \epsilon_{abc} A_\mu^a A_\nu^b A_\rho^c \right) \;, 
\end{equation} 
because, after integrating by parts, we see that the index $\phi$ is forced to be zero, so that the indices $\mu, \nu, \rho$ entering in $J^\phi$ are forced to be spacial indices. On the other hand, the vector kinetic term induces  interaction that are formally of the type 
\begin{eqnarray} 
\sqrt{-g} \, {\cal L} &\supset& - \frac{\sqrt{-g}}{4} F_{\mu \nu}^a F^{\mu \nu,a} 
\supset \; {\cal O } \left( {\rm scalar} \, t_L^2 \right) +  {\cal O } \left( {\rm scalar}^2 \, t_L \right)  +  {\cal O } \left( {\rm scalar}^2 \, t_L^2 \right) \;, 
\end{eqnarray} 
where the scalar modes are initially non-dynamical modes, that are then written in terms of the inflaton perturbations 
through  (\ref{dA-sigma}) and (\ref{Phi-B-sigma}). The  terms of the last two types  give rise to diagrams that are suppressed either because they involve fewer $t_L$ mode functions or the $t_L$ mode function in the zero momentum limit (which is, therefore, not enhanced). This is discussed in more details after eq. (4.2) of  \cite{Papageorgiou:2018rfx}. Therefore, we disregard them. The terms of the first type are included into 
\begin{eqnarray} 
 - \frac{\sqrt{-g}}{4} F_{\mu \nu}^a F^{\mu \nu,a}  \Big\vert_{{\cal O } \left( {\rm scalar} \, t_L^2 \right) } 
&\subset&  \frac{1-\Phi}{2} \, F_{0i}^a \, F_{0i}^a - \partial_i B \, F_{0j}^a \, F_{ij}^a - \frac{\Phi}{4} \, F_{ij}^a \, F_{ij}^a \;. 
\end{eqnarray} 

From this we obtain    
\begin{eqnarray} 
 - \frac{\sqrt{-g}}{4} F_{\mu \nu}^a F^{\mu \nu,a}  \Big\vert_{{\cal O } \left( {\rm scalar} \, t_L^2 \right) } &=& 
  g \, \epsilon^{abc} \, t_{ai}' \, \delta A_0^b \, t_{ci} - \frac{1}{2} \, \Phi \, t_{ai}' \, t_{ai}' 
- \left( \partial_i B \, t_{aj}' - \partial_j B \, t_{ai}' \right) \left( \partial_i t_{aj} + a Q g \epsilon^{aib} \, t_{bj} \right) \nonumber\\ 
  & & - \frac{\Phi}{2} \partial_i t_{aj} \left( \partial_i t_{aj} -   \partial_j t_{ai}  \right) 
 + \Phi \, a Q \,  g \, \epsilon^{abi} \, \partial_i t_{aj} \, t_{bj} \;. 
\end{eqnarray} 

Inserting (\ref{dA-sigma}) and (\ref{Phi-B-sigma}) into this expression provides a direct coupling for the $t_L + t_L \rightarrow \delta \phi$ process, in addition to the production channels considered in the main text. In the remainder of this appendix we estimate the production due to these couplings. 

Based on the results for the $t_L + t_L \rightarrow \delta \chi$ production computed in \cite{Papageorgiou:2018rfx}, and for what is typically obtained for models with perturbations sourced by vector fields, in our estimate we assume that the production takes place mostly at horizon crossing, so that we can substitute $\partial_t \rightarrow a H$ and $\Delta \rightarrow - a^2 H^2$ in the relations  (\ref{dA-sigma}) and (\ref{Phi-B-sigma}) 
\begin{equation}
\delta A_0^a \sim - \partial_a \left( \frac{m_Q}{g} \, \sqrt{\frac{\epsilon_\phi}{2}} \, \frac{\delta \phi}{M_p} \right) \;\;\;,\;\;\; 
a \, H \, B \sim \Phi \sim \sqrt{\frac{\epsilon_\phi}{2}} \, \frac{\delta \phi}{M_p}  \;\;,\;\; 
\end{equation}
where the last of (\ref{slow}) has been used, with $\dot{Q} = 0$, as well as $Q = \frac{H \, m_Q}{g}$. 

Analogously, we replace the derivatives acting on $t_L$ by $a \, H \times m_Q$, where the last factor is due to the (parametric) position of the peak in $t_L$. This results in an interaction that is parametrically of the type 
\begin{equation}
H_{\rm int, \delta \phi \, t \, t} \sim \frac{m_Q^2 \sqrt{\epsilon_\phi} a^2 H^2}{M_p} \, \delta \phi \, t_{ia} \, t_{ia} \;, 
\label{dstt-direct}
\end{equation}
times an order one factor. We denote by $\left\langle \delta \phi  \delta \phi \right\rangle_{\rm direct}$ the two-point correlation function of the inflaton perturbations produced by this vertex, and by ${\cal R}_{\delta \phi,{\rm direct}}$ the ratio between this quantity and the one computed in the rest of this work, namely the one given in eq. (\ref{Rds}). We want to show that this ratio is much smaller than one, so that the terms considered in this appendix can indeed be neglected. We have 
\begin{equation}
\frac{{\cal R}_{\delta \phi,{\rm direct}}}{{\cal R}_{\delta \phi}} = 
\frac{
\frac{\left\langle \delta \phi  \delta \phi \right\rangle_{\rm direct}}{\left\langle \delta \phi \delta \phi \right\rangle_{\rm vacuum}}
}
{
\frac{{\cal C}^2 \,  N_k^2}{9} \, \frac{\left\langle  \delta \chi \delta \chi \right\rangle_{\rm sourced}}{\left\langle  \delta \chi \delta \chi \right\rangle_{\rm vacuum}}\, \left\vert \frac{{\cal D}^{(v)}_\chi}{x \, {\hat \Phi}^{(v)}} \right\vert^2 
 } \,. 
\end{equation} 
Next, we use the fact that $\left\langle  \delta \phi \delta \phi \right\rangle_{\rm vacuum} = \left\vert x \, {\hat \Phi}^{(v)} \right\vert^2$ and that  $\left\langle  \delta \chi \delta \chi \right\rangle_{\rm vacuum} = \left\vert {\cal D}^{(v)}_\chi \right\vert^2$
to write 
\begin{equation}
\frac{{\cal R}_{\delta \phi,{\rm direct}}}{{\cal R}_{\delta \phi}} = 
\frac{9}{{\cal C}^2 \,  N_k^2} \, \frac{\left\langle \delta \phi  \delta \phi \right\rangle_{\rm direct}}{\left\langle  \delta \chi \delta \chi \right\rangle_{\rm sourced}} \,. 
\label{ratio-direct-1}
\end{equation} 

The numerator of this expression is associated to (\ref{dstt-direct}). The corresponding vertex for the denominator is  the direct $\delta \chi \, t \, t$ interaction, that we computed in  \cite{Papageorgiou:2018rfx}. We take eq. (4.6) of that work, and perform the same estimates as those done to obtain  (\ref{dstt-direct}). This leads to 
\begin{equation}
H_{\rm int, \delta \chi \, t \, t} \sim \frac{\lambda}{f} \, a^2 H^2 \, m_Q^2 \, \delta \chi \, t_{ai} \, t_{ai} \;. 
\end{equation}

The last ratio in (\ref{ratio-direct-1}) can be thus estimated by taking that ratio of the squares of the couplings in the corresponding interaction Hamiltonians, and the ratios of the Green functions of the inflaton vs. the axion modes (the amplitudes of the $t_L$ fields cancel in the ratio) 
\begin{equation}
\frac{{\cal R}_{\delta \phi,{\rm direct}}}{{\cal R}_{\delta \phi}} \sim 
\frac{9}{{\cal C}^2 \,  N_k^2} \times 
\left( \frac{
\frac{m_Q^2 \sqrt{ \epsilon_\phi } a^2 H^2}{M_p} 
}{
\frac{\lambda}{f} \, a^2 H^2 \, m_Q^2 
} \right)^2 \times 
\frac{
G_\phi^2 \left( 0.01,\, \frac{m_Q}{2} \right)
}{
G_\chi^2 \left( 0.01,\, \frac{m_Q}{2} \right)
}  \,. 
\label{ratio-direct-2}
\end{equation} 

The arguments of the Green functions are, respectively, the (rescaled) time $x = - k \tau $ at which the perturbations is evaluated and the (rescaled) time at which the source is peaked. We take $0.01$ for the former, just outside the horizon (the fact that the sourced term considered in this paper is active all throughout the super-horizon regime, while the direct coupling is effective only at horizon crossing results in the $1/N_k^2$ suppression accounted for by this factor in (\ref{ratio-direct-2})).  

Using $  {\cal C} \simeq  3 \sqrt{ 2 \, \epsilon_\phi \, \epsilon_B} \, \Lambda$ this ratio simplifies to 
\begin{equation}
\frac{{\cal R}_{\delta \phi,{\rm direct}}}{{\cal R}_{\delta \phi}} \sim 
\frac{1}{2 \, N_k^2 \, m_Q^2 \, \Lambda^4 } \; 
\frac{
G_\phi^2 \left( 0.01,\, \frac{m_Q}{2} \right)
}{
G_\chi^2 \left( 0.01,\, \frac{m_Q}{2} \right)
}  \,. 
\end{equation} 
where the expression for $\epsilon_B$ in (\ref{slow}), and eqs. (\ref{params}) have also been used. 

The Green function for $\phi$ (in the small $x$ limit) is given by 
\begin{equation}
i G_\phi \left( x ,\, x' \right) \equiv \delta \phi \left( x \right) \, \delta \phi^* \left( x' \right) - {\rm c.c.} =  
 -\frac{H^2\left[-x' \cos\left(x'\right)+\sin\left(x'\right)\right]}{k^3} \;. 
\end{equation} 
The Green function for $\chi$ can be rewritten as 
\begin{eqnarray} 
i G_\chi \left( x ,\, x' \right) &\equiv& \delta \chi \left( x \right) \, \delta \chi^* \left( x' \right) - {\rm c.c.} =  -
\frac{2 H^2 x'}{k^2} \, \frac{{\rm Im } \left( X \left( x \right) \,  X^* \left( x' \right) \right)}{\left\vert \sqrt{2 k}\,X \left( x \right) \right\vert} \, 
\left\vert \sqrt{2 k} \, {\cal D}^{(v)}_\chi \right\vert \nonumber\\ 
& = &- \frac{H^2 x'}{k^3 \, \Lambda} \, \sqrt{1+m_Q^2} \, \frac{{\rm Im } \left( \tilde{X}_c \left( x \right) \,  \tilde{X}_c^* \left( x' \right) \right)}{\left\vert \tilde{X}_c \left( x \right) \right\vert} \,  {\widetilde {\cal D}}_\chi^{(v)} \left( m_Q \right) 
\equiv - \frac{H^2 x'}{k^3 \, \Lambda^2} \, {\tilde G}_\chi\left(x'\right)\;, \nonumber\\ 
\end{eqnarray} 
where we recognize the product between the two quantities in eqs. (\ref{greenIR}) and (\ref{axionIR}). We note that the quantity ${\tilde G}_\chi $ does not depend on $\Lambda$. With this in mind, we have 

\begin{equation}
\frac{G_\phi^2 \left(0,\, x' \right)
}{
G_\chi^2 \left(0,\,x' \right)}=\Lambda^4\left(\frac{x' \cos\left(x'\right)-\sin\left(x'\right)}{x'{\tilde G}_\chi\left(x'\right)}\right)^2 \;. 
\end{equation}

The expression for the ratio of the two contributions can then be rewritten as

\begin{equation}
\frac{{\cal R}_{\delta \phi,{\rm direct}}}{{\cal R}_{\delta \phi}} \sim 
\frac{1}{2 \, N_k^2 \, m_Q^2} \; 
\left(\frac{\frac{m_Q}{2} \cos\left(\frac{m_Q}{2}\right)-\sin\left(\frac{m_Q}{2}\right)}{\frac{m_Q}{2}{\tilde G}_\chi\left(\frac{m_Q}{2}\right)}\right)^2 \;. 
\end{equation} 

\begin{figure}[tbp]
\begin{center}
\includegraphics[width=0.7\textwidth,angle=0]{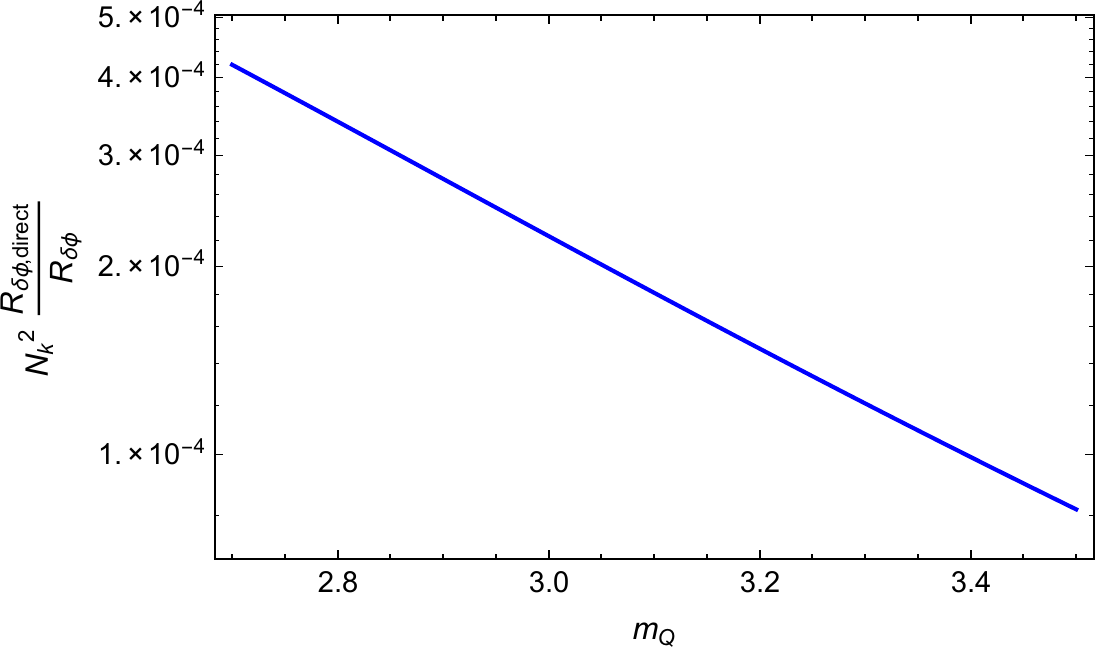}
\end{center}
\caption{Estimate of the ratio between the power of the inflaton perturbations 
produced by the direct interaction considered in this Appendix and the one considered in the rest of this work. This result confirms that the interactions considered in this Appendix can be neglected. 
}
\label{fig:direct}
\end{figure}

The final expression is only a function of the parameter $m_Q$ as well as the number of e-folds $N_k$. In Figure \ref{fig:direct} it is shown that for the range of values of the parameters that we are considering in the present work, the above quantity is always small and therefore the direct coupling can safely be neglected.

\section{Contributions to the Curvature Perturbation}
\label{app:curva}

In this work we have assumed that the inflaton perturbations $\delta \phi$ control the curvature perturbation $\zeta$. In general, one needs to verify that the contribution from the axion perturbations $\delta \chi$ is negligible  \cite{Fujita:2018vmv}. In this appendix we  discuss when this is the case. In spatially flat gauge, the curvature perturbation is written as
\begin{equation}
\zeta = - H \frac{\delta \rho} {\dot \rho} \simeq - H \: \frac{  V' \left( \phi \right) \delta \phi + U' \left( \chi  \right) \delta \chi }{\dot{\rho}}  \equiv \zeta_\phi + \zeta_\chi \;, 
\end{equation}
There are two main contributions to $\zeta$. Requiring that the nonlinear contributions to the inflaton perturbations that we have computed here are smaller than the linear contributions, ${\cal R}_{\delta \phi } \ll 1$,  and using eqs. (\ref{params}) and (\ref{chiSR}) for $U'$, one can write 
\begin{equation}
 \frac{ P_{\zeta_\phi}}{ P_{\zeta_\chi}} \simeq \frac{ 9 H^2 \dot \phi^2 \frac{ H^2}{4\pi^2}} {U^{ ' \, 2}(\chi) \,  
  \left[  P_{\delta \chi^{(v)}}+ P_{\delta \chi^{(s)}} \right]} \simeq \frac{9 H^2 \dot \phi^2 \frac{ H^2}{4\pi^2}} { 9 \Lambda^2 m_Q^2 H^4 Q^2 \, \left[ 1 + {\cal R}_{\delta \chi} \right] \, P_{\delta \chi^{(v)}}} \;. 
\end{equation}

\begin{figure}[tbp]
\begin{center}
\includegraphics[width=0.48\textwidth,angle=0]{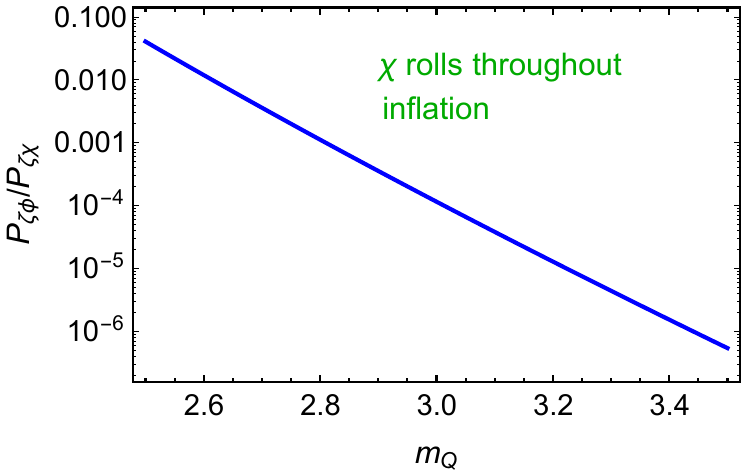}
\end{center}
\caption{
Ratio between the contribution to the curvature power spectrum of the (linear) inflaton and the (nonlinear) axion perturbations, under the assumption that the axion rolls throughout inflation and also that $\epsilon_B\ll\epsilon_\phi$. A much greater ratio is obtained if the axion becomes massive during inflation.  
}
\label{fig:Pdc-vac-s}
\end{figure}

In the parameter space analyzed in the main text the sourced axion perturbations always dominate over the vacuum axion perturbations. Therefore, in our discussion below, we only consider the sourced axion modes. As we now see, this allows to obtain an expression for $  P_{\zeta_\phi} /  P_{\zeta_\chi} $ that only depends on $m_Q$. 

To this end, we note that from eq.~(\ref{axionIR}) one can write 
\begin{eqnarray} 
P_{\delta \chi^{(v)}}^{1/2} &=& \frac{H}{2 \pi} \, \left\vert \sqrt{2 k} \, {\cal D}_\chi^{(v)} \right\vert = 
\frac{H}{2\pi} \; \frac{\sqrt{1+m_Q^2}}{\Lambda}\left( 2.88 - \frac{5.58}{m_Q} + \frac{52.9}{m_Q^2} - \frac{191}{m_Q^3} +  \frac{326}{m_Q^4} \right) \nonumber\\ 
&\simeq & \frac{H}{2\pi} \; \frac{\sqrt{1+m_Q^2}}{\Lambda} \times {\cal O} \left( 1 \right) \;, 
\end{eqnarray} 
and, using  $R_{\delta \chi} \equiv g^2 \, {\widetilde {\cal R}}_{\delta \chi} \left( m_Q \right) $ from eq.  (\ref{eqnRchi}), we have 
\begin{equation}
 \frac{ P_{\zeta_\phi}}{ P_{\zeta_\chi}} \simeq  \frac{ \dot \phi^2} { m_Q^4 \, (1+m_Q^2) \, H^4 \, {\widetilde {\cal R}}_{\delta \chi}  \left( m_Q \right) } \,. 
\end{equation}

Assuming that the measured value of the CMB temperature anisotropies is all due to the inflationary perturbations we then have $P_{\zeta,{\rm measured}} \simeq 2.1 \cdot 10^{-9} \simeq \frac{H^4}{4 \pi^2 \dot \phi^2 \left(1+\frac{\epsilon_B}{\epsilon_\phi}\right)^2}$ (see eq (\ref{eqPzeta})  \cite{Akrami:2018odb}, and so  
\begin{equation}
 \frac{ P_{\zeta_\phi}}{ P_{\zeta_\chi}} \Big\vert_{\chi \; {\rm rolls}} \simeq  \frac{ 1} { m_Q^4 \, (1+m_Q^2) \, 4 \pi^2 \, P_{\zeta, {\rm  measured}} \; \left(1+\frac{\epsilon_B}{\epsilon_\phi}\right)^2 \, {\widetilde {\cal R}}_{\delta \chi}  \left( m_Q \right) } \;. 
\label{Ps-Pc-roll}
\end{equation}

We plot the ratio $P_{\zeta_\phi} /  P_{\zeta_\chi} $ in Figure \ref{fig:Pdc-vac-s}, assuming $\epsilon_B \ll \epsilon_\phi$. We note that this ratio is much smaller than one for nearly all the parameter space we have considered.  If the sourced axion fluctuations dominate over the inflaton fluctuations, then the curvature fluctuations obey a chi-square distribution  (as in the U(1) case \cite{Linde:2012bt}) due to the 2-to-1 sourcing $t_L+t_L \to \delta \chi$. We note that the result (\ref{Ps-Pc-roll}) assumes that the axion rolls all throughout inflation. If the axion becomes massive during inflation,  its energy starts redshifting away as $a^{-3}$, while the energy of the inflaton remains nearly constant. Accordingly, the contribution of the axion to the curvature perturbation strongly decreases. Assume that the CMB modes are produced $N_{\rm CMB} \simeq 60$ e-folds before the end of inflation, while the axion is running. Assume that the axion runs for another $N_k$ e-folds (in Figure  \ref{figparameterspace} we have considered the two cases $N_k = 10$ and $N_k = 50$), and then it reaches a minimum of its potential, where it becomes very massive (with a mass $m > \frac{3}{2} H$). We denote by $N_* = N_{\rm CMB} - N_k$ the number of e-folds of inflation  during which $\phi$ is massive (from the moment it reaches the minimum of the potential to the end of inflation). We then have 
\begin{equation}
\frac{ P_{\zeta_\phi}}{ P_{\zeta_\chi}} \Big\vert_{\rm end \; of \; inflation} \sim {\rm e}^{3 \, N_*} \, 
\frac{ P_{\zeta_\phi}}{ P_{\zeta_\chi}} \Big\vert_{\phi \; {\rm rolls}} \;. 
\label{Ps-Pc-tot}
\end{equation} 

Even assuming that $N_* = 10$ gives an increase of about $13$ orders of magnitude with respect to the result shown in the right panel of Figure \ref{fig:Pdc-vac-s}. In this work we assume that the increase due to (\ref{Ps-Pc-tot}) is such that the observed curvature perturbations are dominated by the inflaton field.

\section{Generalized slow roll relations for $P_\zeta$ and $r_{\rm vac}$}
\label{app:Pzandrvac}

In this Appendix we derive the relation between the contribution $r_{\rm vac}$ of the vacuum GW to the tensor-to-scalar ratio and the slow roll parameters.~\footnote{We are indebted to Azadeh Maleknejad for private communications on this issue.}  We extend the standard relation to the case in which the slow roll parameter $\epsilon_B$ is not necessarily smaller than the parameter $\epsilon_\phi$. 

In the spatially flat gauge we can compute the primordial density perturbation
\begin{equation}
\zeta=-H\frac{\delta \rho}{\dot{\rho}}\simeq\frac{V_{,\phi} \delta\phi}{6 \epsilon_H H^2 M_p^2}\simeq \frac{1}{\sqrt{2 \epsilon_H}}\left(\frac{\epsilon_\phi}{\epsilon_H}\right)^{\frac{1}{2}}\frac{\delta\phi}{M_p} \;, 
\end{equation}
where the first of (\ref{slow}) has been used, with $\epsilon_H$ is defined in eq (\ref{slow-relation})

The power spectrum of scalar perturbations is then found to be 
\begin{equation}
P_\zeta=  H^2 \left \langle \frac{\delta \rho^2}{ \dot{\rho}^2} \right \rangle \simeq \frac{1}{2\epsilon_H}\left(\frac{\epsilon_\phi}{\epsilon_H}\right)\left(\frac{H}{2\pi M_p}\right)^2 \simeq 
\frac{H^2}{8 \pi^2 \, M_p^2} \, \frac{\epsilon_\phi}{\left( \epsilon_\phi + \epsilon_B \right)^2} \;, 
\label{eqPzeta}
\end{equation}
where in the last step we have used the fact that  $\epsilon_B$ is the dominant CNI slow-roll parameter in the sum (\ref{slow-relation}). From the definitions of $\epsilon_B$ and $m_Q$ (respectively, eqs. (\ref{slow}) and (\ref{params}) of the main text), we can write $\frac{H^2}{M_p^2} = \frac{g^2 \, \epsilon_B}{m_Q^4}$. In this way, the expression for the scalar power spectrum can be also written as 
\begin{equation}
P_\zeta \simeq \frac{g^2}{8 \pi^2 \, m_Q^4} \, \frac{\epsilon_\phi \epsilon_B}{\left( \epsilon_\phi + \epsilon_B \right)^2} \;, 
\label{eqPzeta-ephi-eB}
\end{equation} 

We can also use eq. (\ref{eqPzeta}) to obtain a simple expression for  the ratio between the power spectrum of the vacuum tensor perturbations $P_h^{(v)} \simeq  \frac{2}{\pi^2} \frac{H^2}{M_p^2}$: 

\begin{equation}
r_{\rm vac}=\frac{P_h^{(v)}}{P_\zeta}\simeq 16 \frac{\epsilon_H^2}{\epsilon_\phi} \simeq  16 \, \epsilon_\phi \left( 1 + \frac{\epsilon_B}{\epsilon_\phi} \right)^2 \;, 
\end{equation}
We see that the standard relation $r_{\rm vac}=16\epsilon_\phi$ is recovered in the $\epsilon_\phi \gg \epsilon_B$ case.


\begin{thebibliography}{99}


\bibitem{Freese:1990rb} 
  K.~Freese, J.~A.~Frieman and A.~V.~Olinto,
  Phys.\ Rev.\ Lett.\  {\bf 65}, 3233 (1990).
  doi:10.1103/PhysRevLett.65.3233

\bibitem{Pajer:2013fsa} 
  E.~Pajer and M.~Peloso,
  Class.\ Quant.\ Grav.\  {\bf 30}, 214002 (2013)
  doi:10.1088/0264-9381/30/21/214002
  [arXiv:1305.3557 [hep-th]].

\bibitem{Banks:2003sx} 
  T.~Banks, M.~Dine, P.~J.~Fox and E.~Gorbatov,
  JCAP {\bf 0306}, 001 (2003)
  doi:10.1088/1475-7516/2003/06/001
  [hep-th/0303252].

\bibitem{ArkaniHamed:2003wu} 
  N.~Arkani-Hamed, H.~C.~Cheng, P.~Creminelli and L.~Randall,
  Phys.\ Rev.\ Lett.\  {\bf 90}, 221302 (2003)
  doi:10.1103/PhysRevLett.90.221302
  [hep-th/0301218].

\bibitem{Kim:2004rp} 
  J.~E.~Kim, H.~P.~Nilles and M.~Peloso,
  JCAP {\bf 0501}, 005 (2005)
  doi:10.1088/1475-7516/2005/01/005
  [hep-ph/0409138].


\bibitem{Dimopoulos:2005ac} 
  S.~Dimopoulos, S.~Kachru, J.~McGreevy and J.~G.~Wacker,
  JCAP {\bf 0808}, 003 (2008)
  doi:10.1088/1475-7516/2008/08/003
  [hep-th/0507205].


\bibitem{Silverstein:2008sg} 
  E.~Silverstein and A.~Westphal,
  Phys.\ Rev.\ D {\bf 78}, 106003 (2008)
  doi:10.1103/PhysRevD.78.106003
  [arXiv:0803.3085 [hep-th]].


\bibitem{McAllister:2008hb} 
  L.~McAllister, E.~Silverstein and A.~Westphal,
  Phys.\ Rev.\ D {\bf 82}, 046003 (2010)
  doi:10.1103/PhysRevD.82.046003
  [arXiv:0808.0706 [hep-th]].



\bibitem{Kaloper:2008fb} 
  N.~Kaloper and L.~Sorbo,
  Phys.\ Rev.\ Lett.\  {\bf 102}, 121301 (2009)
  doi:10.1103/PhysRevLett.102.121301
  [arXiv:0811.1989 [hep-th]].


\bibitem{Marchesano:2014mla} 
  F.~Marchesano, G.~Shiu and A.~M.~Uranga,
  JHEP {\bf 1409}, 184 (2014)
  doi:10.1007/JHEP09(2014)184
  [arXiv:1404.3040 [hep-th]].

\bibitem{Bachlechner:2014gfa} 
  T.~C.~Bachlechner, C.~Long and L.~McAllister,
  JHEP {\bf 1512}, 042 (2015)
  doi:10.1007/JHEP12(2015)042
  [arXiv:1412.1093 [hep-th]].


\bibitem{Kappl:2015esy} 
  R.~Kappl, H.~P.~Nilles and M.~W.~Winkler,
  Phys.\ Lett.\ B {\bf 753}, 653 (2016)
  doi:10.1016/j.physletb.2015.12.073
  [arXiv:1511.05560 [hep-th]].


\bibitem{Choi:2015aem} 
  K.~Choi and H.~Kim,
  Phys.\ Lett.\ B {\bf 759}, 520 (2016)
  doi:10.1016/j.physletb.2016.05.097
  [arXiv:1511.07201 [hep-th]].

\bibitem{Parameswaran:2016qqq} 
  S.~Parameswaran, G.~Tasinato and I.~Zavala,
  JCAP {\bf 1604}, no. 04, 008 (2016)
  doi:10.1088/1475-7516/2016/04/008
  [arXiv:1602.02812 [astro-ph.CO]].


\bibitem{Anber:2009ua} 
  M.~M.~Anber and L.~Sorbo,
  Phys.\ Rev.\ D {\bf 81}, 043534 (2010)
  doi:10.1103/PhysRevD.81.043534
  [arXiv:0908.4089 [hep-th]].
  
\bibitem{Adshead:2012kp} 
  P.~Adshead and M.~Wyman,
  Phys.\ Rev.\ Lett.\  {\bf 108}, 261302 (2012)
  doi:10.1103/PhysRevLett.108.261302
  [arXiv:1202.2366 [hep-th]].
  
\bibitem{Maleknejad:2012fw} 
  A.~Maleknejad, M.~M.~Sheikh-Jabbari and J.~Soda,
  Phys.\ Rept.\  {\bf 528}, 161 (2013)
  doi:10.1016/j.physrep.2013.03.003
  [arXiv:1212.2921 [hep-th]].
  
\bibitem{Barnaby:2010vf} 
  N.~Barnaby and M.~Peloso,
  Phys.\ Rev.\ Lett.\  {\bf 106}, 181301 (2011)
  doi:10.1103/PhysRevLett.106.181301
  [arXiv:1011.1500 [hep-ph]].


\bibitem{Barnaby:2011vw}
  N.~Barnaby, R.~Namba and M.~Peloso,
  JCAP {\bf 1104} (2011) 009
  [arXiv:1102.4333 [astro-ph.CO]].



\bibitem{Meerburg:2012id} 
  P.~D.~Meerburg and E.~Pajer,
  JCAP {\bf 1302}, 017 (2013)
  doi:10.1088/1475-7516/2013/02/017
  [arXiv:1203.6076 [astro-ph.CO]].


\bibitem{Cook:2011hg} 
  J.~L.~Cook and L.~Sorbo,
  Phys.\ Rev.\ D {\bf 85}, 023534 (2012)
  Erratum: [Phys.\ Rev.\ D {\bf 86}, 069901 (2012)]
  doi:10.1103/PhysRevD.86.069901, 10.1103/PhysRevD.85.023534
  [arXiv:1109.0022 [astro-ph.CO]].

\bibitem{Barnaby:2011qe} 
  N.~Barnaby, E.~Pajer and M.~Peloso,
  Phys.\ Rev.\ D {\bf 85}, 023525 (2012)
  doi:10.1103/PhysRevD.85.023525
  [arXiv:1110.3327 [astro-ph.CO]].


\bibitem{Domcke:2016bkh} 
  V.~Domcke, M.~Pieroni and P.~Bintruy,
  JCAP {\bf 1606}, 031 (2016)
  doi:10.1088/1475-7516/2016/06/031
  [arXiv:1603.01287 [astro-ph.CO]].
  
\bibitem{Garcia-Bellido:2016dkw} 
  J.~Garcia-Bellido, M.~Peloso and C.~Unal,
  JCAP {\bf 1612}, no. 12, 031 (2016)
  doi:10.1088/1475-7516/2016/12/031
  [arXiv:1610.03763 [astro-ph.CO]].

\bibitem{Bartolo:2016ami} 
  N.~Bartolo {\it et al.},
  JCAP {\bf 1612}, no. 12, 026 (2016)
  doi:10.1088/1475-7516/2016/12/026
  [arXiv:1610.06481 [astro-ph.CO]].


\bibitem{Sorbo:2011rz} 
  L.~Sorbo,
  JCAP {\bf 1106}, 003 (2011)
  doi:10.1088/1475-7516/2011/06/003
  [arXiv:1101.1525 [astro-ph.CO]].


\bibitem{Crowder:2012ik} 
  S.~G.~Crowder, R.~Namba, V.~Mandic, S.~Mukohyama and M.~Peloso,
  Phys.\ Lett.\ B {\bf 726}, 66 (2013)
  doi:10.1016/j.physletb.2013.08.077
  [arXiv:1212.4165 [astro-ph.CO]].

\bibitem{Linde:2012bt} 
  A.~Linde, S.~Mooij and E.~Pajer,
  Phys.\ Rev.\ D {\bf 87}, no. 10, 103506 (2013)
  doi:10.1103/PhysRevD.87.103506
  [arXiv:1212.1693 [hep-th]].

\bibitem{Bugaev:2013fya} 
  E.~Bugaev and P.~Klimai,
  Phys.\ Rev.\ D {\bf 90}, no. 10, 103501 (2014)
  doi:10.1103/PhysRevD.90.103501
  [arXiv:1312.7435 [astro-ph.CO]].

\bibitem{Erfani:2015rqv} 
  E.~Erfani,
  JCAP {\bf 1604}, no. 04, 020 (2016)
  doi:10.1088/1475-7516/2016/04/020
  [arXiv:1511.08470 [astro-ph.CO]].


\bibitem{Cheng:2015oqa} 
  S.~L.~Cheng, W.~Lee and K.~W.~Ng,
  Phys.\ Rev.\ D {\bf 93}, no. 6, 063510 (2016)
  doi:10.1103/PhysRevD.93.063510
  [arXiv:1508.00251 [astro-ph.CO]].

\bibitem{McDonough:2016xvu} 
  E.~McDonough, H.~B.~Moghaddam and R.~H.~Brandenberger,
  JCAP {\bf 1605}, no. 05, 012 (2016)
  doi:10.1088/1475-7516/2016/05/012
  [arXiv:1601.07749 [hep-th]].


\bibitem{Domcke:2017fix} 
  V.~Domcke, F.~Muia, M.~Pieroni and L.~T.~Witkowski,
  JCAP {\bf 1707}, 048 (2017)
  doi:10.1088/1475-7516/2017/07/048
  [arXiv:1704.03464 [astro-ph.CO]].


\bibitem{Garcia-Bellido:2017aan} 
  J.~Garcia-Bellido, M.~Peloso and C.~Unal,
  JCAP {\bf 1709}, no. 09, 013 (2017)
  doi:10.1088/1475-7516/2017/09/013
  [arXiv:1707.02441 [astro-ph.CO]].


\bibitem{odd-TTT} 
  J.~L.~Cook and L.~Sorbo,
  JCAP {\bf 1311}, 047 (2013)
  [arXiv:1307.7077 [astro-ph.CO]]; 
%
  M.~Shiraishi, A.~Ricciardone and S.~Saga,
  JCAP {\bf 1311}, 051 (2013)
  [arXiv:1308.6769 [astro-ph.CO]]; 
%
  M.~Shiraishi, M.~Liguori and J.~R.~Fergusson,
  JCAP {\bf 1405}, 008 (2014)
  doi:10.1088/1475-7516/2014/05/008
  [arXiv:1403.4222 [astro-ph.CO]]; 
%
  M.~Shiraishi, M.~Liguori and J.~R.~Fergusson,
  JCAP {\bf 1501}, no. 01, 007 (2015)
  doi:10.1088/1475-7516/2015/01/007
  [arXiv:1409.0265 [astro-ph.CO]]; 
%
  N.~Bartolo {\it et al.},
  arXiv:1806.02819 [astro-ph.CO].


\bibitem{Ferreira:2015omg} 
  R.~Z.~Ferreira, J.~Ganc, J.~Norea and M.~S.~Sloth,
  JCAP {\bf 1604}, no. 04, 039 (2016)
  Erratum: [JCAP {\bf 1610}, no. 10, E01 (2016)]
  doi:10.1088/1475-7516/2016/10/E01, 10.1088/1475-7516/2016/04/039
  [arXiv:1512.06116 [astro-ph.CO]].


\bibitem{Peloso:2016gqs} 
  M.~Peloso, L.~Sorbo and C.~Unal,
  JCAP {\bf 1609}, no. 09, 001 (2016)
  doi:10.1088/1475-7516/2016/09/001
  [arXiv:1606.00459 [astro-ph.CO]].


\bibitem{Maleknejad:2011jw} 
  A.~Maleknejad and M.~M.~Sheikh-Jabbari,
  Phys.\ Lett.\ B {\bf 723}, 224 (2013)
  doi:10.1016/j.physletb.2013.05.001
  [arXiv:1102.1513 [hep-ph]].

\bibitem{Maleknejad:2011sq} 
  A.~Maleknejad and M.~M.~Sheikh-Jabbari,
  Phys.\ Rev.\ D {\bf 84}, 043515 (2011)
  doi:10.1103/PhysRevD.84.043515
  [arXiv:1102.1932 [hep-ph]].

\bibitem{Adshead:2012qe} 
  P.~Adshead and M.~Wyman,
  Phys.\ Rev.\ D {\bf 86}, 043530 (2012)
  doi:10.1103/PhysRevD.86.043530
  [arXiv:1203.2264 [hep-th]].


\bibitem{SheikhJabbari:2012qf} 
  M.~M.~Sheikh-Jabbari,
  Phys.\ Lett.\ B {\bf 717}, 6 (2012)
  doi:10.1016/j.physletb.2012.09.014
  [arXiv:1203.2265 [hep-th]].



\bibitem{Dimastrogiovanni:2012st} 
  E.~Dimastrogiovanni, M.~Fasiello and A.~J.~Tolley,
  JCAP {\bf 1302}, 046 (2013)
  doi:10.1088/1475-7516/2013/02/046
  [arXiv:1211.1396 [hep-th]].


\bibitem{Dimastrogiovanni:2012ew} 
  E.~Dimastrogiovanni and M.~Peloso,
  Phys.\ Rev.\ D {\bf 87}, no. 10, 103501 (2013)
  doi:10.1103/PhysRevD.87.103501
  [arXiv:1212.5184 [astro-ph.CO]].

\bibitem{Adshead:2013qp} 
  P.~Adshead, E.~Martinec and M.~Wyman,
  Phys.\ Rev.\ D {\bf 88}, no. 2, 021302 (2013)
  doi:10.1103/PhysRevD.88.021302
  [arXiv:1301.2598 [hep-th]].

\bibitem{Adshead:2013nka} 
  P.~Adshead, E.~Martinec and M.~Wyman,
  JHEP {\bf 1309}, 087 (2013)
  doi:10.1007/JHEP09(2013)087
  [arXiv:1305.2930 [hep-th]].

\bibitem{Papageorgiou:2018rfx} 
  A.~Papageorgiou, M.~Peloso and C.~Unal,
  JCAP {\bf 1809}, no. 09, 030 (2018)
  doi:10.1088/1475-7516/2018/09/030
  [arXiv:1806.08313 [astro-ph.CO]].
 
\bibitem{Maleknejad:2018nxz} 
  A.~Maleknejad and E.~Komatsu,
  arXiv:1808.09076 [hep-ph].
  
\bibitem{Namba:2013kia} 
  R.~Namba, E.~Dimastrogiovanni and M.~Peloso,
  JCAP {\bf 1311}, 045 (2013)
  doi:10.1088/1475-7516/2013/11/045
  [arXiv:1308.1366 [astro-ph.CO]].

\bibitem{Obata:2014loa} 
  I.~Obata, T.~Miura and J.~Soda,
  Phys.\ Rev.\ D {\bf 92}, no. 6, 063516 (2015)
  Addendum: [Phys.\ Rev.\ D {\bf 95}, no. 10, 109902 (2017)]
  doi:10.1103/PhysRevD.95.109902, 10.1103/PhysRevD.92.063516
  [arXiv:1412.7620 [hep-ph]].

\bibitem{Obata:2016tmo} 
  I.~Obata {\it et al.} [CLEO Collaboration],
  Phys.\ Rev.\ D {\bf 93}, no. 12, 123502 (2016)
  Addendum: [Phys.\ Rev.\ D {\bf 95}, no. 10, 109903 (2017)]
  doi:10.1103/PhysRevD.95.109903, 10.1103/PhysRevD.93.123502
  [arXiv:1602.06024 [hep-th]].

\bibitem{Obata:2016xcr} 
  I.~Obata and J.~Soda,
  Phys.\ Rev.\ D {\bf 94}, no. 4, 044062 (2016)
  doi:10.1103/PhysRevD.94.044062
  [arXiv:1607.01847 [astro-ph.CO]].

\bibitem{Caldwell:2017chz} 
  R.~R.~Caldwell and C.~Devulder,
  Phys.\ Rev.\ D {\bf 97}, no. 2, 023532 (2018)
  doi:10.1103/PhysRevD.97.023532
  [arXiv:1706.03765 [astro-ph.CO]].

\bibitem{DallAgata:2018ybl} 
  G.~Dall'Agata,
  Phys.\ Lett.\ B {\bf 782}, 139 (2018)
  doi:10.1016/j.physletb.2018.05.020
  [arXiv:1804.03104 [hep-th]].

\bibitem{Fujita:2018ndp} 
  T.~Fujita, E.~I.~Sfakianakis and M.~Shiraishi,
  arXiv:1812.03667 [astro-ph.CO].

\bibitem{Dimastrogiovanni:2016fuu} 
  E.~Dimastrogiovanni, M.~Fasiello and T.~Fujita,
  JCAP {\bf 1701}, no. 01, 019 (2017)
  doi:10.1088/1475-7516/2017/01/019
  [arXiv:1608.04216 [astro-ph.CO]].

\bibitem{McDonough:2018xzh} 
  E.~McDonough and S.~Alexander,
  arXiv:1806.05684 [hep-th].


\bibitem{Adshead:2016omu} 
  P.~Adshead, E.~Martinec, E.~I.~Sfakianakis and M.~Wyman,
  JHEP {\bf 1612}, 137 (2016)
  doi:10.1007/JHEP12(2016)137
  [arXiv:1609.04025 [hep-th]].
  
\bibitem{Baumann:2008aq} 
  D.~Baumann {\it et al.} [CMBPol Study Team],
  AIP Conf.\ Proc.\  {\bf 1141}, no. 1, 10 (2009)
  doi:10.1063/1.3160885
  [arXiv:0811.3919 [astro-ph]].
  
\bibitem{Agrawal:2017awz} 
  A.~Agrawal, T.~Fujita and E.~Komatsu,
  Phys.\ Rev.\ D {\bf 97}, no. 10, 103526 (2018)
  doi:10.1103/PhysRevD.97.103526
  [arXiv:1707.03023 [astro-ph.CO]].

\bibitem{Agrawal:2018mrg} 
  A.~Agrawal, T.~Fujita and E.~Komatsu,
  arXiv:1802.09284 [astro-ph.CO].

\bibitem{Maleknejad:2016qjz} 
  A.~Maleknejad,
  JHEP {\bf 1607}, 104 (2016)
  doi:10.1007/JHEP07(2016)104
  [arXiv:1604.03327 [hep-ph]].


\bibitem{Thorne:2017jft} 
  B.~Thorne, T.~Fujita, M.~Hazumi, N.~Katayama, E.~Komatsu and M.~Shiraishi,
  arXiv:1707.03240 [astro-ph.CO].
  

\bibitem{Adshead:2018doq} 
  P.~Adshead, J.~T.~Giblin and Z.~J.~Weiner,
  Phys.\ Rev.\ D {\bf 98}, no. 4, 043525 (2018)
  doi:10.1103/PhysRevD.98.043525
  [arXiv:1805.04550 [astro-ph.CO]].


\bibitem{Agrawal:2018gzp} 
  A.~Agrawal,
  Int.\ J.\ Mod.\ Phys.\ D {\bf 28}, no. 02, 1950036 (2018)
  doi:10.1142/S0218271819500366
  [arXiv:1804.01481 [astro-ph.CO]].

\bibitem{Dimastrogiovanni:2018xnn} 
  E.~Dimastrogiovanni, M.~Fasiello, R.~J.~Hardwick, H.~Assadullahi, K.~Koyama and D.~Wands,
  arXiv:1806.05474 [astro-ph.CO].

\bibitem{Fujita:2018vmv} 
  T.~Fujita, R.~Namba and I.~Obata,
  arXiv:1811.12371 [astro-ph.CO].
  


\bibitem{Ferreira:2014zia} 
  R.~Z.~Ferreira and M.~S.~Sloth,
  JHEP {\bf 1412}, 139 (2014)
  doi:10.1007/JHEP12(2014)139
  [arXiv:1409.5799 [hep-ph]].


\bibitem{Namba:2015gja} 
  R.~Namba, M.~Peloso, M.~Shiraishi, L.~Sorbo and C.~Unal,
  JCAP {\bf 1601}, no. 01, 041 (2016)
  doi:10.1088/1475-7516/2016/01/041
  [arXiv:1509.07521 [astro-ph.CO]].

\bibitem{Fujita:2017jwq} 
  T.~Fujita, R.~Namba and Y.~Tada,
  Phys.\ Lett.\ B {\bf 778}, 17 (2018)
  doi:10.1016/j.physletb.2017.12.014
  [arXiv:1705.01533 [astro-ph.CO]].

\bibitem{Himmetoglu:2009qi} 
  B.~Himmetoglu, C.~R.~Contaldi and M.~Peloso,
  Phys.\ Rev.\ D {\bf 80}, 123530 (2009)
  doi:10.1103/PhysRevD.80.123530
  [arXiv:0909.3524 [astro-ph.CO]].
  
  
\bibitem{Ade:2015ava} 
  P.~A.~R.~Ade {\it et al.} [Planck Collaboration],
  Astron.\ Astrophys.\  {\bf 594}, A17 (2016)
  doi:10.1051/0004-6361/201525836
  [arXiv:1502.01592 [astro-ph.CO]].
  
\bibitem{Ade:2018gkx} 
  P.~A.~R.~Ade {\it et al.} [BICEP2 and Keck Array Collaborations],
  Phys.\ Rev.\ Lett.\  {\bf 121}, 221301 (2018)
  doi:10.1103/PhysRevLett.121.221301
  [arXiv:1810.05216 [astro-ph.CO]].


\bibitem{Abazajian:2016yjj} 
  K.~N.~Abazajian {\it et al.} [CMB-S4 Collaboration],
  arXiv:1610.02743 [astro-ph.CO].

\bibitem{Akrami:2018odb} 
  Y.~Akrami {\it et al.} [Planck Collaboration],
  arXiv:1807.06211 [astro-ph.CO].

\end{thebibliography}
\end{document}